\documentclass[aps,prb,twocolumn, eqsecnum]{revtex4-1}

\usepackage{esint, bm, amssymb, amsthm}

\usepackage{hyperref}

\begin{document}


\title{Emergent non-Hermitian boundary contributions to charge pumping and electric polarization}

\author{K. Kyriakou}
\author{K. Moulopoulos}


\affiliation{University of Cyprus, Department of Physics, 1678, Nicosia, Cyprus}


\date{\today}

\begin{abstract}
The phenomenon of charge pumping and the modern theory of electric polarization are reconsidered by analytically taking into account emergent non-Hermitian contributions. These are accounted for through the use of an extended definition of the velocity operator and are determined by means of a dynamic Hellmann–Feynman theorem (DHFT) that we derive here for the first time. The DHFT introduces generalized Berry curvatures and it is valid for calculating observables non-perturbatively, hence with results valid to all orders of the external fields. 
By using the extended velocity operator we rigorously show how the charge pumping is linked up with the boundaries of the material (with the non-Hermiticity being essential for this connection), and by means of the DHFT we show that the well-known topological quantization of the pumped charge breaks down due to a non-integrable Aharonov-Anandan phase in driven non-equilibrium processes whenever the periodic gauge cannot be applied to the Floquet-Bloch states. Likewise, we show that the electronic polarization change has an additional non-Hermitian contribution, which is overlooked in the modern theory of electric polarization.
The non-Hermitian contribution is by definition a bulk quantity that may equally be evaluated as a boundary quantity due to a symmetric structure that allows the bulk integration to be transformed into a boundary one. This non-Hermitian contribution is very sensitive to the realistic boundary conditions imposed on the wavefunctions and it is therefore expected to be significant in biased insulators where charge accumulation over their boundaries is present during the process that causes the polarization change. Finally, we show how a well-defined surface-charge theorem can be formulated in terms of the boundary non-Hermitian contribution.

\end{abstract}

\pacs{}

\maketitle

\section{Introduction}  \label{s1h}
Boundary effects, combined with the involvement of dynamics and topological concepts  have become the cornerstones of the quantum mechanical description of topological condensed matter systems \cite{{roy2017floquet},{caio2015quantum},{kitagawa2010topological}}.  
The interplay between bulk and boundary contributions within a quantum system \cite{{qi2006general}, {wang2017spontaneous}, {hatsugai2016bulk}, {wang2016bulk},{wang2018symmetric}}, seems to be far less clear than the counterpart in a classical system.  A simple quantum theoretical framework that, at the same time takes into account the dynamics, the bulk and boundary contributions of an observable, together with the use of topological concepts, seems to be absent. 
In the theoretical studies of charge pumping and electric polarization, dynamics and bulk contributions, as well as topological concepts, are always explicitly involved. It would therefore be a desirable theoretical advance, if a simple theoretical method could be given that takes into account all involved aspects, together with explicit boundary contributions, in a unified formulation. The present article proposes precisely such an advance in an analytically careful and rigorous manner.

 The central quantity to be calculated in charge pumping studies is the electrons' collective  center of mass displacement 
 \thinspace  ${ \displaystyle 
 \Delta \! \left\langle \mathbf{r} \right\rangle_{\text{coll}}   
 /  N }$  \thinspace
at zero voltage-bias  (absence of mean force) after a time period $T$  that the Hamiltonian has gone a periodic change, where $N$ is the number of electrons that have been displaced.
The adiabatic (equilibrium) charge pumping was first studied by Thouless  in the seminal work of     
Ref. \onlinecite{thouless1983quantization}  and serves as a simple yet fundamental example of topology in quantum systems. It shares the same topological origin as the integrally quantized Hall conductivity \cite{{thouless1982quantized},{kohmoto1985topological}} and may thus be regarded as a dynamical version of the Integer Quantum Hall Effect (IQHE).
This will actually show up in our derivation of the dynamic Hellmann–Feynman theorem (DHFT), where the IQHE and Thouless charge pumping will appear as parts of a single result; this connection appeared analytically in earlier unpublished work \cite{kyriakou2015hellmann} but without the emergent non-Hermitian contribution that is the central focus of the present work.

The hallmark of this effect of (equilibrium) charge pumping is the transport of a precisely quantized amount of charge during an adiabatic cycle in the parameter space.  
This remarkable phenomenon, has been demonstrated by experiments  in a one-dimensional chain of ultra-cold atoms trapped in an optical lattice \cite{{lu2016geometrical},{nakajima2016topological},{lohse2016thouless}}.
On the other hand, driven non-equilibrium (non-adiabatic) particle processes using the Floquet-Bloch bands \thinspace $ \varepsilon_{a}(\mathbf{k})$ \thinspace and the counterpart Floquet states, were first studied in 
Ref.~\onlinecite{shih1994nonadiabatic}  where it was shown that the quantization of the particle transport breaks down due to emergence of band gaps at the quasienergy bands.   
Similar findings were presented in a recent study  \cite{privitera2018nonadiabatic}, where, 
by a careful Floquet analysis 
of a closed, clean, and non-interacting driven Rice-Mele model in the thermodynamic limit,
it was found that the pumped charge deviates from the topologically  quantized value  by an exponentially small amount, provided that the initial state coincides with the lowest-energy Floquet state.  
Specifically, in  
Refs~\onlinecite{{shih1994nonadiabatic}, {privitera2018nonadiabatic}} 
a Hellmann-Feynman theorem for  time-periodic systems\cite{sambe1973steady} has been used (derived from the eigenvalue equation of the so-called Floquet Hamiltonian) that, however,  does not take into account
non-Hermitian, Berry curvature and Aharonov-Anandan phase contributions; in this respect, charge pumping has been studied solely in terms of the spectral properties of Floquet quasienergy alone.
Along these lines but in  a different context (by studying the Floquet operator spectra) \thinspace Ref.~\onlinecite{kitagawa2010topological}  \,  has provided a topological characterization of periodically driven quantum systems in terms of the quasienergy winding alone.

In the development of the modern theory of electric polarization \cite{{resta1992theory}, {king1993theory},{ortiz1994macroscopic},{resta1994macroscopic},{resta2007theory},{resta2010electrical},{vanderbilt2018berry}} for periodic and extended systems, the charge pumping process has played a central role.  
One of the differences between charge pumping and polarization change  processes,  is the voltage-bias   (non-zero mean force) that is present in the latter process while is absent in the former.
Polarization  changes  ${ \Delta \mathbf{P} }$  
have been theoretically realized to be more fundamental quantities
than the \textquotedblleft absolute\textquotedblright   \thinspace bulk polarization \thinspace ${ \mathbf{P} }$ \thinspace itself, and they can be quantified by using the Berry phase of the electronic wavefunctions.
The modern theory avoids addressing the \textquotedblleft absolute\textquotedblright \thinspace 
polarization of a given equilibrium state, quite in agreement with the experiments, which invariably measure polarization differences.
The polarization difference is defined in Ref.~\onlinecite{resta2007theory} as 
the time-integrated transient macroscopic current that flows through the insulating sample during the switching process
\begin{equation}  \label{p1}
\Delta \mathbf{P}= \int_{0}^{T} \mathbf{J}(t)dt,
\end{equation}
provided that the time-dependent Hamiltonian remains insulating at all times. Although \thinspace Eq.~(\ref{p1}) \thinspace is the fundamental equation of modern theory, its differential form 
\begin{equation}  \label{p2}
\frac{d \mathbf{P}(t)}{dt}= \mathbf{J}(t),
\end{equation}
has the same structure as the one that gives the classical polarization current with respect to bound charges   \thinspace ${ \displaystyle \frac{d \mathbf{P_b}(t)}{dt}= \mathbf{J}_b(t) }$.  \vspace{4pt}
On the other hand, by having in mind that in a periodic band insulator the distinction between \textquotedblleft free\textquotedblright \thinspace   (de-localized)
and 
\textquotedblleft bound\textquotedblright \thinspace (localized)
charges is ambiguous,  one can realize that the  polarization difference 
${ \Delta \mathbf{P} }$  given by \thinspace Eq.~(\ref{p1}) \thinspace  must generally have an extra contribution due to de-localized (bound) Bloch states that may be regarded as the 
\textquotedblleft free\textquotedblright \thinspace charge states contribution. We show that such a contribution indeed exists and is captured by an emerging non-Hermitian term that accounts for such delocalized states.
Therefore, we propose that the collective induced electric polarization difference \thinspace $\Delta \mathbf{P}$ \thinspace
 must be defined (assuming non-interacting particles and by deploying the many-body Slater determinant wavefunction) as
\begin{equation} \label{e36bbh}
\Delta \mathbf{P}
=
\frac{1}{V}
\sum \limits_i   q_i  \  
\Delta 
\left\langle
\psi_i(t) \vert 
\, \mathbf{r} \,
\vert
\psi_i(t)
\right\rangle,
\end{equation}
where $ q_i $ and \thinspace
${ \displaystyle  \Delta
\left\langle
\psi_i(t) \vert 
\, \mathbf{r} \,
\vert
\psi_i(t)
\right\rangle }$  
\thinspace
are the charge and the  displacement of each particle respectively, 
and  $V$ is the volume of the material.
In \thinspace Ref. \onlinecite{kyriakou2018orbital} \thinspace
it was shown that, for a periodic and extended system in the thermodynamic limit, even though each electron's position expectation value 
\thinspace
${ \displaystyle 
\left\langle
\psi_i(t) \vert 
\, \mathbf{r} \,
\vert
\psi_i(t)
\right\rangle }$  
\thinspace
is an undefined quantity,  its displacement 
\thinspace
${ \displaystyle  \Delta
\left\langle
\psi_i(t) \vert 
\, \mathbf{r} \,
\vert
\psi_i(t)
\right\rangle }$  
\thinspace
is a well-defined quantity provided that one uses the extended velocity operator definition
\begin{equation}   \label{displac}
\Delta
\left\langle
\psi_i(t) \vert 
\, \mathbf{r} \,
\vert
\psi_i(t)
\right\rangle
=
\int_0^T  \!  \!
\left\langle 
\psi_i(t) \vert \, \mathbf{v}_{ext} 
\, 
\vert
\psi_i(t)
\right\rangle
dt,
\end{equation}
where the expectation values are assumed to be evaluated over the entire volume $V$ of the system.
The extended velocity operator expectation value  
\thinspace
${  \left\langle 
\psi_i(t) \vert \, \mathbf{v}_{ext} 
\, 
\vert
\psi_i(t)
\right\rangle  }$  
\thinspace
has two separate contributions (originating from the additive structure of the extended operator itself, see below  Eq.~(\ref{e1})); one is due to the standard velocity operator  \thinspace
${ \displaystyle \mathbf{v}= \frac{i}{\hbar}\left[ H(\mathbf{r},t),\mathbf{r} \right]  }$, \thinspace
and another due to an emerging non-Hermitian contribution that is attributed to the boundary velocity operator defined as  \thinspace
${ \mathbf{v}_{b}=\frac{i}{\hbar}\!\left( {H(\mathbf{r},t)}^+ -H(\mathbf{r},t) \right)\!\mathbf{r} }$.
The boundary velocity expectation value has the physical meaning of  a position weighted probability flux through the face (boundary) of the material.
For bulk localized states when the wavefunctions are zero over the material boundaries, the boundary velocity expectation value turns to zero;  on the other hand, bulk delocalized  Bloch states give
non-zero boundary velocity expectation value.
In this framework,  \thinspace Eq.~(\ref{e36bbh}) \thinspace together with \thinspace Eq.~(\ref{displac}) \thinspace  may be regarded as an extended theoretical method within the modern theory of polarization, that simultaneously takes into account both  the \textquotedblleft bound\textquotedblright \thinspace  charge contribution (due to localized states) and the  
\textquotedblleft free\textquotedblright \thinspace charge contribution  (due to delocalized states) described by this emergent non-Hermitian effect,  and as such it has not yet been considered in the modern theory. (We shall see later in 
Eq.~(\ref{e7})  that such a contribution is necessary for a complete study of polarization changes and the associated induced currents.)

In order to calculate exactly the boundary and the extended  velocity operator expectation values (without using time-dependent perturbation theory), quantities that, as we shall see, are involved in the charge pumping and polarization change respectively, we have derived in the present work a dynamical Hellmann-Feynman theorem (DHFT). This theorem  can be applied to arbitrary time-dependent states with arbitrary time-dependent parameters, and it is valid for calculating observables up to all orders of the external fields.
The DHFT explicitly involves generalized Berry curvatures as well as boundary contributions due to an emerging non-Hermitian effect term.
 By using this DHFT,  one  is enabled  to study simultaneously in the simplest manner possible the dynamics of the system, the emerging non-Hermitian contributions and the involvement of  topology, in a unified formalism.
The  DHFT can also be employed to non-equilibrium driven quantum processes, \textit{i.e.} to Floquet states,  where the quest for topological invariants is still in an ongoing research state \cite{{kitagawa2010topological},{dehghani2015out},{lindner2011floquet},{carpentier2015topological},{rudner2013anomalous},{budich2017helical},{oka2009photovoltaic},{privitera2018nonadiabatic},{wang2015interband},{kolodrubetz2018topological}}.

With respect to the new DHFT with the inclusion of the non-Hermitian boundary effect (that is here incorporated for the first time into the general foundations of the theory of pumping and polarization), it is useful  to provide some further details on how this has been accomplished: we first re-examine the Thouless adiabatic particle transport (for an insulator with a fully occupied valence band) by using the extended velocity operator definition and the adiabatic limit of the DHFT  without using (first-order) time-dependent perturbation theory.  
Instead of assuming a sliding potential \thinspace ${ V(\mathbf{r}-\upsilon \, t \, \bm{e}_x )}$  \thinspace
like in the original problem \cite{thouless1983quantization},
we assume a microscopic electric field \thinspace
${ \displaystyle E_x^{(micro)}(x, t)=E(t) cos(2\pi x/a) }$ \thinspace
along $\bm{e}_x $ direction. The microscopic electric field is periodic in space (with period equal to the lattice constant $a$)  as well as over time (with period equal to $T$) that (i) originates from a scalar potential that crucially breaks the bulk mirror symmetry of the Hamiltonian along $\bm{e}_x $ direction and (ii) gives zero macroscopic (spatially averaged) electric field, therefore no voltage-bias is present during the pumping. 
By using the emerging non-Hermitian effect boundary velocity operator \thinspace ${ \mathbf{v}_b }$,  \thinspace 
we show that, for delocalized Bloch states, each electron's displacement over the material boundaries can be calculated by means of
\thinspace
${ \displaystyle
\Delta  \left\langle {\mathbf{r}}\right\rangle \! |_{bound}
=
\int_{0}^{T}   \!  \!  \left\langle  \mathbf{v}_{b}  \right\rangle dt  }$.
Given now that each electron's ground eigenstate evolves in time adiabatically, and by using the DHFT, we find that the collective non-Hermitian electrons' displacement in the $x$ direction and over the boundaries  \thinspace 
${  \sum_i^N  \Delta \left\langle x_i \right\rangle \! |_{bound} \ / \ {N}  }$   \thinspace is quantized in units of lattice constant due to topology. Hence, we demonstrate this way that, in the adiabatic limit, the non-Hermitian boundary term is needed in order to obtain the standard topological invariant (in fact, an omission of this would lead to a paradox in topological arguments - see discussion after Eq.~(\ref{e44fh})).
We then relax adiabaticity and study the  non-equilibrium driven charge pumping. We assume delocalized  Floquet-Bloch states \cite{gomez2013floquet} for the electrons' motion,  and use the extended velocity operator definition together with the non-adiabatic form of  DHFT,  in order to evaluate the non-Hermitian collective displacement of the centre of mass of the electrons \thinspace 
${  \sum_i^N  \Delta \left\langle x_i \right\rangle \! |_{bound} \ / \ {N}  }$.  
We find that, for a fully occupied Floquet-Bloch band \thinspace $ \varepsilon_a $, \thinspace the quantization of the pumped charge (per cycle) 
breaks down due to a non-trivial (non-integrable) Aharonov-Anandan phase \cite{aharonov1987phase}.  Particularly, in non-equilibrium driven processes the  particle transport 
deviates from the topologically quantized values as the difference of the Aharonov-Anandan phases at the edges of the Brillouin zone. This driven particle transport breaking, emerges whenever the  Floquet-Bloch states 
\thinspace
${ \Phi_{\varepsilon_{a}}(\mathbf{r}, t, \mathbf{k}) }$
\thinspace
of the electrons cannot satisfy the requirements of a periodic gauge along the direction of the externally applied electric field, that is 
\thinspace
${ 
\Phi_{\varepsilon_{a}}(\mathbf{r}, t, \mathbf{k})
\neq
\Phi_{\varepsilon_{a}}(\mathbf{r}, t, \mathbf{k}+G_x \bm{e}_x)  
}$, 
\thinspace
where the electric field is assumed to be in \thinspace $\bm{e}_x $  \thinspace direction and 
 $G_x  $ is any reciprocal lattice vector in $\bm{e}_x$.

Subsequently, we extend the standard outcomes of the modern theory of polarization by evaluating the electronic polarization change by means of \thinspace Eqs.~(\ref{e36bbh}) \textendash \thinspace (\ref{displac}) \thinspace
and by using the adiabatic form of DHFT.
In this framework,  we rigorously take into account both contributions, the ones due to localized states as well as the ones due to delocalized states.
By using the DHFT we calculate the collective electrons' displacement 
\thinspace
${ \sum_i^N
\Delta \left\langle  \mathbf{r}_i  \right\rangle   }$ 
\thinspace
as a continuous $\mathbf{k}$-space formula,
which in turn provides the induced electronic polarization change
\thinspace $\Delta \mathbf{P}$ \thinspace
according to \thinspace Eq.~(\ref{e36bbh}).
Surprisingly, we find that the adiabatically induced  electronic polarization  change, has two extra, overlooked,  contributions which are absent from the modern theory \cite{{king1993theory},{ortiz1994macroscopic},{resta1994macroscopic},{resta2007theory},{resta2010electrical},{vanderbilt2018berry}}. The one is attributed to a Berry curvature and is not zero only within ferromagnetic insulators where time-reversal symmetry is broken, and the other, is a non-Hermitian boundary contribution that captures the delocalization of the wavefunctions.
The non-Hermitian contribution is by definition a bulk quantity that, however, may equivalently be evaluated as a boundary quantity due to a symmetry that allows the bulk integration to be transformed into a boundary integration. When one takes into account only bulk localized states, that is, the wavefunctions are zero over the material boundaries, the overlooked non-Hermitian contribution always goes to zero.
This non-Hermitian contribution is very sensitive to the realistic boundary conditions of the wavefunctions during the time-periodic process that causes the polarization change. We therefore expect it to be significant in biased insulators with electron charge accumulation over their boundaries during the process that causes the polarization change (i.e. during the spontaneous polarization change experiments in a ferroelectric material,  or during the induced polarization change experiments in a piezoelectric material).

By using the non-Hermitian contribution to the polarization change, we show how one can extend the surface-charge theorem  of  
 Ref.~\onlinecite{vanderbilt1993electric}. Specifically, we argue that a macroscopic surface charge density change theorem can be formulated by means of 
 \begin{equation}  \label{p4}
 \Delta \sigma_{\text{surf}}= \Delta \mathbf{P_{b}}
 \cdot \hat{\mathbf{e}}_{a},
 \end{equation}
 where, 
 \thinspace
 ${ \Delta \mathbf{P_{b}} }$
 \thinspace is the 
 polarization change that is attributed only to boundary properties of the electrons'  wavefunctions and
 \thinspace $ \hat{\mathbf{e}}_{a} $ \thinspace
 is the unit vector normal to the surface of the material.
 When only bulk localized states are taken into account, the surface charge change \thinspace $\Delta \sigma_{\text{surf}}$  \thinspace \thinspace  Eq.~(\ref{p4}) 
\thinspace turns to zero by definition, as expected.
 
We have organized the article as follows. 
In Sec.~\ref{s2h} we review the extended and the boundary velocity operator definitions. 
In Sec.~\ref{DDHFT} we present the dynamic Hellmann-Feynman theorem (DHFT) that we analytically derive.
By then using the boundary velocity operator definition and the DHFT, we study the adiabatic charge pumping in  Sec.~\ref{sATPh} and the counterpart driven and non-equilibrium 
charge pumping in Sec.~\ref{NAPT04}. We then reconsider and re-study the adiabatic 
electronic polarization change in Sec.~\ref{elpol} and finally, we summarize and conclude in Sec.~\ref{concl}.  Some very useful limiting forms of the dynamic Hellmann-Feynman theorem
are derived in Appendix~\ref{DHFT}, while some helpful discrete symmetries
of the curvatures are given in Appendix~\ref{ab}. It is important to also note that, for completeness, the recently discussed boundary integral theory of emergent non-Hermiticities, widely overlooked throughout the history of quantum mechanics (and first noted in ref. 40), is developed {\it in an alternative way} (and within the formulation developed in the present article) in Appendix A.

\section{Extended velocity operator}  \label{s2h}

The extended velocity operator  ${  \mathbf{v}_{ext}}$  was first introduced in \thinspace Ref. \onlinecite{kyriakou2018orbital} \thinspace in order to take into account an anomaly of the position operator $ \mathbf{r} $  that creates a boundary, non-Hermitian effect term, in the Ehrenfest theorem. Namely, by defining the velocity operator as the operator that its expectation value is always equal with the rate of change of the position operator expectation value, it turns out that the velocity must be defined in an extended theoretical framework 
\begin{equation} \label{e1b}
\left\langle  \mathbf{v}_{ext} \right\rangle 
=
\frac{d}{dt}\!\left\langle {\mathbf{r}}\right\rangle,
\end{equation}
where the extended velocity operator has two terms
\begin{equation} \label{e1}
\mathbf{v}_{ext}=\mathbf{v}+\mathbf{v}_{b},
\end{equation}
and the expectation values in \thinspace Eq.~(\ref{e1b}) \thinspace are evaluated over the entire volume $V$ of the system.
 
The first term
\begin{equation} \label{e2}
\mathbf{v}=\frac{i}{\hbar}\left[ H(\mathbf{r},t),\mathbf{r} \right]
\end{equation}
is the standard velocity operator, while the second term 
\begin{equation} \label{e3}
\mathbf{v}_{b}=\frac{i}{\hbar}\!\left( {H(\mathbf{r},t)}^+ -H(\mathbf{r},t) \right)\!\mathbf{r}
\end{equation}
is a new object that we have termed the boundary velocity operator.

The introduction of this new operator $\mathbf{v}_b $ is rather naturally motivated by \thinspace Refs.~\onlinecite{{esteve1986anomalies},{esteve2002origin}} \thinspace and its expectation value is not zero only whenever the position operator becomes anomalous due to the non-Hermitian effect, in which case there are paradoxes first noted in \thinspace Ref. \onlinecite{hill1973paradox}.  
Specifically,  the position operator becomes an anomalous operator whenever the state \thinspace $\mathbf{r}\Psi(\mathbf{r},t)$ \thinspace does not lie within the given Hilbert space of the Hermitian Hamiltonian \thinspace ${ H(\mathbf{r},t) }$.  As consequence,  whenever periodic boundary conditions (PBCs) are adopted for the wavefunction
\thinspace $ \Psi(\mathbf{r},t)$ \thinspace the boundary velocity expectation value ${ \left\langle \mathbf{v}_{b} \right\rangle }$ is not zero. 
Although the expectation value of the boundary velocity operator \thinspace Eq.~(\ref{e3}) \thinspace given by 
\begin{equation}  \label{e3b} 
\left\langle \mathbf{v}_{b} \right\rangle = 
\frac{i}{\hbar} 
( 
\left\langle 
H(\mathbf{r},t) \Psi(t) \vert\, \mathbf{r} \Psi(t) 
\right\rangle 
-
\left\langle
\Psi(t) \vert \, 
H(\mathbf{r},t) \mathbf{r} \Psi(t) 
\right\rangle 
)  
\end{equation} 
is by definition a bulk quantity, through a space-volume integration in position representation (assuming for the moment a 3D system), it can always and equivalently be evaluated as a boundary quantity due to the symmetric structure of the integrands that allows an integration by parts. (Everything however carries out in an obvious manner to other dimensionalities as well, with the results that follow still being valid but with appropriate adjustments, with boundaries being lines (edges) in a 2D and points (ends) in a 1D system).

In this respect, by working in position representation, for real scalar and vector potentials and after a straightforward integration by parts, the expectation value of \thinspace Eq.~(\ref{e3}) \thinspace can be given in the form 
\begin{equation} \label{e4}
\left\langle\mathbf{v}_{b}\right\rangle=
-\oiint_S \mathbf{r} \, (\mathbf{J}_{pr}(\mathbf{r},t)\! \cdot \!d\mathbf{S})
+
\frac{i\hbar}{2m}
\oiint_S \left|\Psi(\mathbf{r},t)\right|^2\!d\mathbf{S}
\end{equation}
with $S$ being the terminated boundary surface of the system  where the boundary conditions are imposed, and \thinspace ${  \mathbf{J}_{pr}(\mathbf{r},t)\!=\!\text{Re}[\Psi(\mathbf{r},t)^{\dagger} \mathbf{v}\,\Psi(\mathbf{r},t)] }$  \thinspace is the standard local probability current density. 
It is to be noted that, apart from the imaginary part (which will always be cancelled out in the total velocity expectation value as we shall see below), the boundary velocity expectation value \thinspace ${      \left\langle \mathbf{v}_b \right\rangle  }$ \thinspace has the meaning of  a position weighted probability flux through the face of the material.

The general form of \thinspace Eq.~(\ref{e4}) \thinspace can be further reduced for periodic systems. Specifically, by assuming a Bloch wavefunction \thinspace ${ \displaystyle  \Psi_{\mathbf{k}}(\mathbf{r}, t) = \frac{1}{\sqrt{N}} \, e^{ \displaystyle i \mathbf{k.r}} u_{\mathbf{k}}(\mathbf{r}, t) }$ \thinspace 
and cell normalization convention 
\thinspace
${\displaystyle
\left\langle \Psi_{\mathbf{k}}(t)\vert \Psi_{\mathbf{k}}(t)\right\rangle
= 
\left\langle u_{\mathbf{k}}(t)\vert u_{\mathbf{k}}(t)\right\rangle_{cell}
=1}$  
\thinspace 
(where $N$ is the total number of the unit cells enclosed within the volume $V$ of the system), then, 
\thinspace Eq.~(\ref{e3b}) \thinspace truncates into the form 
\begin{equation} \label{e3b1}  
\left\langle \mathbf{v}_{b} \right\rangle =
\frac{i}{\hbar}
\left( 
\left\langle 
H_k \, u_{\mathbf{k}}(t) \vert \, \mathbf{r}  
\, u_{\mathbf{k}}(t) 
\right\rangle _{cell} 
-
\left\langle
u_{\mathbf{k}}(t) \vert \, 
H_k \, \mathbf{r}  \, u_{\mathbf{k}}(t) 
\right\rangle _{cell} 
\right)
\end{equation} 
where \thinspace
${H_{k} }$
\thinspace
is given by
\thinspace ${ H_{k} = e^{ \displaystyle -i \mathbf{k.r}}H(\mathbf{r}, t)e^{ \displaystyle i \mathbf{k.r}}  }$.
In deriving \thinspace Eq.~(\ref{e3b1}) \thinspace
we have used the fact 
that \thinspace ${ u_{\mathbf{k}}(\mathbf{r}, t)  }$ \thinspace lies within the domain of definition of the Hamiltonian  ${H_k}$, that is  \thinspace
${\left\langle 
H_{k} \,  u_{\mathbf{k}}(t) \vert \, u_{\mathbf{k}}(t) \right\rangle 
-
\left\langle u_{\mathbf{k}}(t) \vert \, 
H_{k} \,  u_{\mathbf{k}}(t) \right\rangle = 0 }$.
By then exploiting the symmetry of the integrands and performing integration by parts, \thinspace Eq.~(\ref{e3b1}) \thinspace takes the simplified form
\begin{equation}  \label{e3b2}
\left\langle \mathbf{v}_{b} \right\rangle =
-
\oiint_{cell} \!\!\! \mathbf{r} \, (\, \mathbf{J}_{pr}(\mathbf{r}, t,  \mathbf{k})\! \cdot \!d\mathbf{S} \, ) 
\end{equation}
that is valid for periodic systems.

The first term of \thinspace Eq.~(\ref{e4}) \thinspace can be seen as a position-weighted probability flux through the boundaries of the system, while the second and purely imaginary part, cancels a possible imaginary remnant part of the standard velocity operator expectation which is given by
\begin{equation} \label{e5}
\left\langle\mathbf{v}\right\rangle=\iiint_V \mathbf{J}_{pr}(\mathbf{r},t)dV
-\frac{i\hbar}{2m} \oiint_S \left|\Psi(\mathbf{r},t)\right|^2\!d\mathbf{S}
\end{equation} 
(by adding \thinspace Eq.~(\ref{e4}) \thinspace and \thinspace Eq.~(\ref{e5}), that is taking \thinspace
${  
\left\langle\mathbf{v}_{ext}\right\rangle =
\left\langle\mathbf{v}\right\rangle
+
\left\langle\mathbf{v}_{b}\right\rangle 
}$,
we see that ${ \left\langle\mathbf{v}_{ext}\right\rangle }$
is always a real quantity, as should be expected from Eq.~(\ref{e1b})). Incidentally, these elementary facts (never been pointed out in the literature to the best of our knowledge) exemplify the general inadequacy (for certain boundary conditions) of the standard velocity operator in basic quantum mechanics; most importantly for the present work, they display the necessity of defining and using the augmented velocity operator $\mathbf{v}_{ext}$ as a first example (with more demonstrations of this necessity to be discussed in the more focused solid state applications that follow). A point that we wish to make is that this need for the use of 
$\mathbf{v}_{ext}$ is {\it required}  for serious investigations, especially if one wants to have an analytical machinery, that can address issues with possible subtleties (such as the ones that we will confront with in the present work).

The extended velocity operator guarantees that, each Hamiltonian's stationary eigenstate indexed by $n$,  will always produce zero displacement (irrespectively of the static potentials) for the electron
\begin{equation}  \label{e3c}
 \frac{d}{dt}\!\left\langle {\mathbf{r}}\right\rangle_n
 =
 \left\langle\mathbf{v}_{ext}\right\rangle_n
 =\left\langle\mathbf{v}\right\rangle_n
 +
 \left\langle\mathbf{v}_{b}\right\rangle_n
 =0, 
 \end{equation} as expected from the trivial fact that the position operator expectation value $\left\langle {\mathbf{r}}\right\rangle_n$  is a static quantity with respect to any stationary state (since the phase factors containing time multiply to 1).  However, for a stationary and extended plane wave state of a free electron of mass $m$ with well-defined momentum $\hbar \mathbf{k}$ in a finite volume $V$, \thinspace Eq.~(\ref{e3c}) \thinspace is violated (hence a paradox) whenever the boundary velocity expectation value  ${ \left\langle \mathbf{v}_{b} \right\rangle_n }$  is not taken into account (as the standard velocity ${ \left\langle \mathbf{v}_{n} \right\rangle_n } = \hbar \mathbf{k}/m$ is generally nonzero). 
 This apparent paradox is bypassed within the extended velocity operator definition, as it turns out that the boundary velocity contributes a quantity of equal magnitude and opposite sign to the bulk electrons' velocity $ \left\langle \mathbf{v}\right\rangle_n $  resulting in zero displacement ${\displaystyle \Delta{\left\langle\mathbf{r}\right\rangle}_n=0}$ at every instant $t$ for the assumed stationary state. In this framework,  Eq.~(\ref{e3c}) can be used to form a bulk-boundary correspondence in a general sense for every stationary state, namely ${\left\langle\mathbf{v}\right\rangle_n=-\left\langle\mathbf{v}_{b}\right\rangle_n}$ ; this is an example, therefore, of a bulk formulation that properly takes into account boundary currents that are rigorously related to the bulk band structure.
We will see in what follows that similar line of reasoning applied to (and extending) the Hellmann-Feynman theorem will lead to the same conclusion about various systems of modern interests (see discussions after Eq.~(\ref{e18eh})).

By taking into account Eqs.~\ref{e1b} -  \ref{e1} and 
Eq.~\ref{e4}, we can formally define a non-Hermitian displacement of the electron through the sample boundaries  \ ${ \Delta  \left\langle {\mathbf{r}}\right\rangle \! |_{bound} }$  \ by using the  identification
\begin{equation}  \label{e1c}
\left\langle  \mathbf{v}_{b} \right\rangle 
=
\frac{d}{dt}\!\left\langle {\mathbf{r}}\right\rangle \! |_{bound},
\end{equation}
which results into
\begin{equation}  \label{e1d}
\displaystyle
\Delta  \left\langle {\mathbf{r}}\right\rangle \! |_{bound}
=
\int_{0}^{t}   \left\langle  \mathbf{v}_{b}  \right\rangle dt'.
\end{equation}

With the aid of \thinspace Eq.~(\ref{e1}) \textendash \thinspace (\ref{e3}), the extended velocity operator can be recast in the form
\begin{equation}\label{e6}
\mathbf{v}_{ext}=\frac{i}{\hbar}({H(\mathbf{r},t)}^+\mathbf{r} -\mathbf{r}H(\mathbf{r},t))
\end{equation}
which,  by direct application of Eq.~(\ref{e1}) with respect to the Hamiltonian's instantaneous eigenstates labelled by $n$ and $m$, gives the off-diagonal position matrix elements
\begin{eqnarray}  \label{e7} \nonumber
\!  \!  \!  \!  \!  \!  \!  \!  \!  \!  \!  \!  \! 
\displaystyle \frac{i}{\hbar}(E_m(t)-E_n(t))\left\langle m(t)\vert\mathbf{r}\vert n(t)\right\rangle
&=& 
\\* 
\left\langle m(t)\vert\mathbf{v}\vert n(t)\right\rangle 
&+& 
\left\langle m(t)\vert\mathbf{v}_b\vert n(t)\right\rangle  
\end{eqnarray}
that are proportional to the electron's transition dipole moment. We see, therefore, that the emission and absorption of photons can be rigorously related with boundary properties owing to the off-diagonal boundary velocity matrix elements.  
It is thus essential to point out that, \thinspace Eq.~(\ref{e7}),  without the boundary velocity matrix elements taken into account, is generally inadequate to be  employed as a starting point formula - as often done in the modern theory (i.e. in Ref. \onlinecite{vanderbilt2018berry}) - in order to theoretically study the electronic polarization change and the associated induced current, especially if subtle issues need to be addressed for all types of realistic boundary conditions.

\section{Derivation of the Dynamic Hellmann-Feynman theorem (DHFT)}    \label{DDHFT}

We consider a  real vector  parameter  \thinspace  ${ \mathbf{R}  }$  \thinspace that has an  arbitrary  time-dependence (without  any  adiabatic  approximation involved), namely, \thinspace  ${ \mathbf{R}=\bm{\mathcal{R}}(t,\mathbf{R_o})  }$  \thinspace where \thinspace ${ \mathbf{R_o}}$
\thinspace is the initial value of the parameter satisfying \thinspace  ${ \mathbf{R_o}=\bm{\mathcal{R}}(0,\mathbf{R_o})  }$. Therefore, the parameter satisfies the general equation of motion \thinspace
${ \displaystyle 
	\mathbf{R}=\mathbf{R_o} + 
	\int_{0}^{\displaystyle t} \frac{\partial \mathbf{R}}{\partial t'} dt' }$, \thinspace and its time derivative is given by  \linebreak 
\thinspace  ${ \displaystyle  
	\frac{\partial \mathbf{R}}{ \partial t}
	=
	\frac{\partial \bm{\mathcal{R}}(t,\mathbf{R_o})}{ \partial t}
}$. 
The  theorem that we are about to prove   is  for  a  continuous  vector  parameter \thinspace  ${ \mathbf{R}  }$, \thinspace the initial value \thinspace  ${ \mathbf{R_o}  }$ \thinspace  been therefore assumed to have continuous  values as well.
The  Hamiltonian of the system \thinspace 
${ H(t, \mathbf{R}) }$, \thinspace 
apart from the implicit time-dependence (via the parameters) may  also  have  an  arbitrary  explicit  time-dependence (something that is often not discussed).  The  derivation that is  given  below owes its existence to the  Hamiltonian  being  the  generator  of  time  evolution  of  quantum  states.  We  provide  the  derivation  for  a  single-particle  state  while  the  generalization  to  a  many-particle  system  is  straightforward. A particle's  motion  is  generally  encoded  in its normalized  time-dependent  state \thinspace   ${ \left| \Psi(t, \mathbf{R}) \right\rangle  }$  \thinspace  which  evolves  either  by  the time-dependent  Schrödinger  equation  for  non-relativistic  and  spinless  particle,  or  by  the time-dependent  Dirac equation  for  spinful   particle.  We assume for simplicity a one particle quantum system. The motion of the particle is described  by a general  state, not necessarily an  eigenstate  of  the  Hamiltonian  nor  a  localized   state  (such as a  narrow wave  packet).  The   system is assumed to be closed 
\thinspace 
${ \left\langle 
	\Psi(t,\mathbf{R} )   
	\vert \Psi(t,\mathbf{R} )
	\right\rangle=1 
}$, 
\thinspace 
and the quantum state  time evolution  is  determined  by  the   time-dependent  equation
\begin{equation} \label{e1h}
i \hbar \frac{d}{dt}\left| \Psi(t, \mathbf{R}) \right\rangle
=
H(t, \mathbf{R}) 
\left| \Psi(t, \mathbf{R}) \right\rangle, 
\end{equation}
where  the  Hamiltonian  is  either  of  Schrödinger  or  Dirac  type.
The time derivative in \thinspace Eq.~(\ref{e1h})  \thinspace is the total time derivative given by
\begin{equation} \label{e2h}
\frac{d}{dt}
=
\frac{\partial}{\partial t}
+
\frac{\partial \mathbf{R}}{\partial t} \! \cdot \! 
\bm{\nabla_{\mathbf{R}}}
\end{equation}  
\\
where \thinspace
${ \displaystyle
	\bm{\nabla_{\mathbf{R}}}
	=
	\sum_{i=1}^{3} 
	\mathbf{e}_i
	\frac{\partial}{\partial R_i}
}$. The  initial  value  of  the  parameter 
\thinspace  ${ \mathbf{R_o}  }$ \thinspace 
that implicitly enters \thinspace Eq.~(\ref{e1h}) \thinspace  can be  used  to  label the quantum states  \thinspace   ${ \left| \Psi(t, \mathbf{R}) \right\rangle  }$. 
The  expectation  value   of  the  Hamiltonian   
\begin{equation} \label{e3h}
\left\langle \Psi (t,\mathbf{R}) 
\right|  H(t, \mathbf{R}) \left.  \!
\Psi (t, \mathbf{R})\right\rangle 
= 
E(t, \mathbf{R})
\end{equation}
can  be  viewed as  the  instantaneous  time-dependent  
\textquotedblleft energy\textquotedblright   
\thinspace of  the  particle \thinspace
${ E(t, \mathbf{R}) }$.
Differentiation  with  respect  to  the  parameter \thinspace  ${ \mathbf{R} }$  \thinspace of  both  sides of \thinspace Eq.~(\ref{e3h}) \thinspace  gives
\begin{equation} \label{e4h}
\left\langle \Psi \, 
\right|  
\left( \bm{\nabla_{\mathbf{R}}}
H \right)
\left.  
\Psi \right\rangle 
= 
\bm{\nabla_{\mathbf{R}}} E
-
\left\langle \bm{\nabla_{\mathbf{R}}} \Psi \,  \vert \, 
H \Psi \right\rangle 
-
\left\langle  \Psi \vert \, 
H \, \bm{\nabla_{\mathbf{R}}} \Psi  \right\rangle. 
\end{equation}
Taking now into account that the parameter gradient operator \thinspace  ${ \bm{\nabla_{\mathbf{R}}} }$  \thinspace  can generally be an anomalous operator, that is, the states \thinspace  ${ \left| \bm{\nabla_{\mathbf{R}}} \Psi  \right\rangle  }$  \thinspace may not fall within the domain of the Hermitian Hamiltonian, something that can be expressed by the non-trivial inequality \linebreak
${ \left\langle H \Psi \, \vert \,  
	\bm{\nabla_{\mathbf{R}}}   \Psi \right\rangle 
	= 
	\left\langle \Psi \, \vert \,  
	H^{+} \, 
	\bm{\nabla_{\mathbf{R}}} 
	\Psi
	\right\rangle
	\neq  
	\left\langle \Psi \, \vert \,   
	H \, 
	\bm{\nabla_{\mathbf{R}}} 
	\Psi \right\rangle }$,
we recast \thinspace Eq.~(\ref{e4h}) \thinspace into the form
\begin{eqnarray} \label{e5h}  \nonumber
\left\langle \Psi \, 
\right|  
\left( \bm{\nabla_{\mathbf{R}}}
H \right)
\left. 
\Psi \right\rangle 
&=&  
\bm{\nabla_{\mathbf{R}}} E
-
\left\langle \bm{\nabla_{\mathbf{R}}} \Psi \,  \vert \, 
H \Psi \right\rangle 
-
\left\langle  H \Psi \vert \, 
\bm{\nabla_{\mathbf{R}}} \Psi  \right\rangle 
\\*
&&
+
\left\langle \Psi \, \vert \,  
\left( 
H^+ 
- 
H 
\right) 
\bm{\nabla_{\mathbf{R}}} 
\Psi \right\rangle. 
\end{eqnarray}
By then using \thinspace Eq.~(\ref{e1h}) \thinspace in \thinspace Eq.~(\ref{e5h}) \thinspace we find 
\begin{eqnarray} \label{e6h} \nonumber
\left\langle \Psi \, 
\right|  
\left( \bm{\nabla_{\mathbf{R}}}
H \right) 
\left. 
\Psi \right\rangle 
&=& 
\bm{\nabla_{\mathbf{R}}} E
\, + \,
\left\langle \Psi \, \vert   
\left( 
H^+ 
- 
H 
\right) 
\bm{\nabla_{\mathbf{R}}} 
\Psi \right\rangle
\\* [4pt]  \nonumber
&&
-
i \hbar
\left( 
\left\langle \bm{\nabla_{\mathbf{R}}} \Psi \,  \vert \, 
\frac{d\Psi }{dt} \right\rangle 
-
\left\langle  \frac{d\Psi }{dt} \vert \, 
\bm{\nabla_{\mathbf{R}}} \Psi  \right\rangle
\right).
\\*
\end{eqnarray}
Applying then the total time derivative \thinspace Eq.~(\ref{e2h}) \thinspace on  \thinspace Eq.~(\ref{e6h}) \thinspace we obtain
\begin{eqnarray}  \label{e7h}  \nonumber
\left\langle \Psi \, 
\right|  
\left( \bm{\nabla_{\mathbf{R}}}
H \right) 
\left. 
\Psi \right\rangle 
&=& 
\bm{\nabla_{\mathbf{R}}} E
\, + \,
\left\langle \Psi \, \vert  
\left( 
H^+ 
- 
H 
\right) 
\bm{\nabla_{\mathbf{R}}} 
\Psi \right\rangle
\\*[6pt]  \nonumber
&&
-
i \hbar
\left( 
\left\langle \bm{\nabla_{\mathbf{R}}} \Psi \,  \vert \, 
\frac{\partial \Psi }{ \partial t} \right\rangle 
-
\left\langle  \frac{ \partial \Psi }{\partial t} \vert \, 
\bm{\nabla_{\mathbf{R}}} \Psi  \right\rangle
\right)
\\*  [6pt] \nonumber 
&&
\!  \!  \!  \!  \!  \!  \!  \!  \!  \!  \!  \!  \!  \!  \!  \!  \!  \!  \!  \!  \!  \!  \!  \! 
\!  \!  \!  \!  \!  \!  \!  \!  \!  \! 
-
i \hbar
\left( 
\left\langle \bm{\nabla_{\mathbf{R}}} \Psi \,  \vert \, 
\frac{\partial \mathbf{R}}{\partial t}
\! \cdot \! \bm{\nabla_{\mathbf{R}}} \Psi
\right\rangle 
-
\left\langle  
\frac{\partial \mathbf{R}}{\partial t}
\! \cdot \! \bm{\nabla_{\mathbf{R}}} \Psi
\vert \, 
\bm{\nabla_{\mathbf{R}}} \Psi  \right\rangle
\right). 
\\*  
\end{eqnarray}
Using then the vector identity \ \smallskip  
${  \displaystyle 
	\frac{\partial \mathbf{R}}{\partial t}
	\times ( \mathbf{A} \times \mathbf{B})
	=
	\mathbf{A}
	\left( \frac{\partial \mathbf{R}}{\partial t} \! \cdot \! \mathbf{B} \right)   
	-
	\left( \frac{\partial \mathbf{R}}{\partial t} \! \cdot \! \mathbf{A}  \right)  
	\mathbf{B}
}$  in the last term on the right hand side of
\thinspace Eq.~(\ref{e7h}), \thinspace 
we finally obtain our new dynamic Hellmann-Feynman theorem (DHFT), namely
\begin{eqnarray}  \label{e8h}   \nonumber
\left\langle \bm{O}(t,\mathbf{R}) \right\rangle
&=&
\bm{\nabla_{\mathbf{R}}} E(t, \mathbf{R})
\, + \,
\bm{\mathcal{S}}(t, \mathbf{R}) 
\\* [6pt]
&&
- \,  \hbar \,
\bm{\mathcal{E}}(t, \mathbf{R})
- 
\hbar \,
\frac{\partial \mathbf{R}}{\partial t}
\! \times 
\bm{\mathcal{B}}(t, \mathbf{R}),
\end{eqnarray}
where
\begin{equation} \label{e9bh}
\left\langle \bm{O}(t,\mathbf{R}) \right\rangle
=
\left\langle \Psi(t, \mathbf{R}) \, 
\right|  
\left( \bm{\nabla_{\mathbf{R}}}
H(t, \mathbf{R})  \right)
\left. 
\Psi(t, \mathbf{R}) \right\rangle 
\end{equation}
is the observable in quest, whereas
\begin{equation} \label{e9h}
\bm{\mathcal{B}}(t, \mathbf{R})
= i
\left\langle \bm{\nabla_{\mathbf{R}}}
\Psi(t, \mathbf{R}) \right| 
\times 
\left|  \bm{\nabla_{\mathbf{R}}}
\Psi(t, \mathbf{R}) \right\rangle 
\end{equation} 
 and
\begin{eqnarray} \label{e10h} \nonumber
\bm{\mathcal{E}}(t, \mathbf{R})
&=&
i
\left\langle \bm{\nabla_{\mathbf{R}}} \Psi(t, \mathbf{R}) \,  \vert \, 
\frac{\partial \Psi(t, \mathbf{R}) }{ \partial t} \right\rangle  
\\* [2pt]
&&
-
i
\left\langle  \frac{ \partial \Psi(t, \mathbf{R}) }{\partial t} \vert \, 
\bm{\nabla_{\mathbf{R}}} \Psi(t, \mathbf{R})  \right\rangle
\end{eqnarray}
are generalized Berry curvatures (of both `magnetic' and `electric' type) in the \thinspace  ${ t\! \times \!\mathbf{R} }$ \thinspace space, each of them being a real quantity as manifest from their definition. 

The \thinspace ${  \bm{\mathcal{S}}(t, \mathbf{R})  }$
\thinspace term is an emergent non-Hermitian contribution to the above theorem which is by definition a bulk quantity.  However, as shown in Appendix~\ref{BC},  by working in position representation and assuming real scalar and vector potentials, the \thinspace ${  \bm{\mathcal{S}}(t, \mathbf{R})  }$
\thinspace term is always transformed (due to symmetric form of the integrands) and evaluated by means of a boundary integral over the system's boundaries and 
is given (assuming a 3D system, although this can trivially be generalized) by
\begin{eqnarray}  \label{e11h}  \nonumber
\bm{\mathcal{S}}(t, \mathbf{R}) 
&=& 
\left\langle \Psi(t, \mathbf{R}) \, \vert \,  
\left( 
H(t, \mathbf{R})^+ 
- 
H(t, \mathbf{R}) 
\right)  
\bm{\nabla_{\mathbf{R}}} 
\Psi(t, \mathbf{R}) \right\rangle 
\\* [4pt]  \nonumber
&&
\! \! \! \! \! \!  \! \! \! \! \!  \! \! \! \! \! \! \!  \! \! \! \! \! \! 
\! 
= 
\frac{i \hbar}{2} \oiint_S 
\mathbf{n}\!\cdot\! 
\left( \,
( \mathbf{v} \, \Psi(t, \mathbf{R}) )^{+} 
+ \Psi(t, \mathbf{R})^{+} \, \mathbf{v}  
\, \right) \! 
\, \bm{\nabla_{\mathbf{R}}} \Psi(t, \mathbf{R})
\, dS,
\\* [-2pt]
\end{eqnarray}
where \thinspace 
${ \displaystyle 
	\mathbf{v}=\frac{i}{\hbar}
	\left[ H(t, \mathbf{R}),\mathbf{r} \right]
}$
\thinspace 
is the standard velocity operator and \thinspace ${ \mathbf{n} }$ \thinspace  
is the unit vector that is locally normal to the surface that encloses the system.
As shown in Appendix~\ref{BC}, the boundary integral form of the non-Hermitian contribution 
\thinspace ${  \bm{\mathcal{S}}(t, \mathbf{R})  }$  \thinspace
given by  \ Eq.~(\ref{e11h}), is valid for every energy scale of the particle, that is, either the time evolution is governed by  a (i) spinless non-relativistic Hamiltonian, or (ii) a spinful Dirac Hamiltonian, or a (iii) a spinful Hamiltonian in the non-relativistic limit.
In all above cases,   the non-Hermitian contribution 
\thinspace ${  \bm{\mathcal{S}}(t, \mathbf{R})  }$  is extremely sensitive  to the realistic boundary conditions on the wavefunctions. 
Moreover, \thinspace ${  \bm{\mathcal{S}}(t, \mathbf{R})  }$ \thinspace is a real quantity whenever the operator \thinspace
${ \bm{\nabla_{\mathbf{R}}}H  }$ \thinspace
behaves as a Hermitian operator. This can be shown 
by starting from a closed system assumption \thinspace
${ \left\langle \Psi (t,\mathbf{R}) 
\right|  H \left.  \!
\Psi (t, \mathbf{R})\right\rangle 
-
\left\langle  H  \Psi (t,\mathbf{R}) 
\right|  \left.  \!
\Psi (t, \mathbf{R})\right\rangle 
=0
  }$,  \thinspace
and then differentiate  both  sides with  respect  to  the  parameter \thinspace  ${ \mathbf{R} }$  \thinspace  giving
\begin{equation}  \label{a201}
\bm{\mathcal{S}}(t, \mathbf{R})
=
\bm{\mathcal{S}}(t, \mathbf{R})^{\displaystyle *}
+
\left( \,
\left\langle \Psi \, 
\right|  
\left( \bm{\nabla_{\mathbf{R}}}
H \right)
\left.  
\Psi \right\rangle 
-
\left\langle \left( \bm{\nabla_{\mathbf{R}}}H \right)\,  \Psi \, 
\right|  
\left.  \!
\Psi \right\rangle 
\, \right).
\end{equation}
Interestingly, whenever the assumed Hamiltonian does not have explicit parameter dependence, but the considered states do depend on the parameters (as i.e. it happens for the original $H$ in a periodic system (before passing to the Bloch Hamiltonian), which does not depend on the Bloch wavevector {\bf k}, although its eigenstates, the standard Bloch states, do), then, the non-Hermitian contribution is always a real quantity; in the example given above, the non-Hermitian contribution evaluated with respect to a Bloch state by identifying ${  \mathbf{R}=\mathbf{k} }$ is always a real quantity.  
Eq.~(\ref{a201}) will be crucial in properties such as the surface-charge theorem discussed towards the end of the paper (see Eq.~(\ref{sc})).

Due to the structure of the DHFT, the last term on the right side of  \thinspace Eq.~(\ref{e8h}) \thinspace gives a contribution that is transverse to
the direction of the parameter variation \thinspace ${ \partial \mathbf{R} }$, \thinspace namely,
${ \displaystyle
\left\langle \bm{O}(t,\mathbf{R}) \right\rangle_{tran}
=
- \hbar \,
\frac{\partial \mathbf{R}}{\partial t}
\! \times 
\bm{\mathcal{B}}(t, \mathbf{R})
}$
that satisfies \thinspace 
${ \left\langle \bm{O}(t,\mathbf{R}) \right\rangle_{tran} \cdot \partial \mathbf{R}=0  }$.  This occurs in IQHE, but this term is also expected to contribute significantly to non-linear quantum processes whenever the assumed parameter varies with time, \textit{i.e} within non-linear materials where longitudinal as well as transverse polarization changes occurs simultaneously.

\subsection{Application of the DHFT to a state with ansatz form 
	$\displaystyle \left| \Psi(t, \mathbf{R}) \right\rangle 
	=
	e^{ 
		\displaystyle i \Theta(t, \mathbf{R})}
	\left| 
	\Phi(t, \mathbf{R})
	\right\rangle $.} 
\label{DHFTa}

When the considered quantum state, assumed to be a solution of \thinspace Eq.(\ref{e1h}),  \thinspace
has the ansatz form
\begin{equation}  \label{pp1} 
	\left| \Psi(t, \mathbf{R}) \right\rangle 
	=
	e^{ \displaystyle i \Theta(t, \mathbf{R})}
	\left| 
	\Phi(t, \mathbf{R})
	\right\rangle,
\end{equation}
then, as shown in Appendix \ref{DHFT},
the DHFT  \thinspace Eq.~(\ref{e8h}) \thinspace  retains its structure form but all involved quantities are now evaluated with respect to the quantum state 
$\left| 
\Phi(t, \mathbf{R})
\right\rangle $, 
namely
\begin{eqnarray}  \label{e14hab}  \nonumber  
	\left\langle \bm{O}(t,\mathbf{R}) \right\rangle
	&=&
	\bm{\nabla_{\mathbf{R}}} E_{\Phi}(t, \mathbf{R})
	\, + \,
	\bm{\mathcal{S}}_{\Phi}(t, \mathbf{R}) 
	\\* [4pt]
&&
\! \! \! \! \! \!
	- \hbar \,
	\frac{\partial \mathbf{R}}{\partial t}
	\! \times 
	\bm{\mathcal{B}}_{\Phi}(t, \mathbf{R})
	\, - \,  \hbar \,
	\bm{\mathcal{E}}_{\Phi}(t, \mathbf{R}).
\end{eqnarray}
In other words, the observables in quest 
${ \left\langle \bm{O}(t,\mathbf{R}) \right\rangle }$
given by \thinspace Eq.~(\ref{e9bh}) \thinspace
are insensitive to phases 
\thinspace
${ e^{ \displaystyle i \Theta(t, \mathbf{R})} }$
\thinspace
that do not depend explicitly on the position coordinate 
${ \mathbf{r} }$.  Such simple applications of the DHFT 
are described below:

For an adiabatically evolved eigenstate of a time-dependent Hamiltonian, we replace in \thinspace Eq.(\ref{pp1}),  \thinspace
${ \displaystyle
	\Theta(t, \mathbf{R}) 
	\mapsto
	\Theta_n(t, \mathbf{R}) 
}$
\thinspace
and
\thinspace
${
\left| 
\Phi(t, \mathbf{R})
\right\rangle
\mapsto
\left| 
n(t, \mathbf{R})
\right\rangle
}$, 
where 
\thinspace ${\Theta_n(t, \mathbf{R}) }$ \thinspace is the total (sum of dynamic and geometric) phase and 
\thinspace 
${ \left| 
	n(t, \mathbf{R})
	\right\rangle  }$
\thinspace is the instantaneous eigenstate of the Hamiltonian satisfying the eigenvalue equation 
\thinspace
${
H(t, \mathbf{R}) \left| 
n(t, \mathbf{R})
\right\rangle 
=
E_n(t, \mathbf{R}) \left| 
n(t, \mathbf{R})
\right\rangle.
}$

For a stationary eigenstate of a static Hamiltonian where the assumed parameter is static 
\thinspace  ${ \displaystyle  
\frac{\partial \mathbf{R}(t)}{ \partial t} = 0
}$, \thinspace
we replace in \thinspace Eq.(\ref{pp1}),  \thinspace
${ \displaystyle
	\Theta(t, \mathbf{R}) 
	\mapsto
	\Theta_n(t, \mathbf{R}) 
}$
\thinspace
and
\thinspace
${
	\left| 
	\Phi(t, \mathbf{R})
	\right\rangle
	\mapsto
	\left| 
	n(\mathbf{R})
	\right\rangle
}$, 
where 
\thinspace ${\Theta_n(t, \mathbf{R}) }$ \thinspace is the dynamic phase and 
\thinspace 
${ \left| 
n(\mathbf{R})
\right\rangle  }$
\thinspace is the instantaneous eigenstate of the Hamiltonian satisfying the eigenvalue equation 
\thinspace
${
H( \mathbf{R}) \left| 
n(\mathbf{R})
\right\rangle 
=
E_n(\mathbf{R}) \left| 
n(\mathbf{R})
\right\rangle.
}$
The Berry curvature 
${  \bm{\mathcal{E}}_n (\mathbf{R})    }$
is zero by definition according to  
Eq.~(\ref{e16ha})  owing to
${  \left| 
	n(\mathbf{R})
\right\rangle 
}$
that does not have explicit time dependence,
whereas the Berry curvature 
${  \bm{\mathcal{B}}_n (\mathbf{R})    }$
is not zero but  it does not contribute into the DHFT   \thinspace Eq.~(\ref{e14hab}) due to zero velocity of the parameter.

For a non-equilibrium and nonadiabatic quantum state where
 the assumed parameter may have arbitrary  high magnitude of velocity \thinspace  ${ \displaystyle  
\frac{\partial \mathbf{R}(t)}{ \partial t}, 
}$ \thinspace 
that, however, evolves in time by a time-periodic Hamiltonian
\thinspace
${ H(t+T, \mathbf{R}(t+T))=H(t, \mathbf{R}(t))
}$, 
\thinspace 
where  \thinspace 
${ \mathbf{R}(t+T)=\mathbf{R}(t) }$  \thinspace
and \thinspace ${ T }$ \thinspace is the period of driving, 
 we replace in \thinspace Eq.(\ref{pp1})
 \thinspace
${
	\displaystyle
	\Theta(t, \mathbf{R}) 
	\mapsto
	- \frac{\varepsilon_{a} \, t}{\hbar} 
}$
\thinspace
and
\thinspace  
${
\left| 
\Phi(t, \mathbf{R})
\right\rangle
\mapsto
\left| 
\Phi_a(t, \mathbf{R})
\right\rangle
}$,
\thinspace
where \thinspace 
${ \left| 
	\Phi_a(t, \mathbf{R})
	\right\rangle   }$ \thinspace
is the Floquet mode which is periodic in time \thinspace 
${ \left| 
	\Phi_a(t, \mathbf{R})
	\right\rangle
	=
	\left| 
	\Phi_a(t+T, \mathbf{R})
	\right\rangle
}$
and satisfies the eigenvalue equation
\thinspace 
${
\displaystyle  
( 
H(t,  \mathbf{R})
-i \hbar \frac{d}{dt}
\, ) 
\Phi_a(\mathbf{r}, t, \mathbf{k})
=
\varepsilon_a 
\,
\Phi_a(\mathbf{r}, t, \mathbf{k}),
}$
with ${ \varepsilon_a }$ \thinspace being the static quasienergy.
In conclusion, we argue that the DHFT
\thinspace Eq.~(\ref{e14hab}) \thinspace  can be applied to general topological band theory.

\subsection{The need for the non-Hermitian term  in the DHFT}   \label{statlimith}

The simplest example where one can ascertain the necessity  of presence of the boundary non-Hermitian term \thinspace ${ \bm{\mathcal{S}}(t, \mathbf{R})  }$ \thinspace is a free electron motion in 1D. 
In this respect, we assume the Hamiltonian \linebreak ${\displaystyle  H=-\frac{\hbar^2}{2m} \frac{d^2}{dx^2}  }$ \thinspace which does not have any parameter dependence. We consider an electron's motion described by the eigenstate 
\thinspace ${ \displaystyle \Phi_k (x) = \frac{1}{\sqrt{L}} e^{\displaystyle ikx} }$
\thinspace
of the Hamiltonian
with well defined energy  \thinspace ${\displaystyle 
E(k)=\frac{\hbar^2 k^2 }{2m} }$,
where $L$ is the length of the system. We identify  the static electron's 1D  wave vector $k$  (assumed to take continuous values) as the parameter ${ R \equiv k}$ and examine the necessity of a nonzero ${S(k)}$.

Specifically, by employing the standard static HF theorem without the non-Hermitian term, that is, the 1D analog of \thinspace Eq.~(\ref{a8}) \thinspace derived in Appendix~\ref{a22} without the boundary term ${S(k)}$,  will eventually lead to a paradox since \thinspace ${ \displaystyle 
	\left\langle \frac{dH}{dk} \right\rangle
	=0 
}$ \thinspace while 
\thinspace ${ \displaystyle 
\frac{d E(k)}{dk}
=
\frac{d}{dk}
\left( 
\frac{\hbar^2 k^2 }{2m}
\right) 
=
\frac{\hbar^2 k}{m} \neq 0
}$. 
The  paradox is resolved whenever one takes into account the boundary non-Hermitian term ${S(k)}$. Specifically, the 1D analogue of \thinspace 
Eq.~({\ref{a8}}) \thinspace is given by 
\begin{equation} \label{e18bh}
0 = \frac{dE(k)}{dk} + S(k),
\end{equation}  
where ${S(k)}$ is in 1D truncated to a two point formula (\textit{cf}. Eq.~(\ref{e17ha})) given by
\begin{equation} \label{e18ch}
S(k)
=
\frac{i \hbar}{2} 
\left[ 
\left( \,
( \text{v} \, \Phi_k(x) )^{ \displaystyle *} 
+ \Phi_k(x) ^{\displaystyle * } \, \text{v}  
\, \right) \! 
\, \frac{d \Phi_k (x)}{dk} 
\right]_{x}^{x+L}. 
\end{equation}  
We substitute in \thinspace Eq.~(\ref{e18ch}) \thinspace the standard velocity operator \thinspace ${ \displaystyle \text{v}=  -\frac{i \hbar}{m} \frac{\partial}{\partial x} }$, \thinspace as well as \thinspace
$  \text{v} \, \Phi_k (x) =
\displaystyle 
\frac{1}{\sqrt{L}}
\frac{\hbar k}{m}
e^{\displaystyle ikx}
=
\frac{\hbar k}{m} \Phi_k (x)
$
\thinspace and \thinspace
${ \displaystyle 
	\frac{d \Phi_k (x)}{dk}
	=
	\frac{1}{\sqrt{L}}
	i \, x \,
	e^{\displaystyle ikx}
	=
	i \, x \, \Phi_k (x)
}$. \thinspace
\thinspace Eq.~(\ref{e18ch}) \thinspace then takes the value 
\begin{eqnarray} \label{e18dh} \nonumber 
S(k)
&=&
\frac{i \hbar}{2 } 
\left[ 
\left( \,
\frac{\hbar k}{m} \Phi_k(x) ^{\displaystyle *} 
-
\frac{i \hbar}{m}
\Phi_k(x)^{\displaystyle *} \,  \frac{\partial}{\partial x}
\, \right) \! 
\, 
i \, x \, \Phi_k(x)
\right]_{x}^{x+L}
\\*[6pt]  \nonumber 
&=&
\frac{i \hbar}{2 } 
\left[ 
\frac{i \, \hbar \, k}{m} |\Phi_k (x)|^2 x
+
\frac{\hbar }{m} |\Phi_k (x)|^2 \right. 
\\* \nonumber 
&&
\left. 
+i \, x \,
\Phi_k^{\displaystyle *} \,
( \text{v} \, \Phi_k (x) )
\frac{}{} 
\right] _{x}^{x+L}
\\*[4pt]   \nonumber
&=& 
\frac{i \hbar}{2 } 
\left[
2 
\frac{i \, \hbar \, k}{m} |\Phi_k (x)|^2 x
+
\frac{\hbar}{m} |\Phi_k (x)|^2
\right]_{x}^{x+L}
\\*[4pt]   \nonumber
&=& 
\frac{i \hbar^2 }{ 2 m  L } 
\left[
2 \, i \, k \, x
+
1
\frac{}{} \right]_{x}^{x+L}
\\*[4pt]   \nonumber
&=& 
\frac{i \hbar^2 }{ 2 m L} (2 \, i \, k \, L)
= 
-\frac{\hbar^2 k}{m}, 
\end{eqnarray} 
which together with   
\thinspace ${ \displaystyle 
	\frac{d E(k)}{dk}
	=
	\frac{\hbar^2 k}{m} 
}$ \thinspace 
verifies \thinspace Eq.~(\ref{e18bh}) \thinspace and guarantees the validity of the static HF theorem. Whenever, therefore, the momentum derivative operator ${\displaystyle  \frac{d}{dk} }$ becomes anomalous, the boundary non-Hermitian contribution cannot be excluded from the standard, static HF theorem;
a similar kind of anomaly as in
\thinspace 
Eq.~({\ref{e18bh}}), \thinspace
was noticed in
Ref. \onlinecite{esteve2010generalization} but, notably,
without indicating any usage into conventional band theory.
The above non-Hermitian contribution satisfies the antisymmetry 
\thinspace ${ S(k)= - S(-k) }$,  \thinspace which,
as shown in Appendix~\ref{ab},  is a general property of band theory (recall that the non-Hermitian contribution is always evaluated with respect to a Bloch state,  Eq.~(\ref{e11h})) when the Hamiltonian satisfies mirror symmetry along  $x$.

\subsection{The non-Hermitian contribution  as a probe of the dispersion relation}

If we identify the  wave vector  as a static parameter entering
\thinspace Eq.~(\ref{e14hab}), \thinspace  that is  \thinspace ${ \mathbf{R} = \mathbf{k} }$,  then,  whenever the initial static Hamiltonian of a bulk periodic system does not depend explicitly on the wave vector 
\thinspace ${ \bm{\nabla_{\mathbf{k}}}
	H(\mathbf{k})=0  }$,  since the wave vector is only involved only in the Bloch states 
\thinspace ${  \Psi_{n}(\mathbf{r}, \mathbf{k})  }$, \thinspace
one can in principle use the non-Hermitian boundary term \thinspace ${ \bm{\mathcal{S}}_n (\mathbf{k})  }$ \thinspace (where now the label $n$ refers to consideration of a particular band) as a probe for detecting information about the energy dispersion relation (through the boundary behavior of the Bloch functions), namely
\begin{equation} \label{e18eh}
\bm{\nabla_{\mathbf{k}}} E_n(\mathbf{k})
=
- 
\bm{\mathcal{S}}_n(\mathbf{k}). 
\end{equation} 
Consequently, by defining the bulk localized eigenstates of the Hamiltonian as the ones that satisfy 
\thinspace 
${ \bm{\mathcal{S}}_n (\mathbf{k})=0  }$ \thinspace 
for every crystal momentum ${ \mathbf{k} }$ in the Brillouin zone
(which in the simplest scenario occurs when the Bloch wavefunction and all of its derivatives are zero over the boundaries of the material), then, it is evident from \thinspace Eq.~(\ref{e18eh}) \thinspace that these kind of  states will have flat band dispersion relations 
\thinspace ${ \bm{\nabla_{\mathbf{k}}} E_n(\mathbf{k})
=0 }$, \thinspace indicating the existence of a large degree of degeneracy.
Notably, flat bands are known to be formed in conventional band theory of periodic (perfectly ordered) systems
\cite{{leykam2018artificial}, {rontgen2018compact}, {maimaiti2017compact}, {bodyfelt2014flatbands}, {danieli2015flat}, {ramachandran2017chiral}} and are attributed to compact localized states that are created due to destructive interference, but, their formation still lacks a comprehensive generating principle
\cite{maimaiti2017compact}.
In this framework we argue that, whenever the non-Hermitian contribution of a given band is zero, it signals the creation of a flat band.

Moreover, by employing \thinspace Eq.~(\ref{e18eh}), \thinspace one can theoretically explain the changing of the dispersion relation slope in relation to the localization of the involved states.
For example, in the Integer Quantum Hall Effect, the bulk localized states are described by flat bands (Landau levels)  \thinspace $ E_n(\mathbf{k})\equiv E_n=\textit{constant} $, while the states near the boundaries have dispersion relations with non-zero slopes that are approximately linear
\cite{halperin1982quantized} (this linearity being with respect to guiding center coordinates in real space, which however for the simple Landau problem in a Landau gauge is directly proportional to a wave vector value k, hence the linearity being again in k). This kind of behavior, can qute possibly be theoretically explained when the non-Hermitian boundary term is taken into account.  
Moreover, in Dirac and Weyl semimetals \cite{{armitage2018weyl},{vafek2014dirac},{yan2017topological},{jia2016weyl},{burkov2018weyl}},
linear dispersion relations in k are found near the vicinity of energy-degenerate points in the Brillouin zone which represent the bulk electrons' physics near these degeneracy points in k-space.
Although the electrons' motions are not fully relativistic (apart from correction due to spin-orbit coupling), these 
are theoretically explained by employing various model Hamiltonians ${ H(\mathbf{k}) }$ which are effectively (formally) relativistic. 
We propose that, 
these emerging relativistic linear dispersion relations that are accompanied by exotic boundary behaviors such as Fermi arcs, can alternatively be explained by employing the non-Hermitian boundary term. 
Specifically, by taking a Taylor expansion of the non-Hermitian boundary term
\thinspace ${ \bm{\mathcal{S}}_n (\mathbf{k}) }$ \thinspace near a degeneracy point ${ \mathbf{k}_o }$ we find
\begin{eqnarray} \label{e18fh} \nonumber
\bm{\nabla_{\mathbf{k}}} E_n(\mathbf{k})
&=&
- 
\bm{\mathcal{S}}_n(\mathbf{k}_o)
-
\left( 
(\mathbf{k}-\mathbf{k}_o) \! \cdot \!
\bm{\nabla_{\mathbf{k}}}  
\right) 
\bm{\mathcal{S}}_n(\mathbf{k})
\arrowvert
_{\mathbf{k} = \mathbf{k}_o}
\\* [4pt]
&&
+
O(\mathbf{k}^2)
+ \dots.
\end{eqnarray} 
With the dispersion relation ${ E_n(\mathbf{k}) }$ profile experimentally obtained,
one 
can extract information about the boundary term \thinspace ${ \bm{\mathcal{S}}_n (\mathbf{k}) }$  \thinspace 
by keeping only up to first order terms 
in the Taylor expansion Eq.~(\ref{e18fh}).  By then also 
employing the Bloch functions
\thinspace ${ \Psi_n(\mathbf{r}, \mathbf{k})}$
\thinspace
in \thinspace Eq.~(\ref{e17ha}), \thinspace  further information about the behavior of the wavefunction at the boundary can be extracted.

\section{Thouless adiabatic charge pump}  \label{sATPh}

Let us now re-examine the adiabatic charge pumping of non-interacting spinless electrons in a slightly different context than the original method of Ref~\onlinecite{thouless1983quantization}. Rather than assuming a sliding potential \thinspace ${ V((\mathbf{r}- \mathbf{u} \, t) \, \bm{e}_x )}$,   \thinspace
we assume an electric field along  $\bm{e}_x $  direction which is periodic in space (over $x$ coordinate with period equal to the lattice constant) as well as  over time $t$ (with elementary time period equal to $T$). By then using  the non-Hermitian boundary velocity operator and  the adiabatic limit of DHFT, we here evaluate  the collective, center of mass, electrons' displacement in the  $\bm{e}_x $  direction and over the boundaries \thinspace 
${  \sum_i^N  \Delta \left\langle x_i \right\rangle \! |_{bound} \ / \ {N}  }$  \thinspace
without using (first-order) time-dependent perturbation theory.  
  
By attributing the electric field to a scalar potential only, each spinless electron's wavefunction evolves in time by the Hamiltonian
\begin{equation} \label{e35ah}
H(\mathbf{r}, t)=\frac{1}{2m}
{\mathbf{p}^2 
	+
	e
	\phi(x, t)}	
+ 
V_{crys}(\mathbf{r})
\end{equation}
where $m$ and $-e$ are the mass and charge of the electron respectively, 
$ \mathbf{p} $ is the canonical momentum operator,  and \thinspace $V_{crys}(\mathbf{r})$  \thinspace is the crystal potential.
The applied scalar potential is given by
\begin{equation}   \label{a122}
\phi (x, t) 
=
-
\frac{a}{2 \pi}
E(t) sin(2\pi x/a), 
\end{equation}
where \thinspace  $a$ \thinspace is the primitive cell length in $x$ direction, and satisfies
\thinspace ${ \phi(x + n a, t) = \phi(x, t) }$ \thinspace 
with $n$  an integer,  as well as, 
\thinspace ${ \phi(x, T) = \phi(x, 0)=0 }$ \thinspace  and
\thinspace 
${ \left.  \displaystyle \frac{\partial \phi(x, t)}{\partial t}\right|_{t=0}
=
 \left.  \displaystyle \frac{\partial \phi(x, t)}{\partial t}\right|_{t=T} 
=0  }$.
\thinspace
Therefore, the Hamiltonian \thinspace 
Eq.~(\ref{e35ah}) \thinspace is periodic with respect to the elementary time period $T$ satisfying \thinspace 
${ H(\mathbf{r}, 0)=H(\mathbf{r}, T) }$
\thinspace 
and 
\thinspace 
${ \left.  \displaystyle \frac{\partial H( \mathbf{r}, t)}{\partial t}\right|_{t=0}
=
\left.  \displaystyle \frac{\partial H(\mathbf{r}, t)}{\partial t}\right|_{t=T} 
=0  }$ ,
\thinspace 
and it is also periodic over position coordinates
\thinspace $ H(\mathbf{r} + \mathbf{R}, t)=H(\mathbf{r}, t) $, where $\mathbf{R}$ is a Bravais lattice vector.

Provided that the crystal potential $V_{crys}(\mathbf{r})$ is centromeric, the applied scalar potential crucially breaks the 
space-inversion symmetry of the Hamiltonian as well as the mirror symmetry  along the  $\bm{e}_x $ direction due  to 
\thinspace ${ \phi(x, t) \neq \phi(-x, t) }$. 
This results to the   Berry curvature's 
(defined by Eq.~(\ref{e10h}) and assumed to be evaluated with respect to an instantaneous eigenstate of the Hamiltonian)
breaking of  antisymmetry 
 \thinspace
 ${
 \mathcal{E}_{n, \, x} (t, k_x, k_y, k_z)
 \neq
 -
 \mathcal{E}_{n, \, x} (t, -k_x, k_y, k_z)
 }$, 
\thinspace
which in turn, 
gives non-zero collective charge pumping. 
Analogous mirror symmetry breaking over $x$ direction
is involved into the case of the Thouless sliding scalar potential 
\thinspace ${ V(x-\upsilon \, t, y, z)
\neq 
V(-x-\upsilon \, t, y, z) 
}$.
Due to the applied scalar potential,  each electron is subject to an applied microscopic electric field 
\thinspace 
${ \displaystyle \mathbf{E}^{(micro)}(x, t)=
- \bm{\nabla}\phi(x, t)
=
E(t) cos(2\pi x/a) 
\, \bm{e}_x
}$
along $\bm{e}_x $ direction that,  is periodic over position coordinate $x$ (with period equal to the lattice constant $a$)  and over time ($E(T)=E(0)=0 $).  
The  applied microscopic electric field gives zero macroscopic (spatially averaged) electric field, therefore no voltage-bias is present during the pumping.
Crucially, as we will show later, if an intermediate (middle-) time period symmetry is present, that is,  $E(T)=E(0)=E(T/2) $, the charge pumping turns into zero due to time reversal symmetry.

Each electron's state evolves in time according to the time-dependent Schr\"odinger equation (TDSE) 
\begin{equation} \label{e36ah}
i \hbar \frac{d}{dt} 
\left| \Psi(t) \right\rangle 
=
\left(
\frac{1}{2m}
{\mathbf{p}^2 
	+
	e
	\phi(x, t)}	
+ 
V_{crys}(\mathbf{r})
\right) 
\left| \Psi(t) \right\rangle.
\end{equation}
We assume that each electron is initially in a non-degenerate ground state, \linebreak
\thinspace ${ \left| \Psi(t_o) \right\rangle 
	\equiv 
	\left| \psi_n(t_o) \right\rangle 
}$, \thinspace 
where $n$ labels the band.
The ground state energy band is for all moments \  ${ 0 \leq t \leq T }$ \ separated by a gap from the upper bands.
Each electron's ground state wavefunction satisfies periodic boundary conditions over the material's boundaries
\[
\psi_n(\mathbf{r}+\mathbf{L}, t_o)=\psi_n(\mathbf{r}, t_o),  
\]
where $ \mathbf{L} $ denotes the (vector) length of the material in each direction.
Assuming that the electric field ${ \mathbf{E}(x, t) }$ is slowly changing with time, meaning \thinspace ${ T \rightarrow \infty }$, \thinspace each ground state evolves in time adiabatically 
\[
\Psi_n(\mathbf{r}, t)
=
e^{ \displaystyle i \Theta_n(t)}
\psi_n(\mathbf{r}, t),
\]
where \thinspace ${ \Theta_n(t) }$ \thinspace 
is the total phase of the wavefunction (the sum of the dynamic and the geometric adiabatic phase). Each wavefunction \thinspace ${ \psi_n( \mathbf{r}, t) }$ \thinspace satisfies the instantaneous eigenvalue equation
\begin{equation} \label{e38h}
\left( 
\frac{1}{2m}
{\mathbf{p}^2 
	+
	e
	\phi(x, t)}	
+ 
V_{crys}(\mathbf{r})
\right) 
\psi_n(\mathbf{r}, t)
=
E_n \,
\psi_n(\mathbf{r}, t).
\end{equation} 
By following Ref. 9, we assume that the electron's wavefunction has the form
\begin{equation} \label{e38ah}
\psi_n(\mathbf{r}, t, \mathbf{k})
=
e^{\displaystyle i \mathbf{k} \! \cdot \! \mathbf{r}}
u_n(\mathbf{r}, t, \mathbf{k}),
\end{equation}
and the \thinspace ${ u_n(\mathbf{r}, t, \mathbf{k}) }$ \thinspace  satisfies periodic boundary conditions 
\thinspace $ u_n(\mathbf{r}+\mathbf{L}, t, \mathbf{k})=
u_n(\mathbf{r}, t, \mathbf{k}) $, \thinspace
which constrains the allowed wave vector \thinspace ${ \mathbf{k} }$-values
by the condition \thinspace ${ e^{\displaystyle i \mathbf{k} \! \cdot \! \mathbf{L}}=1 }$. 
In order to apply the DHFT given by  \thinspace Eq.~(\ref{e8h}) \thinspace we need to define a continuous parameter \thinspace 
${ \mathbf{R} }$ \thinspace and its corresponding velocity \thinspace  
${ \displaystyle \frac{\partial \mathbf{R}}{ \partial t}  }$.
The parameter definition is made by using the assumed form of the quantum state.
Namely, by having in mind that, in the presence of translational invariance of the Hamiltonian, the wave vector ${  \mathbf{k} }$ is a conserved quantity of the dynamics, 
we define the parameter ${ \mathbf{R} }$ as the static wave vector \thinspace ${ \mathbf{R} \equiv \mathbf{k} }$ \thinspace which has zero corresponding velocity  
\thinspace  
${ \displaystyle \frac{\partial \mathbf{R}}{ \partial t}
	\equiv
	\displaystyle \frac{\partial \mathbf{k}}{ \partial t}=0  }$.
In order for the application of the theorem to be valid, the parameter \thinspace ${ \mathbf{k} }$ \thinspace has to be a continuous one. Therefore, we assume that the length of the system ${ \mathbf{L} }$ is infinite, which results to infinitesimal minimum spacing between the allowed values of the wave vectors  ${ \Delta \mathbf{k} \rightarrow 0}$.  
By then noting that the Hamiltonian \thinspace
${ H(\mathbf{r}, t) }$ \thinspace
does not depend on any parameter 
\[ \bm{\nabla_{\mathbf{k}}}
	H(\mathbf{r}, t)=0,  
\]
application of the adiabatic form of  DHFT with respect to the state \thinspace 
${ \psi_n(\mathbf{r}, t,  \mathbf{k}) }$,
namely,   Eq.~(\ref{a12})
derived in Appendix~\ref{ADHFT}
\thinspace gives
\begin{equation} \label{e39h}
0
=
\bm{\nabla_{\mathbf{k}}} E_n(t, \mathbf{k})
\, + \,
\bm{\mathcal{S}}_n(t, \mathbf{k}) 
\, - \,  \hbar \,
\bm{\mathcal{E}}_n(t, \mathbf{k}),
\end{equation}
where
\begin{eqnarray} \label{e40h} 
\bm{\mathcal{E}}_n (t, \mathbf{k})
&=&
i
\left\langle \bm{\nabla_{\mathbf{k}}} 
\psi_n(t, \mathbf{k}) \,  \vert \, 
\frac{\partial \psi_n(t, \mathbf{k}) }
{ \partial t} \right\rangle  
\\ \nonumber
&&
-
i
\left\langle  
\frac{ \partial \psi_n(t, \mathbf{k}) }
{\partial t} 
\vert \, 
\bm{\nabla_{\mathbf{k}}} \psi_n (t, \mathbf{R})  \right\rangle
\end{eqnarray}
is the earlier defined electric field type of Berry curvature, 
and
\begin{eqnarray} \label{e41h}   \nonumber
\bm{\mathcal{S}}_n (t, \mathbf{k}) 
&=& 
\left\langle \psi_n(t, \mathbf{k}) \, \vert \,  
\left( 
H(\mathbf{r}, t)^+ 
- 
H(\mathbf{r}, t) 
\right)  
\bm{\nabla_{\mathbf{k}}} 
\psi_n(t, \mathbf{k}) \right\rangle 
\\* [4pt]  
&=&
\frac{i \hbar}{2} 
\oiint_S 
\mathbf{n}\!\cdot\! 
\left( \,
( \mathbf{v} \, \psi_n)^{+} 
+ \psi_n^{+} \, \mathbf{v}  
\, \right) \! 
\, \bm{\nabla_{\mathbf{k}}} \psi_n
\, dS
\end{eqnarray}
is the non-Hermitian contribution,
where $\mathbf{n}$ is the unit vector that is locally normal to the surface $S$.

By then using 
\thinspace Eq.~(\ref{e38ah}) \thinspace
into \thinspace Eq.~(\ref{e41h}), \thinspace
that is, by substituting the identity 
\[ 
\bm{\nabla}_{\mathbf{k}} \psi_n(\mathbf{r}, t, \mathbf{k})
=
i 
\, \mathbf{r} \,
\psi_n(\mathbf{r}, t, \mathbf{k})
\, + \,
e^{\displaystyle i \mathbf{k} \! \cdot \! \mathbf{r}}
\,
\bm{\nabla}_{\mathbf{k}}
u_n(\mathbf{r}, t, \mathbf{k}), 
\]
we find an expression  that relates the non-Hermitian contributions  
\begin{equation}   \label{e80abha}
\bm{\mathcal{S}}_n(t, \mathbf{k})=
\hbar
\left\langle 
\, 
\mathbf{v}_{b}(t, \mathbf{k})
\, 
\right\rangle_n
\, + \,
\widetilde{
	\bm{\mathcal{S}}_{n}}  (t, \mathbf{k}),
\end{equation}
where \thinspace 
${ \left\langle 
\, 
\mathbf{v}_{b}(t, \mathbf{k})
\, 
\right\rangle_n }$ \thinspace  
is the non-Hermitian boundary velocity expectation value  (that is non-zero when the position operator $ \mathbf{r} $ becomes anomalous) given by \thinspace Eq.~(\ref{e3b}). \thinspace
${ \bm{\mathcal{S}}_n(t, \mathbf{k}) }$  \thinspace
and \thinspace 
${ \left\langle 
\, \mathbf{v}_{b}(t, \mathbf{k})
\, \right\rangle_n }$ 
\thinspace
are evaluated with respect to the Bloch state 
\thinspace ${ \psi_n(\mathbf{r}, t, \mathbf{k}) }$, \thinspace 
while
${  \widetilde{\bm{\mathcal{S}}_{ n}} (t, \mathbf{k}) }$
\thinspace
is a non-Hermitian boundary contribution  that is evaluated with respect to the instantaneous cell periodic states  \thinspace ${ u_n(\mathbf{r}, t, \mathbf{k}) }$ \thinspace 
(whose value is not zero when the momentum gradient operator $ \bm{\nabla_{\mathbf{k}}}$ becomes anomalous with respect to the  \thinspace ${ u_n(\mathbf{r}, t, \mathbf{k}) }$)
and is given by
\begin{eqnarray}   \label{e17fha}  \nonumber  
\widetilde{
\bm{\mathcal{S}}_{n}} (t, \mathbf{k}) 
&=&
\left\langle u_n(t, \mathbf{k}) \, \vert \,  
\left( 
H_k(t, \mathbf{k})^+ 
- 
H_k(t, \mathbf{k}) 
\right)  
\bm{\nabla_{\mathbf{k}}} 
u_n(t, \mathbf{k}) \right\rangle
\\*[4pt]
&=&
\frac{i \hbar}{2} 
\oiint_S 
\mathbf{n}\!\cdot\! 
\left( \,
( \mathbf{v}_k \,  u_n)^{+} 
+ u_n^{+} \, \mathbf{v}_k  
\, \right) \! 
\, \bm{\nabla_{\mathbf{k}}} u_n
\, dS,
\end{eqnarray}
where \thinspace 
${ \displaystyle 
\mathbf{v}_k =\frac{i}{\hbar}
\left[ H_k (\mathbf{r}, t, \mathbf{k}),\mathbf{r} \right]
}$
\thinspace
is the standard velocity operator, 
\thinspace
${ H_k(\mathbf{r}, t, \mathbf{k})
=
e^{\displaystyle i \mathbf{k} \! \cdot \! \mathbf{r}}
H(\mathbf{r}, t)
e^{\displaystyle i \mathbf{k} \! \cdot \! \mathbf{r}}
}$, 
and ${ u_n \equiv u_n(\mathbf{r}, t, \mathbf{k}) }$.

By substituting  \thinspace Eq.~(\ref{e80abha}) \thinspace 
into \thinspace Eq.~(\ref{e39h}) \thinspace we find the non-Hermitian boundary velocity of the electron given by
\begin{equation} \label{e39bh}
\left\langle \mathbf{v}_{b}(t, \mathbf{k})
\right\rangle_n 
=
-
\frac{1}{\hbar}
\bm{\nabla_{\mathbf{k}}} E_n(t, \mathbf{k})
  \, - \,
\frac{1}{\hbar}
\widetilde{
\bm{\mathcal{S}}_{n}}  (t, \mathbf{k})
\, +  \,
\bm{\mathcal{E}}_n(t, \mathbf{k}).
\end{equation}

Because we have assumed periodic boundary conditions, \thinspace  
${ \psi_n(\mathbf{r}+\mathbf{L}, t, \mathbf{k})=\psi_n(\mathbf{r}, t, \mathbf{k})   }$
\thinspace
as well as
\thinspace
${ u_n(\mathbf{r}+\mathbf{L}, t, \mathbf{k})=u_n(\mathbf{r}, t, \mathbf{k})   }$, 
\thinspace the quantum state \thinspace
${ \bm{\nabla}_{\mathbf{k}} 
	u_n(\mathbf{r}, t, \mathbf{k}) }$
\thinspace is periodic with respect to space coordinates 
\thinspace
${ \bm{\nabla}_{\mathbf{k}} 
	u_n(\mathbf{r}+\mathbf{L}, t, \mathbf{k})
	=
	\bm{\nabla}_{\mathbf{k}} 
	u_n(\mathbf{r}, t, \mathbf{k}) }$, 
\thinspace
it therefore belongs within the domain of definition of the Hamiltonian  
\thinspace 
${ H_k(\mathbf{r}, t, \mathbf{k})
}$ \thinspace which results into
\begin{equation}  \label{bto} 
\widetilde{
\bm{\mathcal{S}}_{n}} (t, \mathbf{k}) 
=
0. 
\end{equation}
This can be shown by expanding the cell periodic functions in a Fourier series over all reciprocal lattice vectors ${ \mathbf{G} }$, that is, \linebreak 
${ u_n(\mathbf{r}, t,  \mathbf{k})
= \sum_{\mathbf{G}} 
C_{n}(t, \mathbf{k},\mathbf{G}) \  e^{\displaystyle -i\mathbf{G}\!\cdot\! \mathbf{r}} 
}$,  which gives   
\thinspace  
${ \bm{\nabla}_{\mathbf{k}}
u_{n}(\mathbf{r}, t, \mathbf{k})
=
\sum_{\mathbf{G}} 
\bm{\nabla}_{\mathbf{k}}
C_{n}(t, \mathbf{k},\mathbf{G}) \  
e^{\displaystyle -i\mathbf{G}\!\cdot\! \mathbf{r}} 
}$.  Substitution of \thinspace ${ u_n(\mathbf{r}, t,  \mathbf{k}) }$ \thinspace and \thinspace
 ${ \bm{\nabla}_{\mathbf{k}}u_n(\mathbf{r}, t,  \mathbf{k}) }$ 
\thinspace in the definition of the non-Hermitian  term, namely \thinspace
${
\widetilde{
\bm{\mathcal{S}}_{n}} (t, \mathbf{k})
=
\left\langle H_k \, u_n \, \vert \,  
\bm{\nabla_{\mathbf{k}}} 
u_n \right\rangle
-
\left\langle u_n \, \vert \,  
H_k  \,
\bm{\nabla_{\mathbf{k}}} 
u_n \right\rangle
}$,
\thinspace 
results in  \thinspace
${ \widetilde{
\bm{\mathcal{S}}_{n}} (t, \mathbf{k})=0 }$
\thinspace
(for spinful electrons, although, when spin-Hall effect is present, this non-Hermitian term may not be zero).

On the other hand, the quantum state \thinspace 
${ \mathbf{r} \,
\psi_n(\mathbf{r}, t, \mathbf{k}) }$
\thinspace
lies outside of the domain of definition of periodic wavefunctions and leaves a residue in the DHFT resulting in
\begin{equation}  \label{bto2}
\left\langle \mathbf{v}_{b}(t, \mathbf{k}) \right\rangle_n 
\neq
0.
\end{equation}
Eq.~(\ref{e80abha}) together with Eq.~(\ref{bto}) indicates that the non-Hermitian term \thinspace
${ \bm{\mathcal{S}}_n(t, \mathbf{k}) }$,  \thinspace
evaluated with respect to the Bloch state 
\thinspace ${ \psi_n(\mathbf{r}, t, \mathbf{k}) }$, \thinspace 
is never zero when periodic boundary conditions are adopted for the wavefunctions.
To the best of our knowledge, it is the first time that the particle transport is explicitly related with a 
non-Hermitian boundary velocity 
\thinspace ${ \left\langle \mathbf{v}_{b}(t, \mathbf{k})
	\right\rangle_n }$, \thinspace 
together  with  the electric field type of  Berry curvature  \thinspace ${ \bm{\mathcal{E}}_n(t, \mathbf{k}) }$ \thinspace that originates from the DHFT.  

The electron's non-Hermitian boundary velocity in $x$ direction is therefore given by
\begin{equation} \label{e43h}
\left\langle \text{v}_{b}(t, \mathbf{k})
\right\rangle_{n, \, x} 
=
-
\frac{1}{\hbar}
\frac{\partial E_n(t, \mathbf{k})}{\partial k_x} 
\, + \,
\mathcal{E}_{n, \, x} (t, \mathbf{k})
\end{equation}
where the electric curvature is given from
\begin{eqnarray} \label{e43bh}       \nonumber
\mathcal{E}_{n, \, x} (t, \mathbf{k})
&=&
i
\left\langle 
\frac{ \partial
	\psi_n(t, \mathbf{k})}
{\partial k_x}
\,  \vert \, 
\frac{\partial \psi_n(t, \mathbf{k}) }
{ \partial t} \right\rangle  
\\*[2pt]   
&&
-
i
\left\langle  
\frac{ \partial \psi_n(t, \mathbf{k}) }
{\partial t} 
\vert \, 
\frac{ \partial
	\psi_n(t, \mathbf{k})}
{\partial k_x}
\right\rangle 
\end{eqnarray}
By defining the collective electrons' displacement (for a  fully occupied band)
 through the sample boundaries
in the $x$ direction, as
${
\sum_i^N
\Delta \left\langle  x_i  \right\rangle_n \! | _{bound}
=
\sum \limits_{\mathbf{k}  }
\int \limits_0^T
\! 
\left\langle \text{v}_{b}(t, \mathbf{k})
\right\rangle_{n, \, x}  dt
}$, \thinspace then,  for a cubic crystal with lattice constant $a$ and in the thermodynamic limit, this displacement is given by 
\begin{eqnarray} \label{e44bh}  \nonumber
\sum_i^N
\Delta \left\langle  x_i  \right\rangle \! | _{bound}
&=&
-
\frac{V}{(2 \pi)^3}
\int \limits_0^T dt
\iiint \limits_{BZ}
\upsilon_{n, x} (t, \mathbf{k})
d^3k
\\* [4pt]   \nonumber 
&&
+
\frac{V}{(2 \pi)^3}
\int \limits_0^T dt
\iiint \limits_{BZ}
\mathcal{E}_{n, \, x} (t, \mathbf{k})
d^3k.
\\* [-4pt]
\end{eqnarray}
where 
\thinspace 
${  \displaystyle 
\upsilon_{n, x} (t, \mathbf{k})
=
\frac{1}{\hbar}	
\frac{\partial E_n(t, \mathbf{k})}{\partial k_x}}$ 
\thinspace 
is the \thinspace  $x$ \thinspace component of the group velocity.
\\

The group velocity contribution is always truncated into a boundary contribution over the Brillouin edges
\[
\int \limits_0^T dt
\iiint \limits_{BZ}
\bm{\upsilon}_n (t, \mathbf{k})d^3k
=
\frac{1}{\hbar}
\int \limits_0^T dt
\oiint \limits_{BZ}
E_n(t, \mathbf{k}) \, 
\mathbf{n}_k \, 
d^2k,	
\]
where ${ \mathbf{n}_k }$ is the unit vector that is locally normal to the Brillouin zone boundaries,
without any gauge phase ambiguities involved.
On the contrary, the electric field type of Berry curvature term of  Eq.~(\ref{e44bh}) is not generally amenable to further simplifications (unless some symmetries are operating, which is thoroughly discussed next).

\subsection{Symmetry Considerations}
If the time-dependent Hamiltonian is space-inversion symmetric, then, 
\thinspace
${  E_n(t, \mathbf{k}) =
E_n(t, -\mathbf{k})
}$ \thinspace
and the group velocity \thinspace
${ \displaystyle
	\bm{\upsilon}_n (t, \mathbf{k})=
\frac{1}{\hbar}
\bm{\nabla_{\mathbf{k}}} E_n(t, \mathbf{k}) }$ \thinspace
will satisfy  \thinspace
${ \displaystyle
\bm{\upsilon}_n (t, \mathbf{k})
=
-
\frac{1}{\hbar}
\bm{\nabla_{(-\mathbf{k})}} E_n(t, -\mathbf{k})
=
- \bm{\upsilon}_n (t, - \mathbf{k})
}$, \thinspace
resulting in  \thinspace
${ \displaystyle
\iiint_{BZ}	 \! \!
\bm{\upsilon}_n (t, \mathbf{k}) d^3k=0
}$,  \thinspace which means that the first term on the right side of \thinspace Eq.~(\ref{e44bh}) \thinspace integrates to zero. 
Similarly, if the Hamiltonian is mirror symmetric along $x$, then, 
\thinspace ${  E_n (t, -k_x, k_y, k_z) =  E_n (t, k_x, k_y, k_z)   }$ \thinspace resulting in \thinspace
${ \displaystyle \int \limits_{ -\pi /a }^{ + \pi /a }
\frac{\partial E_n(t, \mathbf{k})}{\partial k_x}
\, dk_x = 0 }$,
which also indicates that the first term on the second side of \thinspace Eq.~(\ref{e44bh}) \thinspace integrates to zero.
On the other hand, due to the assumed scalar potential \thinspace Eq.~(\ref{a122}), \thinspace space-inversion symmetry and mirror symmetry along $x$ are broken, therefore, the first term on the second side of \thinspace Eq.~(\ref{e44bh}) \thinspace cannot be zero due to space inversion or mirror symmetry.
If the scalar potential satisfies  ${ \phi(x, -t)^{ \displaystyle *}=\phi(x, t) }$, then, the Hamiltonian is time-reversal symmetric and the energy satisfies  \thinspace
${ E_n(t, \mathbf{k}) = E_n(-t, -\mathbf{k}) }$, \thinspace
resulting in   \thinspace
${ \displaystyle
\bm{\upsilon}_n (t, \mathbf{k})
=
- \bm{\upsilon}_n (-t, - \mathbf{k})
}$ \thinspace
for the group velocity. 
When the eigenvalue  of  the \thinspace \textquotedblleft energy\textquotedblright  \thinspace    $ E_n $
in the instantaneous  eigenvalue equation
Eq.~(\ref{e38h}) \thinspace  
does not have explicit time-dependence
\thinspace 
${ \displaystyle 
\left\langle \psi_n) \right| 
\frac{\partial H(\mathbf{r}, t)}{\partial t} \left|   \psi_n(t, \mathbf{k}) \right\rangle  
=
\frac{ \partial E_n(t, \mathbf{k})}{\partial t}
=0
}$,  \thinspace 
then, the system does not absorb energy from the environment. In this framework the group velocity does not have explicit time-dependence, thus,  due to the assumed time-reversal symmetry it satisfies
\thinspace
${ \displaystyle
\bm{\upsilon}_n (\mathbf{k})
=
- \bm{\upsilon}_n ( - \mathbf{k})
}$,  \thinspace
and as a result,
the group velocity on the second side of \thinspace Eq.~(\ref{e44bh}) \thinspace integrates to zero. On the other hand, when the energy has explicit time-dependence
the group velocity on the second side of \thinspace Eq.~(\ref{e44bh}) \thinspace does not necessarily integrate to zero unless an extra, intermediate (middle) time period symmetry of the Hamiltonian is present, namely, 
\thinspace ${ H(\mathbf{r}, t)=H(\mathbf{r}, t+T)=H(\mathbf{r}, t+T/2) }$.
The latter symmetry indicates that the wavefunctions differs by a phase
${ \psi_n(\mathbf{r}, 0)= e^{\displaystyle i \Theta_{1k}}\psi_n(\mathbf{r}, T)=
e^{\displaystyle i \Theta_{2k}}\psi_n(\mathbf{r}, T/2)  }$,
which guarantees that the integration  \thinspace
${ 
\int \limits_0^T dt
\iiint \limits_{BZ}
\bm{\upsilon}_n (t, \mathbf{k})
d^3k
=
\int \limits_{-T/2}^{T/2} dt
\iiint \limits_{BZ}
\bm{\upsilon}_n (t+T/2, \mathbf{k})
d^3k
 }$ \thinspace
turns into zero due to
(i)  \thinspace
${ \bm{\upsilon}_n (t+T/2, \mathbf{k})=\bm{\upsilon}_n (t, \mathbf{k}) 
 }$ \thinspace and
(ii)  
${ \displaystyle
	\bm{\upsilon}_n (t, \mathbf{k})
	=
	- \bm{\upsilon}_n (-t, - \mathbf{k})
}$.
Counter to a time-reversal symmetric Hamiltonian, when the assumed (real) scalar potential breaks time-reversal symmetry (\textit{i.e.} owing to \thinspace ${ E(-t) \neq E(t) }$), \thinspace then, the group velocity of the time-reversal breaking state 
in \thinspace Eq.~(\ref{e44bh}) \thinspace does not necessarily integrate to zero (unless we are assuming  compact localized states with a flat bands).
\\

On the other hand,  when the translation invariant Hamiltonian has translation-inversion symmetry (see Appendix \ref{ab}), namely,  it remains invariant under the inversions ${ (i,  \mathbf{k}) \mapsto(-i,  -\mathbf{k})  }$,  \thinspace this symmetry dictates that \thinspace\thinspace ${ E_n(t, \mathbf{k}) = E_n(t, -\mathbf{k}) }$ \thinspace and the group velocity integrates to zero, hence, for the assumed scalar potential \thinspace Eq.~(\ref{a122})\thinspace  the first term on the right side of \thinspace Eq.~(\ref{e44bh}) \thinspace is zero.

Interestingly,  if an additional confinement potential  ${ V_{c}(y,z) }$  that localize the electrons near to the $x$  axis is assumed in the Hamiltonian \thinspace Eq.~(\ref{e35ah}),  \thinspace then, the translation symmetry is broken along  $y$ and  $z$ coordinates and the energy band of the 3D system becomes \thinspace ${  E_n  \equiv E_n (t, k_x)     }$, \thinspace therefore, \thinspace${  E_n (t, -k_x)  =  E_n (t, k_x)   }$\thinspace due to time-reversal symmetry. This is precisely  the assumption that was made in  Ref~\onlinecite{thouless1983quantization} where the adiabatic particle transport of a 3D system was found quantized.

As ofr the Berry curvature  \thinspace ${ \mathcal{E}_{n, \, x} (t, \mathbf{k})  }$,  \thinspace
as shown in Appendix \ref{ab}, it does not satisfy the antisymmetry relation
\thinspace
${
\mathcal{E}_{n, \, x} (t, k_x, k_y, k_z)
=
-
\mathcal{E}_{n, \, x} (t, -k_x, k_y, k_z)
}$ \thinspace
due to the mirror symmetry that is broken along \thinspace $x$, 
therefore, 
the integration
\thinspace
${
\displaystyle \int \limits_{ -\pi /a }^{ + \pi /a }	
\mathcal{E}_{n, \, x} (t, \mathbf{k})
\, dk_x
}$ 
\thinspace
 does not turn into zero.
 On the other hand, if time-reversal symmetry is present
 \thinspace
 ${	
 \mathcal{E}_{n, \, x} (t, \mathbf{k})
 =
-
\mathcal{E}_{n, \, x} (-t, -\mathbf{k}) 
 }$ 
 \thinspace
 together with an assumed
 intermediate time period symmetry of the Hamiltonian
 \thinspace ${ H(\mathbf{r}, t)=H(\mathbf{r}, t+T)=H(\mathbf{r}, t+T/2) }$,  \thinspace then, 
 \thinspace
 ${	
 \mathcal{E}_{n, \, x} (t, \mathbf{k})
 =
 \mathcal{E}_{n, \, x} (t+ T/2, \mathbf{k})
 }$ 
 \thinspace and the Berry curvature integrates to zero according to
 \thinspace
 $
 \int \limits_0^T dt
 \iiint \limits_{BZ}
 \mathcal{E}_{n, \, x} (t, \mathbf{k})	
 d^3k
 =
 \int \limits_{-T/2}^{T/2} dt
 \iiint \limits_{BZ}
  \mathcal{E}_{n, \, x} (t+T/2, \mathbf{k})
 d^3k
  =
 \int \limits_{-T/2}^{T/2} dt
 \iiint \limits_{BZ}
 \mathcal{E}_{n, \, x} (t, \mathbf{k})
 d^3k
 =0
 $.  (For consequences of all this see below.)

 \subsection{Topological arguments}
 The integration of the Berry curvature 
${  \mathcal{E}_{n, \, x} (t, \mathbf{k})   }$
over the manifold spanned by \thinspace 
${ dk_x \, dt }$ 
\thinspace
is quantized due to topology and it is equal to
\begin{equation} \label{e44ch}
\left( \,
\int \limits_0^T
\int \limits_{ -\pi /a }^{ + \pi /a }
\mathcal{E}_{n, \, x} (t, \mathbf{k})
\, dt \, dk_x
\right)
=
2 \pi \, 
C_1^{(n)}, 
\ \ \ \
\ \ 
C_1^{(n)} 
\in \mathbb{Z}
\end{equation}
where ${ C_1^{(n)} }$ is the first Chern number \cite{nakahara2003geometry}. 
The first Chern number defines the mapping from the parameter space \thinspace ${ (t, \mathbf{k}) }$ \thinspace to the complex projective space of normalized Bloch states 
${ \Psi_n(\mathbf r, t, \mathbf{k})  }$. Non-zero Chern number indicates that the mapping is topologically non-trivial. 
Specifically, due to the assumed periodic boundary conditions at every instant, 
\linebreak  
${ \psi_n(\mathbf{r}+\mathbf{L}, t, \mathbf{k})=\psi_n(\mathbf{r}, t, \mathbf{k})   }$
\thinspace
as well as
\thinspace
${ u_n(\mathbf{r}+\mathbf{L}, t, \mathbf{k})=u_n(\mathbf{r}, t, \mathbf{k}) }$, \
it implies that the Bloch states are generally periodic over the parameter \thinspace ${ \mathbf{k} }$ \thinspace up to a phase \thinspace
$ \displaystyle 
\psi_n(\mathbf{r}, t, \mathbf{k} + \mathbf{G})= e^{ \displaystyle i \Theta_{\! _G}} \,
\psi_n(\mathbf{r}, t, \mathbf{k})  
\thinspace
$
\thinspace
where ${ \mathbf{G} }$ is a reciprocal lattice vector. Similarly, because we have assumed time-periodic Hamiltonian, the Bloch states are periodic over  \thinspace ${ t }$ \thinspace up to a phase 
\thinspace
${ \displaystyle
\psi_n(\mathbf{r}, t + T, \mathbf{k})
= e^{ \displaystyle i  \Theta_{_T} } 
\,
\psi_n(\mathbf{r}, t, \mathbf{k}). 
}$
\thinspace
The last two equations imply that the Bloch states \thinspace ${ \Psi_n(\mathbf r, t, \mathbf{k})  }$ 
\thinspace
are not in general single-valued quantities in 
the parameter space \thinspace ${  t \times \mathbf{k}  }$. \thinspace
Non-zero first Chern number ${ C_1^{(n)} }$  indicates that the Bloch states \thinspace ${ \Psi_n(\mathbf r, t, \mathbf{k})  }$  \thinspace
cannot be everywhere single-valued in the parameter space domain \thinspace  ${ 0 \leq t \leq T }$  \thinspace and
\thinspace  ${ \displaystyle -\frac{\pi}{a} \leq k_{x} \leq \frac{\pi}{a} }$.  \thinspace

\subsection{Quantization of the non-Hermitian displacement}

Assuming, either a  time-reversal symmetric Hamiltonian and a system that does not absorb energy from the environment, or,  bulk compact localized states with flat bands,
then,  by employing \thinspace Eq. (\ref{e44ch})  \thinspace in \thinspace Eq. (\ref{e44bh}), \thinspace it turns out that the collective electrons' displacement through the system's boundaries in the $x$ direction is given by
\begin{eqnarray} \label{e44dh} 
\sum_i^N
\Delta \left\langle  x_i  \right\rangle \! | _{Boundary}
&=&
\frac{V}{(2 \pi)^2}
 \, C_1^{(n)}  
\int \limits_{ -\pi /a }^{ + \pi /a }
\! \! \!  dk_y    \!  \!
\int \limits_{ -\pi /a }^{ + \pi /a }
\! \! \!   dk_z  
\nonumber
\\*  [4pt]
& =&
\, C_1^{(n)}   
N  \alpha,
\end{eqnarray}
where $N$ is the total number of unit cells of the material.
Eq.~(\ref{e44dh}) \thinspace
implies that, after one period $T$ of adiabatic driving, the  displacement of the center of mass of the electrons
in the $x$ direction and over the boundaries,
is quantized in units of lattice constant $a$ and is given by 
\begin{equation} \label{e44eh} 
 \displaystyle
 \frac{1}{ N}
 \sum_{ i}^N
\Delta \left\langle  x_i  \right\rangle \! | _{bound}
=
\, C_1^{(n)}  \, \alpha,
\end{equation}
meaning that the center of mass  performs a quantized rigid displacement over the boundaries along the $x$ direction, or equivalently, the electrons behaves as an incompressible fluid in $\mathbf{k}$-space.
Accordingly, the collective electrons' displacement over $y$ and $z$ directions are zero
\begin{equation} \label{e44fh} 
\sum_i^N \Delta \left\langle y_i  \right\rangle \! | _{bound}=
\sum_i^N \Delta \left\langle z_i  \right\rangle \! |_{bound}=0,
\end{equation}
due to the mirror inversion symmetries (assumed to be present) that implies \thinspace 
${
\mathcal{E}_{n, \, y} (t, k_x, k_y, k_z)
=
-
\mathcal{E}_{n, \, y} (t, k_x, -k_y, k_z)
}$
as well as
${
\mathcal{E}_{n, \, z} (t, k_x, k_y, k_z)
=
-
\mathcal{E}_{n, \, z} (t, k_x, k_y, -k_z)
}$
which in turn gives zero displacements.

We conclude that, if the non-Hermitian term  \thinspace
${  \bm{\mathcal{S}}_n (t, \mathbf{k})  }$  \thinspace
is not taken into account in the adiabatic form of DHFT \thinspace Eq.~(\ref{e39h}), \thinspace
then, this will eventually lead to an apparent paradox since the curvature flux must always give zero value,  
\thinspace
${  0 =
\int \limits_0^T
\int \limits_{ -\pi /a }^{ + \pi /a }
\mathcal{E}_{n, \, x} (t, \mathbf{k})
\, dt \, dk_x
}$  due to the above assumed symmetries. 
On the other hand, by keeping track of the emerging non-Hermitian contribution  \thinspace
${  \bm{\mathcal{S}}_n (t, \mathbf{k})  }$,  \thinspace
the apparent paradox is resolved, and, 
owing to  \thinspace
${ \widetilde{
\bm{\mathcal{S}}_{n}} (t, \mathbf{k}) 
=
0 }$
\thinspace
(due to the cell-periodic wavefunctions that satisfy periodic boundary conditions over the material boundaries at every instant),
the flux of 
the non-Hermitian term \thinspace
${ \bm{\mathcal{S}}_n(t, \mathbf{k})
=
\hbar
\left\langle 
\, 
\mathbf{v}_{b}
\, 
\right\rangle_n  }$ \thinspace  (due to Eq.~(\ref{e80abha}))
 turns out a topological invariant.

\section {Non-adiabatic Charge Pump}   \label{NAPT04}

We now employ the Floquet states and counterpart Floquet-Bloch bands \thinspace $ \varepsilon_a(\mathbf{k})  $ \thinspace in order to theoretically study the non-equilibrium and driven charge pumping with the DHFT.
Each spinless electron's wavefunction evolves in time by the Hamiltonian,
${ \displaystyle 
H(\mathbf{r}, t)=\frac{1}{2m}
{\mathbf{p}^2 
	+
	e
	\phi(x, t)}	
+ 
V_{crys}(\mathbf{r}),
}$
where $m$ and $e$ are the mass and charge of the electron respectively, 
$ \mathbf{p} $ is the canonical momentum operator,  and \thinspace $V_{crys}(\mathbf{r})$  \thinspace is the crystal potential.
The applied scalar potential is given by
\begin{equation}  \label{a123}
 \phi (x, t) 
 =
\frac{a}{2 \pi}
E(t) \,
 cos(2\pi x/a + \theta), 
\end{equation}
where \thinspace  $a$ \thinspace is the primitive cell length in $x$ direction,  ${  E(t) }$ is the amplitude of the microscopic 
electric field, and $\theta $ is a dimensionless constant.
The scalar potential satisfies
\thinspace ${ \phi(x + n a, t) = \phi(x, t) }$ \thinspace 
where $n$  is  an integer,  as well as 
\thinspace ${ \phi(x, T) = \phi(x, 0)=0 }$.
Whenever  $ \theta $  is zero (or more generally $ \theta = 2 \pi n $),  the Hamiltonian has mirror symmetry along  $\bm{e}_x $ direction, otherwise the mirror symmetry is broken.
The macroscopic (spatially averaged) electric field is zero, thus, no voltage-bias is present during the pumping.
The time-dependent amplitude of the electric field 
$ E(t) $ is periodic ${ E(t+T)=E(t) }$,  but,  has arbitrary large velocity   $ \displaystyle  \frac{dE(t)}{dt}   $, therefore, no adiabatic time-evolution can be implied. Instead, we use the non-equilibrium Floquet states and apply the DHFT  of Eq.~(\ref{e14hab}). 

 Each electron's wavefunction evolves in time according to the TDSE
\begin{equation}  \label{e44agh} 
i \hbar \frac{d}{dt} \Psi(\mathbf{r}, t, \mathbf{k})
=
H(\mathbf{r}, t)\, \Psi(\mathbf{r}, t, \mathbf{k}),
\end{equation}
and the solution of  \thinspace Eq.~(\ref{e44agh})  \thinspace is considered to be a non-equilibrium Floquet state 
\thinspace
${ \displaystyle
	\Psi_a(\mathbf{r}, t, \mathbf{k})  
	=
	e^{-  \displaystyle \frac{i}{\hbar}  \varepsilon_{a}(\mathbf{k}) \, t} 
	\Phi_a(\mathbf{r}, t, \mathbf{k}) 
}$. 
The Floquet modes
\thinspace  
${ \Phi_a(\mathbf{r}, t, \mathbf{k}) 
}$
\thinspace are periodic in time 
\thinspace  
${ \Phi_a(\mathbf{r}, t, \mathbf{k}) 
	=
	\Phi_a(\mathbf{r}, t+T, \mathbf{k})
}$
\thinspace
and 
satisfy the eigenvalue equation
\begin{eqnarray}  \label{e44gh}
\left( 
H(\mathbf{r}, t)
-i \hbar \frac{d}{dt}
\right) 
\Phi_a(\mathbf{r}, t, \mathbf{k}) 
&=&
H_F(\mathbf{r}, t)
\Phi_a(\mathbf{r}, t, \mathbf{k}) 
\nonumber
\\  [2pt] 
& =&
\varepsilon_a(\mathbf{k})
\Phi_a(\mathbf{r}, t, \mathbf{k}), 
\end{eqnarray}
where
\thinspace ${ H_F(\mathbf{r}, t) }$ \thinspace is the so-called Floquet Hamiltonian, and
\thinspace ${\varepsilon_a(\mathbf{k}) }$ \thinspace is the (static) quasienergy that is restricted into the interval 
\thinspace 
${ \displaystyle 
	-
	\pi \hbar / T
	\leq
	\varepsilon_{a}(\mathbf{k})
	\leq
	\pi \hbar / T
}$ \thinspace 
called 
first Floquet-zone which contains all the physically non-equivalent quantum states.
Furthermore,  we assume that each Floquet mode is expressed 
with respect to the Floquet-Bloch state \cite{gomez2013floquet}
\begin{equation} \label{e38aah} 
\Phi_a(\mathbf{r}, t, \mathbf{k})
=
e^{\displaystyle i \mathbf{k} \! \cdot \! \mathbf{r}}
u_a(\mathbf{r}, t, \mathbf{k}),
\end{equation}
where \thinspace ${ u_a(\mathbf{r}, t, \mathbf{k}) }$ \thinspace
is periodic both in ${ \mathbf{r} }$ and $t$. 
Whenever the Floquet Hamiltonian \thinspace ${ H_F(\mathbf{r}, t) }$ \thinspace is time-reversal invariant the quasienergy spectrum is symmetric with respect to crystal momentum inversion
\thinspace ${\varepsilon_a(\mathbf{k})=\varepsilon_a(-\mathbf{k}) }$.  Moreover, when the  Floquet Hamiltonian satisfies mirror symmetry, i.e. along  $x$,  then, for the specific direction it satisfies 
\thinspace ${ \varepsilon_a( k_x, k_y, k_z)
=
\varepsilon_a( -k_x, k_y, k_z)	
}$.

The parameter definition is made by using the assumed form of the quantum state.
Due to the presence of translational invariance of the Hamiltonian, the wave vector ${  \mathbf{k} }$ is a conserved quantity of the dynamics. We therefore define the parameter ${ \mathbf{R} }$ as the static wave vector \thinspace ${ \mathbf{R} \equiv \mathbf{k} }$ \thinspace which has zero corresponding velocity  
\thinspace  
${ \displaystyle \frac{\partial \mathbf{R}}{ \partial t}
	\equiv
	\displaystyle \frac{\partial \mathbf{k}}{ \partial t}=0  }$.

 The Hamiltonian  \thinspace
 ${ H(\mathbf{r}, t) }$ \thinspace
 does not depend on any parameter \thinspace 
 ${ \bm{\nabla_{\mathbf{k}}}
 H(\mathbf{r}, t)=0  }$, therefore, application of the  DHFT  \thinspace Eq.~(\ref{e14hab}) \thinspace with respect to the Floquet state, 
\thinspace
${ \displaystyle
	\Psi_a(\mathbf{r}, t, \mathbf{k})  
	=
	e^{-  \displaystyle \frac{i}{\hbar}  \varepsilon_{a}(\mathbf{k}) \, t} 
	\Phi_a(\mathbf{r}, t, \mathbf{k}) 
}$, 
\thinspace gives
 \begin{equation} \label{e39ah}
 0
 =
 \bm{\nabla_{\mathbf{k}}} E_a(t, \mathbf{k})
 \, + \,
 \bm{\mathcal{S}}_a(t, \mathbf{k}) 
 \, - \,  \hbar \,
 \bm{\mathcal{E}}_a(t, \mathbf{k}),
 \end{equation}
 where the \textquotedblleft energy\textquotedblright \thinspace is given by
 \begin{eqnarray} \label{e18ash}
 E_a(t, \mathbf{k})
 &=&
 \left\langle \Phi_a(t, \mathbf{k})
 \right|  H(t)   
 \left| 
 \Phi_a(t, \mathbf{k}) \right\rangle
\nonumber
\\  [2pt]
 &=& 
 \varepsilon_a(\mathbf{k})
 +
 i \hbar
 \left\langle \Phi_a(t, \mathbf{k}) \, \vert \, 
 \frac{d  \Phi_a(t, \mathbf{k}) }{dt}
 \right\rangle,
 \end{eqnarray}
the non-Hermitian term from
\begin{equation}  \label{p5}
\bm{\mathcal{S}}_a (t, \mathbf{k}) 
=
\left\langle \Psi_a(t, \mathbf{k})  \vert 
\left( 
H(\mathbf{r}, t)^+ 
- 
H(\mathbf{r}, t) 
\right)  
\bm{\nabla_{\mathbf{k}}} 
\Psi_a(t, \mathbf{k}) \right\rangle, 
\end{equation} 
and the electric field type of curvature by  
\begin{eqnarray} \label{e43bha}   \nonumber 
\bm{\mathcal{E}}_a (t, \mathbf{k})
&=&
i
\left\langle \bm{\nabla_{\mathbf{k}}} 
\Phi_a(t, \mathbf{k}) \,  \vert \, 
\frac{\partial \Phi_a(t, \mathbf{k}) }
{ \partial t} \right\rangle  
\\ \nonumber
&&
-
i
\left\langle  
\frac{ \partial \Phi_a(t, \mathbf{k}) }
{\partial t} 
\vert \, 
\bm{\nabla_{\mathbf{k}}} \Phi_a (t, \mathbf{R})  \right\rangle.
\end{eqnarray} 
By then using the identity
 \begin{eqnarray}    \nonumber 
 \bm{\nabla}_{\mathbf{k}} \Psi_a(\mathbf{r}, t, \mathbf{k})
 &=&
 -\frac{i}{\hbar}
  \bm{\nabla_{\mathbf{k}}} \varepsilon_a(\mathbf{k}) 
  \ t \
  \Psi_a(\mathbf{r}, t, \mathbf{k})
\  +  \ 
 i 
 \, \mathbf{r} \,
 \Psi_a(\mathbf{r}, t, \mathbf{k})
 \\*[4pt]   \nonumber 
&& 
 + \,
 e^{-  \displaystyle \frac{i}{\hbar}  \varepsilon_{a}(\mathbf{k}) \, t} 
 e^{\displaystyle i \mathbf{k} \! \cdot \! \mathbf{r}}
 \,
 \bm{\nabla}_{\mathbf{k}}
 u_a(\mathbf{r}, t, \mathbf{k}), 
 \end{eqnarray}
we find the equation that relates the non-Hermitian terms
\begin{equation}   \label{e80abhab}
\bm{\mathcal{S}}_a(t, \mathbf{k})=
\hbar
\left\langle 
\, 
\mathbf{v}_{b}(t, \mathbf{k})
\, 
\right\rangle_a
\, + \,
\widetilde{
\bm{\mathcal{S}}_{a}}  (t, \mathbf{k}),
\end{equation} 
where
\thinspace   
${  \left\langle 
\, 
\mathbf{v}_{b}(t, \mathbf{k})
\, 
\right\rangle_a }$
\thinspace 
is the boundary velocity evaluated with respect to  the Floquet mode 
\thinspace
${ \Phi_a(\mathbf{r}, t, \mathbf{k}) }$ 
\thinspace 
and 
\thinspace
${  \widetilde{
\bm{\mathcal{S}}_{a}}  (t, \mathbf{k}) }$ 
\thinspace is evaluated with respect to the Floquet-Bloch state
\begin{equation}  \label{a200}
\widetilde{
\bm{\mathcal{S}}_{a}}(t, \mathbf{k})
=
\left\langle u_a(t, \mathbf{k})  \vert 
\left( 
H_k(\mathbf{r}, t)^+ 
- 
H_k(\mathbf{r}, t) 
\right)  
\bm{\nabla_{\mathbf{k}}} 
u_a(t, \mathbf{k}) \right\rangle.
\end{equation}

By substituting  \thinspace Eq.~(\ref{e80abhab}) \thinspace into 
\thinspace Eq.~(\ref{e39ah}) \thinspace
we find the boundary velocity of the electron
\begin{equation} \label{e39bha}
 \left\langle \mathbf{v}_{b}(t, \mathbf{k})
 \right\rangle_a 
 =
 -
 \frac{1}{\hbar}
 \bm{\nabla_{\mathbf{k}}} E_a(t, \mathbf{k})
 \, - \,
 \frac{1}{\hbar}
 \widetilde{
 	\bm{\mathcal{S}}_{a}}  (t, \mathbf{k})
 \, +  \,
 \bm{\mathcal{E}}_a(t, \mathbf{k}),
 \end{equation}
and, by substituting  \thinspace Eq.~(\ref{e18ash}) \thinspace
into  \thinspace Eq.~(\ref{e39bha}), \thinspace
we finally find the non-Hermitian boundary velocity in $x$ direction for a single Floquet mode that is given by
 \begin{eqnarray} \label{e43ha}  \nonumber 
 \left\langle \text{v}_{b}(t, \mathbf{k})
 \right\rangle_{a, \, x} 
 &=&
 -
 \frac{1}{\hbar}
 \frac{\partial \varepsilon_a(\mathbf{k})}{\partial k_x} 
  \ -  \
 i
 \frac{\partial}{\partial k_x} 
 \left\langle \!  \Phi_a(t, \mathbf{k}) \, \vert \, 
 \frac{d \Phi_a(t, \mathbf{k}) }{dt}    
  \! \right\rangle
 \\*[4pt]
 &&
  \, - \,
 \frac{1}{\hbar}
 \widetilde{
 \mathcal{S}_{a, \, x}}  (t, \mathbf{k})
 \, + \,
 \mathcal{E}_{a, \, x} (t, \mathbf{k}).
 \end{eqnarray}
In this respect, each electron's displacement   
over the material's boundaries in the $x$ direction after the time of period $T$ is given by
\begin{eqnarray}   \nonumber   \label{e800}
\Delta \left\langle  x (\mathbf{k})  \right\rangle_a \! | _{bound}
&=&
-
\frac{1}{\hbar}
\frac{\partial \varepsilon_a(\mathbf{k})}{\partial k_x} T
\ -  \
\frac{\partial \beta_a (\mathbf{k})}{\partial k_x}
\\*[4pt]   \nonumber
&&
 - \, 
\frac{1}{\hbar}
\int_0^T  \!  \!  \!  
\widetilde{
\mathcal{S}_{a, \, x}}  (t, \mathbf{k})
dt
\ + \
\int_0^T  \!  \!  \!  
\mathcal{E}_{a, \, x} (t, \mathbf{k})
dt
\\*
\end{eqnarray}
where, interestingly, the quantity  \thinspace ${ \beta_a(\mathbf{k})  }$ \thinspace showing up is the non-adiabatic Aharonov-Anandan phase of the Floquet mode
\begin{equation}   \label{p3}
\beta_a(\mathbf{k})=
\int_0^T
i 
\left\langle \Phi_a(t, \mathbf{k}) \, \vert \, 
\frac{d \Phi_a(t, \mathbf{k}) }{dt}
\right\rangle
dt.
\end{equation} 

For a spinless electron and time-reversal symmetric Floquet Hamiltonian,
the Floquet modes satisfy
\thinspace
${ \Phi_a(\mathbf{r}, t, \mathbf{k})
=
\Phi_a(\mathbf{r}, -t, -\mathbf{k})^{\displaystyle *}
}$
(up to a  constant 
\thinspace
${ U(1)}$ phase  \thinspace 
${ e^{ \displaystyle i \lambda}
}$ \thinspace which we neglected for simplicity). In this respect, it is easy to show that  the Floquet modes satisfy
${ \displaystyle 
i 
\left\langle \Phi_a(t, \mathbf{k}) \, \vert \, 
\frac{d \Phi_a(t, \mathbf{k}) }{dt}
\right\rangle
=
i 
\left\langle \Phi_a(-t, -\mathbf{k}) \, \vert \, 
\frac{d \Phi_a(-t, -\mathbf{k}) }{d(-t)}
\right\rangle, 
}$ 
which results to  an Aharonov-Anandan phase that is symmetric under the inversion of crystal momentum
\thinspace ${ \beta_a(\mathbf{k})=\beta_a(-\mathbf{k})  }$.
Analogously, for a space-inversion  symmetric Floquet Hamiltonian
\thinspace
(${ \Phi_a(\mathbf{r}, t, \mathbf{k})
=
\Phi_a(-\mathbf{r}, t, -\mathbf{k})
}$
up to a  constant 
\thinspace
${ U(1)}$ phase),  the Aharonov-Anandan phase satisfies
\thinspace ${ \beta_a(\mathbf{k})=\beta_a(-\mathbf{k})  }$, while, for a mirror-symmetric Floquet Hamiltonian,  i.e. along $x$, it satisfies
${
\beta_a( k_x, k_y, k_z)
=
\beta_a( -k_x, k_y, k_z)
}$.

Interestingly, whenever the Floquet Hamiltonian satisfies  time-reversal symmetry but the mirror symmetry is broken along $x$, then,  
\thinspace 
$
\varepsilon_a( k_x, k_y, k_z)
-
\varepsilon_a( -k_x, -k_y, -k_z)
= 0
$
\thinspace  (due to time-reversal symmetry)
while
${
 \varepsilon_a( k_x, k_y, k_z)
-
\varepsilon_a( -k_x, k_y, k_z)
\neq 0
}$  \thinspace
(due to breaking of the mirror symmetry), for every crystal momentum within the first Floquet-Brillouin zone. This implies that there is at least one discontinuity  of the Floquet quasienergy  band (which extends in the interval 
\thinspace 
${ \displaystyle 
-
\pi \hbar / T
\leq
\varepsilon_{a}(\mathbf{k})
\leq
 \pi \hbar / T
}$) \thinspace 
while traversing the $ k_x $ momenta of the first Brillouin zone, which results to an apparent winding (along the quasienergy axe) of the Floquet band. 
Provided that  there is no gap in the 
first Floquet-Brillouin zone, each apparent winding (discontinuity)
contributes a quantized  amount of 
\thinspace ${ \displaystyle  | \Delta  \varepsilon_{a}(\mathbf{k}) |
=   2 \pi \hbar / T   }$   
resulting into \thinspace
${ \displaystyle 
\int_{BZ}
\frac{\partial \varepsilon_a (\mathbf{k})}{\partial k_x}
dk_x
=
n \, 
2 \pi \hbar / T
}$ 
\thinspace   where $n$ is  the integer number of windings.

Similarly,  when the Floquet Hamiltonian satisfies  time-reversal symmetry but the mirror symmetry is broken along $x$, then, the Aharonov-Anandan phase satisfies
\thinspace 
$
\beta_a( k_x, k_y, k_z)
-
\beta_a( -k_x, -k_y, -k_z)
= 0
$
\thinspace  (due to time-reversal symmetry),
but,  \thinspace
${
\beta_a( k_x, k_y, k_z)
-
\beta_a( -k_x, k_y, k_z)
\neq 0
}$  \thinspace
(due to breaking of the mirror symmetry)  for every crystal momentum within the first Floquet-Brillouin zone.  
This implies that there is at least one discontinuity of the Aharonov-Anandan phase \thinspace ${ \beta_a(\mathbf{k})  }$ \thinspace with unspecified jump while traversing the $ k_x $ momenta of the first Brillouin zone.
Specifically, although the Aharonov-Anandan phase values (modulo $ 2 \pi $) lie in the interval ${ 0 \le \beta_a(\mathbf{k}) \le 2 \pi }$,
the jump is not necessarily equal to $ 2 \pi $
because the phase is not coming out as a  product of an eigenvalue equation, 
in contrast to the Floquet energy 
\thinspace ${\varepsilon_a(\mathbf{k}) }$ \footnote{Each Floquet energy  gives the total phase of the quantum state after the time period of $T$, that is
\thinspace  $ \displaystyle \frac{1}{\hbar}  \varepsilon_a(\mathbf{k}) T  $,  \thinspace
therefore, all different quasienergy eigenvalues define the projective Hilbert space of rays of the single valued Floquet modes} 
which is the eigenvalue of the Floquet Hamiltonian.
In this respect, for mirror-breaking  (along $x$)  Floquet Hamiltonians, we cannot naively assume that 
\thinspace
${ \displaystyle 
	\int_{BZ}
	\frac{\partial \beta_a (\mathbf{k})}{\partial k_x}
	dk_x
	=0 
}$, 
\thinspace 
and the Aharonov-Anandan phase must be taken into account when
calculating the collective non-Hermitian displacement of the electrons.
We conclude that, for mirror-breaking Floquet Hamiltonians,
the Aharonov-Anandan phase behaves as a non-integrable quantity with respect to crystal momentum, 
and as a consequence  it contributes to charge pumping an amount proportional to its non quantized winding.

\subsection{Charge pumping from the lowest-energy Floquet state}
We assume that each independent electron is initially in the lowest-energy non-equilibrium Floquet state  
\thinspace
${ \displaystyle
	\Psi_a(\mathbf{r}, t, \mathbf{k})  
	=
	e^{-  \displaystyle \frac{i}{\hbar}  \varepsilon_{a}(\mathbf{k}) \, t} 
	\Phi_a(\mathbf{r}, t, \mathbf{k}) 
}$,  therefore, the period-averaged 
\textquotedblleft energy\textquotedblright  \ 
\thinspace  
$ \displaystyle
E_a(\mathbf{k})
=
\frac{1}{T} \int_0^T \! \!  
\left\langle \Phi_a(t, \mathbf{k})
\right|  H(t)   
\left| 
\Phi_a(t, \mathbf{k}) \right\rangle  dt
$  \  
is assumed to have the lowest value.
By employing \thinspace Eq.~(\ref{e43ha}) and for a fully occupied quasienergy band, the non-Hermiticity-related electrons' displacement 
in the $x$ direction and over the boundaries of the material (in the thermodynamic limit)  is given by
\begin{widetext}
\begin{eqnarray}   \label{e440bh}   \nonumber
\frac{(2 \pi)^3 }{V}
\sum_i^N
\Delta \left\langle  x_i  \right\rangle \! | _{bound} \,
&=&
-
\int \limits_{ -\pi /a }^{ + \pi /a }
\!  dk_y  \! 
\int \limits_{ -\pi /a }^{ + \pi /a }
\!  dk_z  
\left( 
\frac{T}{ \hbar}
\int \limits_{ -\pi /a }^{ + \pi /a }
\frac{\partial \varepsilon_a(\mathbf{k})}{\partial k_x} 
\!  dk_x
\right) 
\  -  \
\int \limits_{ -\pi /a }^{ + \pi /a }
\!  dk_y  \! 
\int \limits_{ -\pi /a }^{ + \pi /a }
\!  dk_z  
\left( 
\int \limits_{ -\pi /a }^{ + \pi /a }
\frac{\partial \beta_a(\mathbf{k})}{\partial k_x} 
\!  dk_x
\right)     
\\*[8pt]   \nonumber
&&
\! \! \!
-
\int \limits_{ -\pi /a }^{ + \pi /a }
\!  dk_y  \! 
\int \limits_{ -\pi /a }^{ + \pi /a }
\!  dk_z  
\left( \,
\int \limits_0^T
\int \limits_{ -\pi /a }^{ + \pi /a }
\widetilde{
\mathcal{S}_{a, \, x}}  (t, \mathbf{k})
\, dt \, dk_x
\right) 
\  +  \
\int \limits_{ -\pi /a }^{ + \pi /a }
\!  dk_y  \! 
\int \limits_{ -\pi /a }^{ + \pi /a }
\!  dk_z  
\left( \,
\int \limits_0^T
\int \limits_{ -\pi /a }^{ + \pi /a }
\mathcal{E}_{a, \, x} (t, \mathbf{k})
\, dt \, dk_x
\right) \! \!. 
\\* 
\end{eqnarray} 
\end{widetext} 
We are assuming periodic boundary conditions, \thinspace  
${ \psi_a(\mathbf{r}+\mathbf{L}, t, \mathbf{k})=\psi_n(\mathbf{r}, t, \mathbf{k})   }$
\thinspace
as well as
\thinspace
${ u_a(\mathbf{r}+\mathbf{L}, t, \mathbf{k})=u_n(\mathbf{r}, t, \mathbf{k})   }$, thus, the non-Hermitian contribution is in this case zero
\begin{equation}  \label{btoa} 
\widetilde{\mathcal{S}_{a, x}} (t, \mathbf{k}) 
=
0. 
\end{equation}

\subsubsection{Mirror-symmetric Floquet Hamiltonian}
For a mirror-symmetric Floquet Hamiltonian along $x$ we assume
 \thinspace ${  \theta = 0 }$  in \thinspace Eq.~(\ref{a123}).
By employing the eigenvalue equation \thinspace Eq.~(\ref{e44gh}),  \thinspace
the quasienergy and the Aharonov-Anandan phase satisfies, 
\thinspace ${ \varepsilon_a( k_x, k_y, k_z)
=
\varepsilon_a( -k_x, k_y, k_z)	
}$
\thinspace
and 
${
\beta_a( k_x, k_y, k_z)
=
\beta_a( -k_x, k_y, k_z)
}$
\thinspace 
respectively; 
while the Berry curvature satisfies
\thinspace
${
\mathcal{E}_{a, \, x} (t, k_x, k_y, k_z)
=	-
\mathcal{E}_{a, \, x} (t, -k_x, k_y, k_z)
}$
\thinspace
as shown in Appendix~\ref{ab}.
In this respect,  the collective non-Hermitian pumped charge per cycle is trivially zero, 
\[
\displaystyle  
\frac{1}{N}
\sum_i^N
\Delta \left\langle  x_i  \right\rangle \! | _{bound} \, =0.
\]

\subsubsection{Mirror-breaking  Floquet Hamiltonian}

For a mirror-symmetric  along $x$ Floquet Hamiltonian we assume
\thinspace ${  \theta \neq 2 \pi n }$   in \thinspace Eq.~(\ref{a123}).
Due to mirror symmetry breaking, 
\thinspace $ \varepsilon_a( k_x, k_y, k_z)
\neq
\varepsilon_a( -k_x, k_y, k_z)	
$,
\thinspace
as well as
\thinspace
$
\beta_a( k_x, k_y, k_z)
\neq
\beta_a( -k_x, k_y, k_z)
$
\thinspace 
and
\thinspace
$
\mathcal{E}_{a, \, x} (t, k_x, k_y, k_z)
\neq	
-
\mathcal{E}_{a, \, x} (t, -k_x, k_y, k_z)
$.
\thinspace 
The collective, non-adiabatic and formally originating from non-Hermiticity 
pumped charge of the center of mass of the electrons is therefore given by
\begin{widetext}
\begin{eqnarray}   \label{e4bh}   \nonumber
\frac{1}{N}
\sum_i^N
\Delta \left\langle  x_i  \right\rangle \! | _{bound} \,
=
-
\frac{a^3 \, T}{ (2 \pi)^3  \,  \hbar}
\int \limits_{ -\pi /a }^{ + \pi /a }
\!  dk_y  \!  \!   \! 
\int \limits_{ -\pi /a }^{ + \pi /a }
\!  dk_z  \!  \!   \! 
\int \limits_{ -\pi /a }^{ + \pi /a }
\frac{\partial \varepsilon_a(\mathbf{k})}{\partial k_x} 
\!  dk_x
 \  -  \ 
\frac{a^3}{ (2 \pi)^3 }
\int \limits_{ -\pi /a }^{ + \pi /a }
\!  dk_y  \! \!   \!
\int \limits_{ -\pi /a }^{ + \pi /a }
\!  dk_z  \! \!   \!
\int \limits_{ -\pi /a }^{ + \pi /a }
\frac{\partial \beta_a(\mathbf{k})}{\partial k_x} 
\!  dk_x
\ +  \
 C_1^{(n)}  \, \alpha,
\end{eqnarray} 
\end{widetext}
where we have used that 
\thinspace
${ \left( \,
\int \limits_0^T
\int \limits_{ -\pi /a }^{ + \pi /a }
\mathcal{E}_{a, \, x} (t, \mathbf{k})
\, dt \, dk_x
\right)
=
2 \pi \, 
C_1^{(a)} }$
\thinspace
with 
\thinspace
${  C_1^{(a)} }$
\thinspace
the first Chern number.
Non-zero first Chern number ${ C_1^{(a)} }$  indicates that the Floquet states \thinspace ${ \Phi_a(\mathbf r, t, \mathbf{k})  }$  \thinspace
cannot be everywhere single-valued in the parameter space domain \thinspace  ${ 0 \leq t \leq T }$  \thinspace and
\thinspace  ${ \displaystyle -\frac{\pi}{a} \leq k_{x} \leq \frac{\pi}{a} }$. \thinspace
Namely, although the Floquet states are periodic with respect to time 
by definition, \thinspace ${ \Phi_a(\mathbf r, t, \mathbf{k})=\Phi_a(\mathbf r, t+T, \mathbf{k})  }$,  \thinspace
non-zero Chern number indicates that they don't satisfy the periodic gauge along $x$, that is,
\thinspace ${ \Phi_a(\mathbf r, t, \mathbf{k}) \neq
\Phi_a(\mathbf r, t, \mathbf{k} + G_x  \, \mathbf{e}_x )  }$  \thinspace
where $G_x$ is any reciprocal lattice vector in $\mathbf{e}_x$ direction.

If we assume that there is no gap present in the
first Floquet-Brillouin zone, then, the collective, non-adiabatic and non-Hermiticity-related pumped charge of the center of mass of the electrons is given by 
\begin{eqnarray}   \label{e4bah}    \nonumber
\frac{1}{N}
\sum_i^N
\Delta \left\langle  x_i  \right\rangle \! | _{bound} \,
&=&
-
n \, a  \ +  \
C_1^{(n)}  \alpha
\\*  \nonumber
&&
\! \! \! \! \! \! \! \! \! \! \! \! \! \! \! \! \!
 -  
\frac{a^3}{ (2 \pi)^3 } \!
\int \limits_{ -\pi /a }^{ + \pi /a }
\!  dk_y  \!   \!
\int \limits_{ -\pi /a }^{ + \pi /a }
\!  dk_z  \!   \!
\int \limits_{ -\pi /a }^{ + \pi /a }
\frac{\partial \beta_a(\mathbf{k})}{\partial k_x} 
\!  dk_x,
\\*[-4pt]
\end{eqnarray}
where the first term gives a quantized contribution in terms of the integer number \thinspace $n$ \thinspace of windings of the Floquet-Brillouin zone and the last term gives a generically  non-quantized contribution in terms of the non-integrable Aharonov-Anandan phase.  
Interestingly, a naive application of the periodic gauge on the Floquet modes along 
\thinspace $\mathbf{e}_x$ \thinspace direction,  that is, 
\thinspace
${ 
\Phi_a(\mathbf{r}, t, \mathbf{k})
=
\Phi_a(\mathbf{r}, t, \mathbf{k}+G_x  \mathbf{e}_x)  
}$, 
\thinspace
results to the cancellation of the Aharonov-Anandan contribution 
to the pumped charge
\begin{eqnarray}    \nonumber
\int \limits_{ -\pi /a }^{ + \pi /a }
\frac{\partial \beta_a(\mathbf{k})}{\partial k_x} 
\!  dk_x
&=&
\beta_a(\mathbf{k}+G_x  \mathbf{e}_x)
-
\beta_a(\mathbf{k})
\\*  \nonumber
&&
\!  \! \!  \! \!  \!  \!  \! \!  \! \!  \!  \!  \! \!  \! \!  \!  \!  \! \!  \! \!  \!  \!  \! \!  \! \!  \!  \!  \! \!  \! 
=
i
\int_0^T
\left\langle \Phi_a(t, \mathbf{k}+G_x  \mathbf{e}_x ) \, \vert \, 
\frac{d \Phi_a(t, \mathbf{k} +G_x  \mathbf{e}_x) }{dt}
\right\rangle dt
\\*[6pt]  \nonumber 
&&
\!  \! \!  \! \!  \!  \!  \! \!  \! \!  \!  \!  \! \!  \! \!  \!  \!  \! \!  \! \!  \!  \!  \! \!  \! \!  \!  
 - 
i
\int_0^T 
\left\langle \Phi_a(t, \mathbf{k}) \, \vert \, 
\frac{d \Phi_a(t, \mathbf{k}) }{dt}
\right\rangle
dt
\\*[6pt]  \nonumber 
&&
\!  \! \!  \! \!  \!  \!  \! \!  \! \!  \!  \!  \! \!  \! \!  \!  \!  \! \!  \! \!  \!  \!  \! \!  \! \!  \!  \!  \!  \! 
= 0.
\end{eqnarray}

In a slightly different context,  presumably a more intuitive one,  by employing \thinspace Eq.~(\ref{e39bha}), 
\thinspace the pumped charge  can be expressed  in terms of the dynamical phase as 
\begin{eqnarray}   \label{e4bbh}   \nonumber
\frac{1}{N}
\sum_i^N
\Delta \left\langle  x_i  \right\rangle \! | _{bound} \,
&=&
C_1^{(n)}  \, \alpha
\\*  \nonumber
&&
\!  \!   \!  \!  \!   \!  \!  \!   \!  \!  \!   \!
+ 
\frac{a^3}{ (2 \pi)^3 } \!
\int \limits_{ -\pi /a }^{ + \pi /a }
\!  dk_y  \!   \!
\int \limits_{ -\pi /a }^{ + \pi /a }
\!  dk_z  \!  \!   \!
\int \limits_{ -\pi /a }^{ + \pi /a }
\frac{\partial D_a(\mathbf{k})}{\partial k_x} 
\!  dk_x,
\\*[-4pt]
\end{eqnarray}
where the dynamical phase is given by \thinspace
$ \displaystyle
D_a(\mathbf{k})
=
-
\frac{1}{\hbar}
\int_0^T \! \!  
\left\langle \Phi_a(t, \mathbf{k})
\right|  H(t)   
\left| 
\Phi_a(t, \mathbf{k}) \right\rangle  dt
$.  
If the Hamiltonian is mirror-inversion symmetric  along  $x$, then, 
${  C_1^{(n)}=0 }$, and,  
due to the dynamical phase that satisfies 
$ \displaystyle
D_a(k_x, k_y, k_z)
=
D_a(-k_x, k_y, k_z)
$,
the pumped charge turns into zero. On the other hand, for the present mirror-reversal breaking case 
$ \displaystyle
D_a(k_x, k_y, k_z)
\neq
D_a(-k_x, k_y, k_z)
$,   
we don't a priori have any topological argument that the difference of the dynamical phase over the edges of the first Brillouin zone should be quantized. 
In conclusion, for a mirror-breaking Hamiltonian, the quantization of the pumped charge breaks down due to a non-integrable Aharonov-Anandan phase even  if no gaps exist in the Floquet-Brillouin zone, provided that the insulator is a topologically nontrivial one, where the periodic gauge cannot be taken (the well-known obstruction in satisfying it).

\subsection{Trivialization of the non-adiabatic displacement}
The boundary velocity Eq.~(\ref{e43ha}) can be simplified by assuming that, 
\thinspace
${ \displaystyle  
\frac{ \partial}{\partial k_x}
\frac{ \partial }{\partial t} 
\left| \Phi_a(t, \mathbf{k}) \right\rangle
-
\frac{ \partial }{\partial t} 
\frac{ \partial}{\partial k_x}
\left| \Phi_a(t, \mathbf{k}) \right\rangle
=
0
}$,  \thinspace
which is the integrability condition that guarantees that the Floquet modes  \thinspace  
${ \left| \Phi_a(t, \mathbf{k}) \right\rangle }$ 
\thinspace
are analytic (single valued) functions of \thinspace $t$ \thinspace and \thinspace $k_x$  in the  \thinspace  ${  t\! \times \!  k_x   }$ \thinspace 
space.  Therefore, 
either the material is a conventional band insulator if we take into account the entire  \thinspace  ${  t\! \times \!  k_x   }$ \thinspace 
space, or we are assuming a fiber bundle theory with patches.
In this framework, the Berry curvature takes the form
\begin{eqnarray} \label{e43bhac}   \nonumber 
\mathcal{E}_{a, x}(t, \mathbf{k})
&=&
i
 \frac{\partial}{\partial k_x} 
\left\langle 
\Phi_a(t, \mathbf{k}) \,  \vert \, 
\frac{\partial \Phi_a(t, \mathbf{k}) }
{ \partial t} \right\rangle  
\\*[4pt]  \nonumber
&&
-
i
\frac{ \partial}{\partial t} 
\left\langle  
\Phi_a(t, \mathbf{k}) 
\vert \, 
\frac{\partial \Phi_a (t, \mathbf{k})}{\partial k_x} 
\right\rangle
\end{eqnarray} 
and the boundary velocity 
\thinspace Eq.~(\ref{e43ha}) \thinspace truncates into
\begin{eqnarray} \label{e43hb}  \nonumber 
\left\langle \text{v}_{b}(t, \mathbf{k})
\right\rangle_{a, \, x} 
&=&
-
\frac{1}{\hbar}
\frac{\partial \varepsilon_a(\mathbf{k})}{\partial k_x} 
\ - \
i
\frac{\partial}{\partial t}
\left\langle \!  \Phi_a(t, \mathbf{k}) \, \vert \, 
\frac{\partial \Phi_a(t, \mathbf{k}) }{\partial k_x}     
\! \right\rangle
\\*[4pt]
&&
\, - \,
\frac{1}{\hbar}
\widetilde{
	\mathcal{S}_{a, \, x}}  (t, \mathbf{k}).
\end{eqnarray}
Therefore, 
by assuming  that \thinspace  ${ \widetilde{\mathcal{S}_{a, x}} (t, \mathbf{k}) 
=0   }$,  \thinspace
the center of mass,  non-adiabatic displacement over the boundaries Eq.~(\ref{e440bh})  takes the form
\begin{widetext}
\begin{eqnarray}   \label{e4bha}   \nonumber
\frac{1}{N}
\sum_i^N
\Delta \left\langle  x_i  \right\rangle \! | _{bound} \,
=
-
\frac{a^3 \, T}{ (2 \pi)^3  \,  \hbar}
\int \limits_{ -\pi /a }^{ + \pi /a }
\!  dk_y  \!  \!   \! 
\int \limits_{ -\pi /a }^{ + \pi /a }
\!  dk_z  \!  \!   \! 
\int \limits_{ -\pi /a }^{ + \pi /a }
\frac{\partial \varepsilon_a(\mathbf{k})}{\partial k_x} 
\!  dk_x
\  -  \
\frac{a^3 }{ (2 \pi)^3 }
\int \limits_{ -\pi /a }^{ + \pi /a }
\!  dk_y  \!  \!   \! 
\int \limits_{ -\pi /a }^{ + \pi /a }
\!  dk_z  \!  
\int \limits_{ 0}^{ T}
\frac{\gamma_a(t, \mathbf{k})}{dt} 
\!  dt
\end{eqnarray}
\end{widetext}
where  \thinspace 
${ \gamma_a(t, \mathbf{k})  }$
is a first Brillouin zone accumulated phase given by
\begin{equation}
\gamma_a(t, \mathbf{k})
=
i \!
\int \limits_{ -\pi /a }^{ + \pi /a }
\! \! \!
A_x(t, \mathbf{k}) \,
dk_x, 
\end{equation}
where
${ \displaystyle  
A_x(t, \mathbf{k})=i
\left\langle \!  \Phi_a(t, \mathbf{k}) \, \vert \, 
\frac{\partial \Phi_a(t, \mathbf{k}) }{\partial k_x}     
\! \right\rangle  }$
\thinspace 
is the Berry connection of the Floquet modes.
The Floquet modes \thinspace ${ \left|  \Phi_a(t, \mathbf{k}) \right\rangle   }$ \thinspace are periodic with respect to the time period $T$,   \thinspace ${ \left|  \Phi_a(t+T, \mathbf{k}) \right\rangle   
=  \left|  \Phi_a(t, \mathbf{k}) \right\rangle  }$.  \thinspace
If we further assume that the momentum derivative of the Floquet modes
\thinspace 
${ \displaystyle  \left|  \frac{\partial \Phi_a(t, \mathbf{k}) }{\partial k_x}    \! \right\rangle    }$
\thinspace  is also periodic with respect to the time period  $T$, 
that is,  \thinspace
${ \displaystyle  \left|  \frac{\partial \Phi_a(t+T, \mathbf{k}) }{\partial k_x}    \! \right\rangle 
=
\left|  \frac{\partial \Phi_a(t, \mathbf{k}) }{\partial k_x}    \! \right\rangle
 }$, 
\thinspace
which is true whenever the Floquet modes are analytic functions of \thinspace ${  k_x }$   in the \thinspace  ${  t\! \times \!  k_x   }$ \thinspace space, \thinspace then, it is evident that   \thinspace
$  \displaystyle 
\left\langle \!  \Phi_a(T, \mathbf{k}) \, \vert \, 
\frac{\partial \Phi_a(T, \mathbf{k}) }{\partial k_x}     
\! \right\rangle
=
\left\langle \!  \Phi_a(0, \mathbf{k}) \, \vert \, 
\frac{\partial \Phi_a(0, \mathbf{k}) }{\partial k_x}     
\! \right\rangle
$ \thinspace
which results into
${ \displaystyle   \int \limits_{ 0}^{ T}
\frac{\gamma_a(t, \mathbf{k})}{dt} 
\!  dt = 0 }$, \thinspace thus, the  pumped charge is then given only in terms of the winding of the Floquet band. This is the formal case that is encountered in the literature\cite{{shih1994nonadiabatic},{privitera2018nonadiabatic}}, recovered here as a special instance of our more general theory.
\\
\\

\section {Electronic polarization}  \label{elpol}

By using the DHFT and the extended velocity operator definition, we now calculate the adiabatic electronic polarization change of a fully occupied valence band ${ \Delta \mathbf{P}_n }$.  In this framework, we calculate the collective (adiabatic) electrons' displacement \thinspace
${ \sum_i^N
\Delta \left\langle  \mathbf{r}_i  \right\rangle   }$ \thinspace   over the boundaries
(while a crystal is modified or deformed) that gives the induced dipole moment change of the material, which in turn provides the quantum electronic induced polarization change when divided by the volume of the material.

One of the main differences between polarization change  and charge pumping processes,  is the voltage-bias   (non-zero mean force)  that is present in the polarization phenomena while is absent in pumping counterparts. In this manner, the presence of  voltage-bias  enforces  a breaking of periodic Born-von K\'{a}rm\'{a}n  boundary conditions between the edges of the material, hence
${ \left| \Psi_n(\mathbf{r}+ \mathbf{L},t)\right|^2  
	\neq
	\left| \Psi_n(\mathbf{r},t)\right|^2
 }$, although the wavefunction may retain its Bloch form in the bulk.
According to Ref.~\onlinecite{resta2007theory}, similar breaking of the periodic  Born-von K\'{a}rm\'{a}n  boundary conditions emerges in piezoelectric phenomena due to strain, which is not merely a  perturbing term in the Hamiltonian.
  
For an applied electric filed that is now spatially uniform, each spinless electron's quantum state evolves in time according to the TDSE 
\begin{equation} \label{e36abh}
i \hbar \frac{d}{dt} 
\Psi(\mathbf{r}, t) 
=
\left( 
\frac{1}{2m}
 \mathbf{p}^2
-
e \mathbf{r} \! \cdot \! \mathbf{E}(t)
+
 V_{crys}(\mathbf{r}, t)
\right) 
\Psi(\mathbf{r}, t),
\end{equation}
where \thinspace
${ \mathbf{E}(t)
}$ \thinspace
is the applied homogenous electric field 
that is also assumed to be time-periodic and to satisfy \thinspace
${ \mathbf{E}(0)=\mathbf{E}(T)=0
}$.
The crystal potential  ${  V_{crys}(\mathbf{r}, t) }$  is assumed to have time-dependence,  either as a result of the externally applied electric field, or due to applied strain, both deforming the crystal's ions orderly arrangement.
For example, when computing the spontaneous polarization $ \mathbf{P}_s $ in a ferroelectric material such as ${ \text{PbTiO}_3 }$, one can let  
${  V_{crys}(\mathbf{r}, 0) }$  refer to the centrosymmetric cubic crystal structure and 
${  V_{crys}(\mathbf{r}, T) }$  to the noncentrosymetric, ferroelectric crystal structure.

Due to the lack of translational invariance of the Hamiltonian, the wave vector ${  \mathbf{k} }$ is not a conserved quantity of the dynamics. 
We assume the ansatz 
\begin{equation} \label{e38abh}
\psi(\mathbf{r}, t, \mathbf{k})
=
e^{\displaystyle i \mathbf{k}(t) \! \cdot \! \mathbf{r}}
u(\mathbf{r}, t, \mathbf{k}),
\end{equation}
where the wave vector ${ \mathbf{k}(t) }$
is a continuous  time-dependent vector parameter. 
We define the time-dependent parameter ${ \mathbf{R}(t) }$ as equal to the time-dependent wave vector \thinspace ${ \mathbf{R}(t) \equiv \mathbf{k}(t) }$, \thinspace
and then replace 
\thinspace Eq.~(\ref{e38abh})  \thinspace  into  
\thinspace Eq.~(\ref{e36abh}).
The time-dependent wave-vector (parameter) is defined to satisfy the equation of motion
\begin{equation}   \label{345}
\frac{\partial \mathbf{k}(t)}{ \partial t}
=
\frac{e}{\hbar}  \mathbf{E}(t)
\end{equation}
which guarantees that, if the electron is initially in a Bloch state, it will continue to be in a Bloch state after the electric field is turned on. 
Therefore, the parameter evolves in time according to
\begin{equation}   
\mathbf{k}(t)
=
\mathbf{k}_o
+
\frac{e}{\hbar}
\int_0^{\displaystyle t}
\mathbf{E}(t')
dt'
\end{equation}
where  ${ \mathbf{k}_o }$ is the initial value of the parameter.
Substituting \thinspace Eq.~(\ref{e38abh})  \thinspace into 
\thinspace Eq.~(\ref{e36abh}), \thinspace  and by using 
\thinspace Eq.~(\ref{345}) \thinspace
gives
\begin{equation}  \label{e36abbh}  
i \hbar \frac{d}{dt} 
u(\mathbf{r}, t, \mathbf{k}) 
=
H_k(\mathbf{r}, t, \mathbf{k})
u(\mathbf{r}, t, \mathbf{k}),
\end{equation}
where the quantum state  
\thinspace  ${ u(\mathbf{r}, t, \mathbf{k}) }$
\thinspace
evolves over time by the Hamiltonian
\begin{equation}  \label{e36abch}
H_k(\mathbf{r}, t, \mathbf{k})
=
\frac{1}{2m}
\left( \mathbf{p} + 
\hbar \mathbf{k}(t)
\right)^2 
\, + \,
V_{crys}(\mathbf{r}, t).
\end{equation}

We assume that each electron is initially in a ground, Bloch state,  and that the electric field  turns on and changes very slowly over time \thinspace ${ T \rightarrow \infty }$, \thinspace resulting to adiabatic time-evolution, namely
\[
u(\mathbf{r}, t, \mathbf{k})
\equiv
e^{ \displaystyle i \Theta_n(t, \mathbf{k})}
u_n(\mathbf{r}, t, \mathbf{k}),
\]
where \thinspace ${ \Theta_n(t, \mathbf{k}) }$ \thinspace 
is the total phase of the wavefunction (the sum of the dynamic and the geometric adiabatic phase) and 
\thinspace ${ u_n(\mathbf{r}, t, \mathbf{k}) }$ \thinspace
is the instantaneous eigenstate of the Hamiltonian  
\thinspace ${
	H_k(\mathbf{r}, t, \mathbf{k})
}$ \thinspace
satisfying the eigenvalue equation
\begin{equation}
H_k(\mathbf{r}, t, \mathbf{k})
u_n(\mathbf{r}, t, \mathbf{k})
=
E_n(t, \mathbf{k})
u_n(\mathbf{r}, t, \mathbf{k}).
\end{equation}
We apply the adiabatic form of the 
DHFT Eq.~(\ref{e14hab}) 
on \thinspace Eq.~(\ref{e36abbh})  \thinspace where the assumed Hamiltonian is
\thinspace ${
H_k(\mathbf{r}, t, \mathbf{k})
}$.
The gradient of the Hamiltonian with respect to the wave vector gives 
\thinspace
${ \displaystyle
\bm{\nabla}_{\mathbf{k}}
H_k(\mathbf{r}, t, \mathbf{k})
=
\hbar
\left( 
\frac{i}{\hbar}
[H_k(\mathbf{r}, t, \mathbf{k}) , \mathbf{r}
]
\right) 
=
\hbar
\mathbf{v} 
}$,
\thinspace
where $ \mathbf{v} $ is the standard velocity operator.
Therefore, we find
\begin{eqnarray} \label{e14eh}   
\hbar
\left\langle \mathbf{v}  \right\rangle_n 
&=&
\bm{\nabla_{\mathbf{R}}} E_n(t, \mathbf{k})
\, + \,
\widetilde{
\bm{\mathcal{S}}_{n}} (t, \mathbf{k}) 
\\*[6pt]   \nonumber 
&&
\, - \,  \hbar \,
\bm{\mathcal{E}}_n(t, \mathbf{k})
\, - \,  \hbar \,
\frac{\partial \mathbf{k}}{\partial t}
\! \times 
\bm{\mathcal{B}}_{n}(t, \mathbf{k}),
\end{eqnarray}
where all involved quantities are evaluated with respect to the time-dependent eigenstates \thinspace    
${ u_n(\mathbf{r}, t, \mathbf{k}) }$.   \thinspace 
Specifically, the curvatures are given by

\begin{eqnarray}   \nonumber
\bm{\mathcal{E}}_n (t, \mathbf{k})
&=&
i
\left\langle \bm{\nabla_{\mathbf{k}}} 
u_n(t, \mathbf{k}) \,  \vert \, 
\frac{\partial u_n(t, \mathbf{k}) }
{ \partial t} \right\rangle  
\\*[4pt]
&-&
i
\left\langle  
\frac{ \partial u_n(t, \mathbf{k}) }
{\partial t} 
\vert \, 
\bm{\nabla_{\mathbf{k}}} u_n (t, \mathbf{k})  \right\rangle
\end{eqnarray}
and
\begin{equation}  \nonumber
\bm{\mathcal{B}}_n(t, \mathbf{k})
= i
\left\langle \bm{\nabla_{\mathbf{k}}}
u_n(t, \mathbf{k}) \right| 
\times 
\left|  \bm{\nabla_{\mathbf{k}}}
u_n(t, \mathbf{k}) \right\rangle, 
\end{equation} 
and the non-Hermitian contribution is given from
\begin{equation}
\widetilde{
	\bm{\mathcal{S}}_{n}} (t, \mathbf{k}) 
=
\left\langle u_n(t, \mathbf{k}) \, \vert \,  
\left( 
H_k(t, \mathbf{k})^+ 
- 
H_k(t, \mathbf{k}) 
\right)  
\bm{\nabla_{\mathbf{k}}} 
u_n(t, \mathbf{k}) \right\rangle.
\end{equation}
We note that, when the instantaneous eigenstates \thinspace    
${ u_n(\mathbf{r}, t, \mathbf{k}) }$   \thinspace 
do not have explicit time-dependence the electric type of Berry curvature 
\thinspace  ${\bm{\mathcal{E}}_n (t, \mathbf{k}) }$  \thinspace
is by definition zero.

A crucial difference with respect to the theoretical approach that was made in the study of the adiabatic topological charge pumping (Thouless pump) is that the states \thinspace ${ u_n(\mathbf{r}, t, \mathbf{k}) }$ \thinspace  are assumed to be cell periodic in the bulk of the material at every instant, but need not satisfy periodic boundary conditions over the boundaries of the material, therefore 
\thinspace
${  \displaystyle
\widetilde{
	\bm{\mathcal{S}}_{n}} (t, \mathbf{k}) 
 }$
\thinspace
is not zero, due to the presence of voltage-bias or strain.
We can then use \thinspace Eq.~(\ref{e80abha})  \thinspace that relates the non-Hermitian contributions, that is, 
${  \displaystyle 
\bm{\mathcal{S}}_n(t, \mathbf{k})=
\hbar
\left\langle 
\, 
\mathbf{v}_{b}(t, \mathbf{k})
\, 
\right\rangle_n
\, + \,
\widetilde{
	\bm{\mathcal{S}}_{n}}  (t, \mathbf{k}),
}$
where \thinspace 
${ \left\langle 
\, 
\mathbf{v}_{b}(t, \mathbf{k})
\, 
\right\rangle_n }$ \thinspace  
is the non-Hermitian boundary velocity expectation value  (that is non-zero when the position operator $ \mathbf{r} $ becomes anomalous) 
and  \thinspace 
${ \bm{\mathcal{S}}_{ n} (t, \mathbf{k}) }$
\thinspace
is evaluated with respect to the instantaneous Bloch state, namely,
\begin{eqnarray}    \nonumber    \label{nhcS1}
\bm{\mathcal{S}}_n (t, \mathbf{k}) 
&=&
\left\langle \psi_n(t, \mathbf{k}) \, \vert \,  
\left( 
H(\mathbf{r}, t)^+ 
- 
H(\mathbf{r}, t) 
\right)  
\bm{\nabla_{\mathbf{k}}} 
\psi_n(t, \mathbf{k}) \right\rangle
\\* [6pt]   
&=&
\frac{i \hbar}{2} \oiint_S 
\mathbf{n}\!\cdot\! 
\left( \,
( \mathbf{v} \, \psi_{n} )^{ \dagger} 
+ \psi_{n}^{ \dagger } \, \mathbf{v}  
\, \right) \! 
\, \bm{\nabla_{\mathbf{k}}}  \psi_{n}
\, dS,	
\end{eqnarray}
where \thinspace ${\psi_n  \equiv \psi_n(\mathbf{r}, t, \mathbf{k})}$, \thinspace $\mathbf{v}$ is the standard velocity operator and $\mathbf{n}$ is the unit vector that is locally normal to the surface $S$ that encloses the material. The non-Hermitian contribution \thinspace 
${ \bm{\mathcal{S}}_{ n} (t, \mathbf{k}) }$
\thinspace is always a real quantity according to \thinspace Eq.~(\ref{a201}).
In this framework
\thinspace 
Eq.~(\ref{e14eh}) \thinspace
takes the form
\begin{eqnarray} \label{e14eha}   
\hbar
\left\langle \,
\mathbf{v} 
\, \right\rangle_n
+
\hbar
\left\langle \,
\mathbf{v}_b 
\, \right\rangle_n
&=&
\bm{\nabla_{\mathbf{k}}} E_n(t, \mathbf{k})
\, + \,
\bm{\mathcal{S}}_{n} (t, \mathbf{k}) 
\\*[6pt]   \nonumber 
&&
\, - \,  \hbar \,
\bm{\mathcal{E}}_n(t, \mathbf{k})
\, - \,  \hbar \,
\frac{\partial \mathbf{k}}{\partial t}
\! \times 
\bm{\mathcal{B}}_{n}(t, \mathbf{k}).
\end{eqnarray}
By then using the extended velocity  operator definition
\thinspace
${  \displaystyle
\left\langle \,
\mathbf{v}_{ext} 
\, \right\rangle_n
=
\left\langle \,
\mathbf{v} 
\, \right\rangle_n
+
\left\langle \,
\mathbf{v}_b 
\, \right\rangle_n
=
\displaystyle  
\frac{d}{dt} \!
\left\langle \,
\mathbf{r}
\, \right\rangle_n,
}$
\thinspace
we obtain the electron's mean position time derivative that is given by
\begin{eqnarray} \label{e14eha2}   
\frac{d}{dt} \!
\left\langle \,
\mathbf{r}
\, \right\rangle_n
&=&
\bm{\upsilon}_{n} (t, \mathbf{k})
\, + \,
\frac{1}{\hbar}
\bm{\mathcal{S}}_{n} (t, \mathbf{k}) 
\\*[6pt]   \nonumber 
&&
\, - \,  
\bm{\mathcal{E}}_n(t, \mathbf{k})
\, - \,  
\frac{\partial \mathbf{k}}{\partial t}
\! \times 
\bm{\mathcal{B}}_{n}(t, \mathbf{k}).
\end{eqnarray}
where  \thinspace
${ \displaystyle
\bm{\upsilon}_n (t, \mathbf{k})=
\frac{1}{\hbar}
\bm{\nabla_{\mathbf{k}}} E_n(t, \mathbf{k}) }$ \thinspace
is the group velocity.

Therefore,  each electron's adiabatic displacement after the cyclic evolution of the Hamiltonian over a period of time $ T $ is given by
\begin{widetext}
\begin{equation}  \label{e76adh} 
\Delta
\left\langle
u_n(t, \mathbf{k}) \vert 
\, \mathbf{r} \,
\vert
u_n(t, \mathbf{k})
\right\rangle
=
\int_0^T
\left( 
\bm{\upsilon}_n (t, \mathbf{k})
\, + \,
\frac{1}{\hbar}
\bm{\mathcal{S}}_n(t, \mathbf{k}) 
\, - \, 
\bm{\mathcal{E}}_n(t, \mathbf{k})
\, - \, 
\frac{\partial \mathbf{k}}{\partial t}
\! \times 
\bm{\mathcal{B}}_{n}(t, \mathbf{k})
\right) 
dt.
\end{equation}
By then using \thinspace Eq.(\ref{e36bbh}) \thinspace in the thermodynamic limit, the induced electronic polarization change for a single fully occupied valence band is given by
\begin{equation}  \label{e76aeh}
\Delta \mathbf{P}_n
=
\frac{e}{(2 \pi)^3}
\int \limits_0^T
dt 
\iiint \limits_{BZ}
\left( 
\bm{\upsilon}_n (t, \mathbf{k})
\, + \,
\frac{1}{\hbar}
\bm{\mathcal{S}}_n(t, \mathbf{k})
\, - \, 
\bm{\mathcal{E}}_n(t, \mathbf{k})
\, - \, 
\frac{\partial \mathbf{k}}{\partial t}
\! \times 
\bm{\mathcal{B}}_{n}(t, \mathbf{k})
\right)
d^3k.	
\end{equation}
\end{widetext}
Some remarks are then in order regarding the way we have transformed the sum of \thinspace Eq.~(\ref{e36bbh}) \thinspace into an integration with respect to the time-dependent parameter ${ \mathbf{k}(t) }$  in
 \thinspace Eq.~(\ref{e76aeh}). 
Mathematically speaking, 
the equation of motion of the time-dependent parameter in the
DHFT
can always be defined arbitrarily. 
Each elementary volume in parameter space (assumed to be occupied by a single spinless electron) expands or dilates according to the equation
${ \displaystyle d^3k(t)
=
\text{Det}J(t,t_o) \
d^3k_o }$
where \thinspace ${ \text{Det}J(t,t_o) }$ \thinspace 
denotes the determinant of the Jacobian matrix of the transformation that satisfies \thinspace ${ \text{Det}J(t_o, t_o)=1 }$. 
The Jacobian matrix,  captures the transformation of the initial value of the parameter 
to a subsequent one.
The time evolution of the Jacobian matrix 
 is solely determined by the way that one defines the velocity of the parameter and is given by \thinspace
${ \displaystyle
\frac{d}{dt}\text{Det}J(t,t_o)
=
\left( 
\bm{\nabla_{\mathbf{k}}} 
\! \cdot \! \frac{\partial \mathbf{k}}{\partial t}
\right)  \!
\text{Det}J(t,t_o).
}$ \thinspace
In this framework, the equation of motion of the elementary volume of the parameter is determined by
\thinspace
${   \displaystyle
\frac{d}{dt} \left( d^3k(t) \right) 
=
\left( 
\bm{\nabla_{\mathbf{k}}} 
\! \cdot \! \frac{\partial \mathbf{k}}{\partial t}
\right) 
d^3k(t)
}$. \thinspace
Owing to our definition of the parameter  \thinspace Eq.~(\ref{345}),  \thinspace  
${ \displaystyle
\bm{\nabla_{\mathbf{k}}} 
\! \cdot \! \frac{\partial \mathbf{k}}{\partial t}=0  }$,  which indicates that the elementary volume per particle in parameter space remains constant \thinspace ${ d^3k_o=d^3k(t)  }$, \thinspace
so, no modification is needed when we transform the sum into an integration over time-dependent wavevectors.

Comparing \ Eq.~(\ref{e76aeh}) \ to the polarization change formula
of the Modern's Theory of Polarization  \cite{{king1993theory},{vanderbilt1993electric}},
we have rigorously found three extra contributions: one due to the group velocity 
\thinspace ${ \bm{\upsilon}_n (t, \mathbf{k}) }$, \thinspace
another one due to the non-Hermitian boundary contribution
\thinspace ${ \bm{\mathcal{S}}_{n} (t, \mathbf{k}) }$, \thinspace
and the last one due to the Berry curvature
\thinspace ${ \bm{\mathcal{B}}_{n} (t, \mathbf{k}) }$.

A question  now emerges: is each one of these extra contributions non-zero separately? 
To answer the question based on simple criteria, we must rely on the localization of the assumed electrons'  wavefunctions (in general the realistic boundary conditions of the Bloch eigenstates) and the discrete symmetries of the assumed states.
Due to localization, for example, in the simplest model, when encountering only bulk localized states, in the sense that each electron completely avoids the boundaries of the material, the non-Hermitian contribution is zero  \thinspace ${ \bm{\mathcal{S}}_{ n} (t, \mathbf{k})=0 }$  \thinspace according to \thinspace Eq.~(\ref{nhcS1}).
On the other hand, in order to take into account the consequences of the involved discrete symmetries,  we present 
in \linebreak Table \ref{T1}  \thinspace
the transformations of the quantities entering into  Eq.~(\ref{e76aeh})  as a result of,  time-reversal symmetry (T.R.S.),  space-inversion symmetry (S.I.S.) and mirror symmetry along $x$ (M.S.);  
\thinspace each symmetry is derived in Appendix~\ref{ab}. 

\begin{table}[h]
\caption{\label{T1}{Discrete symmetries for quantities involved in the polarization change.}}
\begin{ruledtabular}
\begin{tabular}{c   c   c    c }
			
\\[-10pt]		
			
Quantity & T.R.S.  & S.I.S. & M.S. 
			
\\[6pt]
			
\hline 	
			
\\[-4pt]
			
${ \bm{\upsilon}_n (t, \mathbf{k}) }$    &   
${ -\bm{\upsilon}_n (-t, -\mathbf{k}) }$    & 
${ -\bm{\upsilon}_n (t, -\mathbf{k}) }$  &  
${ -\upsilon_{n, x} (t, -k_x, k_y, k_z) }$

\\[8pt]
			
${ \bm{\mathcal{S}}_{ n} (t, \mathbf{k}) }$   & 
${ -\bm{\mathcal{S}}_{ n} (-t, -\mathbf{k}) }$   & 
${ -\bm{\mathcal{S}}_{ n} (t, -\mathbf{k}) }$ & 
${ -\mathcal{S}_{ n, x}(t, -k_x, k_y, k_z) }$

\\[8pt]
			
${ \bm{\mathcal{E}}_{ n} (t, \mathbf{k}) }$   & 
${ -\bm{\mathcal{E}}_{ n} (-t, -\mathbf{k}) }$   & 
${ -\bm{\mathcal{E}}_{ n} (t, -\mathbf{k}) }$ & 
${ -\mathcal{E}_{ n, x}(t, -k_x, k_y, k_z) }$

\\[8pt]

${ \bm{\mathcal{B}}_{ n} (t, \mathbf{k}) }$   & 
${ -\bm{\mathcal{B}}_{ n} (-t, -\mathbf{k}) }$   & 
${ \bm{\mathcal{B}}_{ n} (t, -\mathbf{k}) }$ & 
${ \mathcal{B}_{ n, x}(t, -k_x, k_y, k_z) }$

\\[4pt] 

\end{tabular}	
\end{ruledtabular} 
\end{table}

For example, in a simple model where no applied external electric field is present, thus
\thinspace  
${ \displaystyle
	\frac{\partial \mathbf{k}}{\partial t}=0  }$, but the crystal potential 
is time-dependent \thinspace ${  V_{crys}(\mathbf{r}, t) }$  \thinspace due to  applied time-periodic strain, if the crystal potential remains space-inversion symmetric  at any instant, then,  the change of electronic polarization over  the time interval  $T$ is zero  \thinspace
${ \Delta \mathbf{P}_n=0  }$ \thinspace 
as a result of 
\thinspace Eq~(\ref{e76aeh}) \thinspace
and the discrete symmetries of the involved quantities presented in
\thinspace Table \ref{T1}.
Further polarization change arguments, based on discrete-symmetries, are discussed below.

\subsubsection{Time-reversal symmetry}

When the Hamiltonian 
\thinspace ${  H_k(\mathbf{r},t, \mathbf{k}) }$ \thinspace given by \thinspace Eq.~(\ref{e36abch}) \thinspace is time-reversal symmetric, namely, is invariant under the inversions \thinspace
${ (i, \mathbf{r}, t, \mathbf{k}(t)) \longmapsto
(-i, \mathbf{r}, -t, -\mathbf{k}(t))  }$, 
\thinspace
and provided that the Hamiltonian ${  H(\mathbf{r},t)}$ \thinspace of \thinspace Eq.~(\ref{e36abh}) \thinspace has an 
intermediate time period symmetry  \thinspace
\thinspace ${ H(\mathbf{r}, 0)=H(\mathbf{r}, T)=H(\mathbf{r}, T/2) }$, \thinspace
(although 
\thinspace ${ \displaystyle  \frac{dH(\mathbf{r}, 0)}{dt} \neq \frac{dH(\mathbf{r}, T/2)}{dt}  }$), \thinspace
implying that the wavefunctions differ by a phase
${ \psi_n(\mathbf{r}, 0)= e^{\displaystyle i \Theta_{1k}}\psi_n(\mathbf{r}, T)=
e^{\displaystyle i \Theta_{2k}}\psi_n(\mathbf{r}, T/2)  }$, leads to zero polarization change 
\thinspace ${ \Delta \mathbf{P}_n=0}$ \thinspace for the time interval of  \thinspace $T$.
We point out that, the inversions actually assumed are \thinspace
${ (i, \mathbf{r}, t, \mathbf{k}_o) \longmapsto
	(-i, \mathbf{r}, -t, -\mathbf{k}_o)  }$
\thinspace where 
\thinspace
${ \mathbf{k}_o  }$ 
\thinspace is the initial value of the wavevector.  The time evolution of the wavevector \thinspace
${ \displaystyle \mathbf{k}(t)
=
\mathbf{k}_o
+
\frac{e}{\hbar}
\int_0^{\displaystyle t}
\mathbf{E}(t')
dt' }$, \thinspace
indicates that \thinspace
${ \mathbf{k}(t) \longmapsto -\mathbf{k}(t) }$
\thinspace  when  
\thinspace
${ \mathbf{E}(-t)=\mathbf{E}(t) }$,
\thinspace  which is true due to the assumed time-reversal symmetry.
This, for example, gives  zero polarization change for the whole hysteresis loop for an assumed time-reversal electric field 
\begin{eqnarray}
\displaystyle
\mathbf{E}(t)
=
\left\lbrace 
\begin{array}{c c r}
\mathbf{E}_o  \,
\left| 
 sin\left(
\frac{2 \pi}{T} t
\right)  
\right|  
&  \mbox{for}  &   0 \le t \le   \frac{T}{2}
\\*[8pt]  \nonumber
- \mathbf{E}_o \,
\left| 
sin\left(
\frac{2 \pi}{T} t
\right)  
\right| 
&  \mbox{for}  &   \frac{T}{2} \le t \le  {T},
\end{array}
\right.  
\end{eqnarray}
while for the half hysteresis loop
\thinspace 
${   0 \le t \le  T/2 }$,
\thinspace
this gives non-zero polarization change along the direction of the electric field. 
Time-reversal symmetry, can for example be broken when the material is magnetically ordered; in this framework, each now spinful electron's Hamiltonian  
\thinspace  Eq.~(\ref{e36abch}) \thinspace
may assume to have an extra term \thinspace  ${  -J_{ex} \, \bm{\sigma} \! \cdot \!  \hat{\mathbf{M}} }$, where   ${  J_{ex} }$ is the exchange coupling strength between the electron spin and the magnetization and \thinspace  ${  \hat{\mathbf{M}} }$ \thinspace is the magnetization unit vector.
\\

\subsubsection{Space-inversion and mirror symmetries}

Space-inversion symmetry can for example be broken by an  externally applied electric field or by a time-dependent strain that deforms the crystal. 
The time-dependent electric field breaks the space-inversion symmetry of the Hamiltonian: under  the inversions  \thinspace
${ (i, \mathbf{r}, t, \mathbf{k}_o) \longmapsto
(i, -\mathbf{r}, t, -\mathbf{k}_o)  }$, 
\thinspace  the time-dependent wavevector 
\thinspace
${ \displaystyle \mathbf{k}(t)
	=
	\mathbf{k}_o
	+
	\frac{e}{\hbar}
	\int_0^{\displaystyle t}
	\mathbf{E}(t')
	dt' }$  \thinspace
does not transform like 
 \thinspace
${  \mathbf{k}(t) \longmapsto  -\mathbf{k}(t)  }$
\thinspace due to the electric field 
\thinspace
${ \mathbf{E}(t) }$
\thinspace 
that is not inverted, therefore, the Hamiltonian \thinspace ${  H_k(\mathbf{r},t, \mathbf{k}) }$ \thinspace does not remain invariant.
Similarly, an assumed time-dependent strain that deforms the crystal breaks the space-inversion symmetry of the Hamiltonian for the time interval that
\thinspace ${  
V_{crys}(\mathbf{r}, t) 
\neq
V_{crys}(-\mathbf{r}, t) 
}$.

Likewise, mirror inversion symmetry is broken in the direction of the externally applied field and in the directions that the crystal potential is modified either due to lower equilibrium symmetry of the crystal or as a result of deformation by an externally applied strain.

\subsection{Longitudinal and transverse polarization change} 

Assuming that the externally applied electric field  
is in the  $x$  direction  \thinspace 
${ \displaystyle
	\frac{\partial \mathbf{k}(t)}{ \partial t}
	=
	\frac{e}{\hbar} E(t) \,  \mathbf{e}_x
}$,
\thinspace
the polarization change \thinspace Eq.~(\ref{e76aeh}) \thinspace  can  be separated into two normal contributions,  the longitudinal  one in the $x$ direction 
\thinspace ${\Delta {P}_{n, x}= 
\Delta \mathbf{P}_n  \!  \cdot   \mathbf{e}_x  
}$  \thinspace 
and the transverse one  normal to the electric field 
\thinspace ${\Delta {P}_{n, y}= 
	\Delta \mathbf{P}_n  \!  \cdot   \mathbf{e}_y  
}$.

The longitudinal polarization change in the $x$ direction, where the mirror symmetry is broken, 
is given by
\begin{widetext}
\begin{equation}  \label{e76aeh1}
\Delta {P}_{n, x}
=
\frac{e}{ (2 \pi)^3}
\int \limits_0^T  dt
\iiint \limits_{BZ}
\upsilon_{n, x} (t, \mathbf{k})
d^3k 
\  \  +  \  
\frac{e}{ \hbar (2 \pi)^3}
\int \limits_0^T  dt
\iiint \limits_{BZ}
\mathcal{S}_{n, x}(t, \mathbf{k})
d^3k 
\ - \
\frac{e}{ (2 \pi)^3}
\int \limits_0^T  dt
\iiint \limits_{BZ}
\mathcal{E}_{n, \, x} (t, \mathbf{k})
d^3k, 
\end{equation}
where the last term on the right side is the 
 Modern's Theory of Polarization well-known contribution \cite{{king1993theory},{vanderbilt1993electric}}.
On the other hand, the transverse polarization change 
is given by
\begin{eqnarray}  \label{e76aeh2}   \nonumber 
\Delta {P}_{n, y}
&=&
\frac{e}{ (2 \pi)^3}
\int \limits_0^T  dt
\iiint \limits_{BZ}
\upsilon_{n, y} (t, \mathbf{k})
d^3k 
\  \  +  \  
\frac{e}{ \hbar (2 \pi)^3}
\int \limits_0^T  dt
\iiint \limits_{BZ}
\mathcal{S}_{n, y}(t, \mathbf{k})
d^3k 
\ - \
\frac{e}{ (2 \pi)^3}
\int \limits_0^T  dt
\iiint \limits_{BZ}
\mathcal{E}_{n, \, y} (t, \mathbf{k})
d^3k
\\* [4pt]  
&& +
\frac{e^2}{  \hbar (2 \pi)^3}
\int \limits_0^T  dt  \, 
\left( 
E(t) 
\iiint \limits_{BZ}  
\mathcal{B}_{n, \, z} (t, \mathbf{k})
d^3k
\right)  
\end{eqnarray}
\end{widetext}
and has an extra bulk contribution, that is  \thinspace 
${  
\frac{e^2}{  \hbar (2 \pi)^3}
\int \limits_0^T  dt  \, 
\left( 
E(t) 
\iiint \limits_{BZ}  
\mathcal{B}_{n, \, z} (t, \mathbf{k})
\right), 
 }$
\thinspace
to be compared to the longitudinal counterpart \thinspace Eq.~(\ref{e76aeh1}).
Assuming that, during the polarization change process, the mirror symmetry remains unbroken along the transverse  $y$ direction, the first three contributions on the right side of \thinspace Eq.~(\ref{e76aeh2}) \thinspace integrate to zero according to Table \ref{T1}. On the other hand, the last contribution does not turn into zero due to mirror symmetry; instead, it integrates to zero due to time-reversal symmetry if it is present. Therefore, in magnetically ordered materials, where time-reversal symmetry is broken, the last contribution is not zero (although mirror symmetry may be present along $y$ direction).
The above transverse polarization change formula, is plausible to be compared to the one found by  Ref.~\onlinecite{goryo2002polarization}
where the electrons were subject to static and homogeneous electric and magnetic fields.

\subsection{Experimental setups and the non-Hermitian contribution}
It is helpful to investigate the existence of the    
non-Hermitian contributions in two important polarization change experiments.  Namely, in the polarization change experiment performed with a ferroelectric material and in the induced polarization change experiment performed with a piezoelectric material.

\subsubsection{Spontaneous polarization change in a ferroelectric material}
In an insulating ferroelectric material, the spontaneous polarization change is experimentally evaluated by applying a time-periodic bias ${ V(\mathbf{r}, t) }$ at the opposite boundaries of the material (which is in a capacitor configuration)  that results to a time-varying current ${ I(t) }$ that flows through a shorten ammeter. Taking into account the hysteresis-loop of the experiment, the measured switching current during the reversal of the polarization is used in order to calculate the spontaneous polarization change at zero electric field ${ \mathbf{E}=0 }$.
Therefore, in the spontaneous polarization change experiments, we have both time-varying externally applied electric field and time-varying realistic boundary conditions (non-periodicones) over the boundaries of the ferroelectric material.  In this framework (assuming non-magnetically ordered material), both integrands containing   \thinspace
${  \bm{\mathcal{S}}_n(t, \mathbf{k}) }$ \thinspace and \thinspace
${  \bm{\mathcal{E}}_n(t, \mathbf{k}) }$ \thinspace on the right side of  \thinspace Eq.~(\ref{e76aeh}) \thinspace are not zero and have to be taken into account. 
We note that, the space-coordinate integration for calculating  \thinspace
${  \bm{\mathcal{E}}_n(t, \mathbf{k}) }$ \thinspace
is taken over the entire volume $V$ of the material and
similarly, the non-Hermitian contribution 
 \thinspace
${  \bm{\mathcal{S}}_n(t, \mathbf{k}) }$ \thinspace
can equally be evaluated either as a bulk quantity 
(by the definition
$
\bm{\mathcal{S}}_n (t, \mathbf{k}) 
=
\left\langle H(\mathbf{r}, t) \, \Psi_n(t, \mathbf{k})
\, \vert \,  
\bm{\nabla_{\mathbf{k}}} 
\Psi_n(t, \mathbf{k}) \right\rangle
-
\left\langle  \Psi_n(t, \mathbf{k})
\, \vert \,  
H(\mathbf{r}, t) \,
\bm{\nabla_{\mathbf{k}}} 
\Psi_n(t, \mathbf{k}) \right\rangle
$
) where again the space-coordinate integration is taken over the entire volume $V$, or equivalently as a boundary quantity given by 
\thinspace Eq.~(\ref{nhcS1}).

\subsubsection{Induced polarization change in a piezoelectric material}    \label{piezoelct}
In an insulating piezoelectric material, strains are externally applied  to the material without any externally applied electric field 
${ \mathbf{E}=0 }$. The material is inserted into a shorted capacitor and the surface charges are removed by the electrodes passing through a shorten ammeter. While the crystal is strained the transient current flowing trough the material is measured. 
In this respect, we can employ a Hamiltonian like \thinspace Eq.~(\ref{e36abch}) \thinspace with zero externally applied scalar potential  \thinspace ${ \phi(x, t) = 0 }$,  \thinspace but, with time-dependent effective deformation energy term owing to time-dependent strains in order to describe the piezoelectric phenomena. The Hamiltonian 
\thinspace 
${ 
H_k(\mathbf{r},  \mathbf{k}, \varepsilon_{ij}(t) ) 
}$  \thinspace
will have implicit time-dependence due to the assumed  time-dependent  
\thinspace ${  \varepsilon_{ij}(t) }$  \thinspace stain tensor coefficients.
 In this fashion, the assumed instantaneous eigenvalue equation to be satisfied is 
 \thinspace 
 $
 H_k(\mathbf{r},  \mathbf{k},  \varepsilon_{ij}(t)) 
 \  u_n ( \mathbf{r}, \mathbf{k}, \varepsilon_{ij}(t))  
 =
 E_n(\mathbf{k},  \varepsilon_{ij}(t)) 
 \ u_n (\mathbf{r},  \mathbf{k},  \varepsilon_{ij}(t)),  
 $  \thinspace
and because the instantaneous eigenfunctions 
\thinspace 
${ \ u_n (\mathbf{r},  \mathbf{k},  \varepsilon_{ij}(t)),  
}$  \thinspace
do not have explicit  time-dependence, results to zero curvature 
 \thinspace
${ \bm{\mathcal{E}}_n(\mathbf{k},  \varepsilon_{ij}(t))=0  }$
\thinspace
by definition  Eq.~(\ref{e16ha}).
Therefore,  the  polarization change due to the piezoelectric phenomena can be theoretically modelled solely by means of the non-Hermitian boundary term  
\thinspace ${ \bm{\mathcal{S}}_{n} (t, \mathbf{k}) }$, which can be viewed as a novel result of the present work.

\subsection{Non-Hermitian contribution and the surface-charge theorem}

Surface-charge theorem, first introduced in 
Ref.~\onlinecite{vanderbilt1993electric}, 
states that, subject to four conditions\cite{vanderbilt2018berry},
the macroscopic surface charge density is given by  \thinspace
${ \sigma_{\text{surf}}:  =  \mathbf{P} 
\cdot  \hat{\mathbf{e}}_{a}  }$,
\thinspace 
where the symbol 
${  \textquoteleft := \text{\textquoteright}  }$ 
\thinspace 
carries the meaning
that the left-hand side is equal to one
of the values on the right-hand side
(the formal polarization ${ \mathbf{P} }$ is assumed to be a lattice-valued
quantity with lattice spacings ${e \mathbf{R}/ V_{cell} }$),
and 
\thinspace
${  \hat{\mathbf{e}}_{a}  }$ 
\thinspace
is the unit vector that is normal to the surface of the material on which the local charge density change is to be evaluated.
Instead of 
\thinspace
${ \sigma_{\text{surf}}:  =  \mathbf{P} 
\cdot  \hat{\mathbf{e}}_{a}  }$,
\thinspace 
the electric polarization change  \thinspace  $ \Delta \mathbf{P} $,  \thinspace which is a well-defined 
single-valued quantity,  
can be connected with the macroscopic surface charge density change 
\thinspace
 $ \Delta \sigma_{\text{surf}} $ 
\thinspace
in a singled-valued manner like
\begin{equation}
\Delta \sigma_{\text{surf}}= \Delta \mathbf{P}
\cdot \hat{\mathbf{e}}_{a}.
\end{equation}

The goal now is to derive a specific formula for the  
polarization surface charge density change 
${ \Delta \sigma_{\text{surf}} }$.
We are assuming that, when encountering only bulk localized states (where the electrons completely avoid the external boundaries of the material), the surface charge density change \thinspace
$ \Delta \sigma_{\text{surf}} $ 
\thinspace must turn into zero by definition.

In virtue of  Eqs.~(\ref{e36bbh})  and  (\ref{displac}), 
as well as the definition of the extended velocity operator 
Eq.~(\ref{e1}), the polarization change $ \Delta \mathbf{P} $  must have two explicit contribution, namely, one due to the standard bulk displacement of the electrons
\begin{equation} \label{e36bbh2}
\Delta \mathbf{P}_{1}
=
\frac{1}{V}
\sum \limits_i   q_i  \ 
\left(  
\int_0^T  \!  \!
\left\langle 
\psi_i(t) \vert \, \mathbf{v}
\, 
\vert
\psi_i(t)
\right\rangle
dt
\right), 
\end{equation}
where ${ \mathbf{v} }$ is the standard bulk velocity operator  defined by \thinspace Eq.~(\ref{e2}),  \thinspace and a second one, due to the non-Hermitian contribution
\begin{equation} \label{e36bbh3}
\Delta \mathbf{P}_{2}
=
\frac{1}{V}
\sum \limits_i   q_i  \ 
\left(  
\int_0^T  \!  \!
\left\langle 
\psi_i(t) \vert \, \mathbf{v}_{b}
\, 
\vert
\psi_i(t)
\right\rangle
dt
\right) 
\end{equation}
where ${ \mathbf{v}_{b} }$ is the boundary velocity operator  defined by \thinspace Eq.~(\ref{e3}).

With respect to an adiabatically evolved eigenstate of the Hamiltonian, by applying the adiabatic form of the DHFT Eq.~(\ref{e14hab}),
the bulk velocity operator expectation value \thinspace ${ \left\langle \mathbf{v}  \right\rangle_n  }$  can be evaluated by means of Eq~(\ref{e14eh}).  
It has three explicit bulk contributions and one non-Hermitian given by  \thinspace 
${ \displaystyle \frac{1}{\hbar} \widetilde{
\bm{\mathcal{S}}_{n}} (t, \mathbf{k})   }$  \thinspace
due to the momentum gradient operator 
 $ \bm{\nabla_{\mathbf{k}}}$  that becomes anomalous.

By taking into account the adiabatic time evolution of the Hamiltonian's eigenstate, as well as its Bloch form, the expectation value of the boundary velocity operator 
\thinspace ${ \left\langle \mathbf{v}_b  \right\rangle_n  }$  \thinspace is given by \thinspace Eq.(\ref{e80abha}), that is 
${
\displaystyle 
\mathbf{v}_{b}(t, \mathbf{k})
=
\frac{1}{\hbar}
\left(
\bm{\mathcal{S}}_n(t, \mathbf{k})
-
\widetilde{
\bm{\mathcal{S}}_{n}}  (t, \mathbf{k})
 \right) 
}$, 
where both terms \thinspace ${\bm{\mathcal{S}}_n(t, \mathbf{k})}$ \thinspace 
 and  
\thinspace ${ \widetilde{\bm{\mathcal{S}}}_n(t, \mathbf{k})}$ \thinspace are non-Hermitian contributions (with use of the two different bases participating in the Bloch forms).
By summing now the two contributions from \thinspace  Eq.~(\ref{e36bbh2}) \thinspace and 
\thinspace  Eq.~(\ref{e36bbh3}), \thinspace
the contributions depending on 
 \thinspace 
${ \bm{\mathcal{S}}_{n} (t, \mathbf{k})   }$  \thinspace
cancel one each other and 
we end up with \thinspace Eq.~(\ref{e76aeh}) \thinspace
for the collective electronic polarization change.

In this framework,
we argue that, the macroscopic surface charge density change 
of  a single fully occupied valence band is given by
\begin{equation}  \label{sc}
\Delta \sigma_{\text{surf}}
=
\Delta \mathbf{P_{(b)}}_n \cdot \mathbf{e}_a
=
\frac{e}{\hbar (2 \pi)^3}
\int \limits_0^T
dt 
\iiint \limits_{BZ}
\bm{\mathcal{S}}_n(t, \mathbf{k})
 \cdot   \mathbf{e}_a
\, \,  \, 
d^3k.
\end{equation}
We point out that  ${ \Delta \sigma_{\text{surf}} }$  \thinspace that is evaluated by means of the non-Hermitian        ${ \bm{\mathcal{S}}_n(t, \mathbf{k}) }$, is always a purely real quantity (as it is supposed to be)  although the non-Hermtian contributions are in general complex quantities.
The reason being that  ${ \bm{\mathcal{S}}_n(t, \mathbf{k}) }$,  which incorporates the instantaneous Bloch eigenstates (by virtue of  Eq.~(\ref{nhcS1})),  turns out to be purely real due to
${ \bm{\nabla_{\mathbf{k}}}H(\mathbf{r},t)=0 }$  in  Eq.~(\ref{a201}).

The latter surface charge density change given by our proposed surface-charge theorem  (i)  is a well-defined quantity,
(ii) is extremely sensitive to the realistic boundary conditions of the electrons (incorporated in the Bloch states), and (iii) it turns zero value when encountering only bulk localized states.
This surface charge density change is evaluated only in terms of the emerging non-Hermitian boundary contribution in an intuitively plausible manner, that is, it relies only on the boundary properties of the electrons' wavefunctions as expected
(although the boundary properties are intertwined with the bulk properties due to the boundary non-Hermitian contribution that can equally be evaluated by bulk volume integration).

\section{Conclusions}   \label{concl}
In this work we have re-examined the charge pumping and polarization change processes.  
By using the extended velocity operator definition and a dynamical extension of the Hellmann-Feynman theorem (DHFT) that we have derived, we have shown how the  charge pumping can be linked to the boundaries of the system  by means of a  displacement originating from an emergent  
non-Hermiticity of the Hamiltonian,  that can be evaluated in terms of a Berry curvature owing to the DHFT. 
Similarly,  by using the extended velocity operator and the DHFT, we have extended the modern theory of polarization by finding two overlooked contributions. One is an emergent boundary non-Hermitian contribution, and the other depends on generalized Berry curvatures.

Boundary non-Hermitian contributions are involved in both extended velocity operator definition and DHFT.
A common characteristic of these non-Hermitian contributions is that they are both by definition bulk quantities,  that, however, may equally be evaluated as  boundary quantities due to a symmetry that allows the bulk integrations to be transformed into  boundary integrations.
These non-Hermitian boundary contributions are always overlooked within von Neumann's theory of self-adjoint operators. On the other hand, if one lays on the Schr\"odinger (or Dirac) dynamics, and insists on the unitary  time evolution of closed systems, the non-Hermitian boundary contributions  have to be taken into account a priori in order to have self-consistent coupling between time-evolution and observables' operators. This route leads to an extended definition of operators which gives  contributions to observables so far widely overlooked.

We have also studied the  non-equilibrium, driven charge pumping processes by assuming delocalized  Floquet-Bloch states  for the electrons' motion. By   using the extended velocity operator definition together with the non-adiabatic form of  DHFT,  we have evaluated the collective displacement of the center of mass of the electrons \thinspace 
${  \sum_i^N  \Delta \left\langle x_i \right\rangle \! |_{bound} \ / \ {N}  }$  \, that has an emergent non-Hermitian origin.  
We have found that, for a fully occupied Floquet-Bloch band \thinspace $ \varepsilon_a $, \thinspace the quantization of the pumped charge (per cycle) 
breaks down due to a non-trivial (non-integrable) Aharonov-Anandan phase \cite{aharonov1987phase}, that emerges whenever the  Floquet-Bloch states 
cannot satisfy the periodic gauge along the direction of the externally applied electric field.

We then studied the general polarization change processes, and found that the adiabatic ground state polarization change  
has two extra, overlooked,  contributions which are absent from the modern theory \cite{{king1993theory},{ortiz1994macroscopic},{resta1994macroscopic},{resta2007theory},{resta2010electrical},{vanderbilt2018berry}}. The one is attributed to a Berry curvature and is not zero only within ferromagnetic insulators and the other is a non-Hermitian boundary contribution. When one takes into account only bulk localized states, that is, the wavefunctions are zero over the material boundaries, the overlooked non-Hermitian contribution always turns into zero.
This non-Hermitian contribution is very sensitive to the realistic boundary conditions of the wavefunctions during the time-periodic process that causes the polarization change. We therefore expect it to be significant in biased insulators with electron charge accumulation over their boundaries during the process that causes the polarization change (i.e. during the spontaneous polarization change experiments in a ferroelectric material,  or during the induced polarization change experiments in a piezoelectric material).

Finally,  we have demonstrated how one can extend  the surface-charge theorem  of  
Ref.~\onlinecite{vanderbilt1993electric}. 
We have shown how one can evaluate a well defined  surface charge change in terms of a boundary, non-Hermitian contribution. When only bulk localized states are taken into account, the surface charge change   \thinspace  Eq.~(\ref{p4}) 
\thinspace turns into zero by definition as expected. Overall, our work demonstrates that the widely overlooked emergent non-Hermiticities can be crucial to be included for the consistent and complete description of key properties and phenomena in the physics of condensed matter, especially in view of the fact that they are intertwined with geometric (of a generalized Berry type) and topological quantities, something that has carefully and rigorously been shown in the present article.

\appendix

\section{Boundary integral form of the non-Hermitian contribution $ \mathbf{S}(t, \mathbf{R}) $ 
in position representation.}
\label{BC}

We assume a 3D system and Cartesian coordinates, where the parameter gradient operator is given by \thinspace
${ \displaystyle
	\bm{\nabla_{\mathbf{R}}}
	=
	\sum_{i=1}^{3} 
	\mathbf{e}_i
	\frac{\partial}{\partial R_i}
}$.
In order to analytically evaluate the expectation value of the non-Hermitian boundary contribution \thinspace ${  \bm{\mathcal{S}}(t, \mathbf{R})  }$,
\thinspace we need an explicit form of the Hamiltonian \thinspace 
${ H(t, \mathbf{r}, \mathbf{R}) }$ \thinspace that governs the time evolution. Therefore, we assume a general form Hamiltonian \thinspace ${  H(t, \mathbf{r}, \mathbf{R}) }$ \thinspace  that, by setting the appropriate limits, may capture, (i) a spinless non-relativistic motion,  (ii) a fully relativistic spinful motion, or finally (iii) a spinful motion in the non-relativistic limit.  For these purposes and for the clarity of the derivation,  it is convenient to separate the Hamiltonian into
\begin{eqnarray} \label{a2S}  \nonumber
H(t, \mathbf{r}, \mathbf{R}) &=&
\lambda_1   H_1(t, \mathbf{r}, \mathbf{R})
\ + \
\lambda_2
H_2(t, \mathbf{r}, \mathbf{R})
\\* \nonumber
&&
\! \! \! \! \! \! \! \! \! \! \! \! \! \! \! \! 
+ \
\lambda_3
H_3(t, \mathbf{r}, \mathbf{R})
\ + \
\lambda_4
H_4(t, \mathbf{r}, \mathbf{R})
\ + \
\lambda_5 
H_5(t, \mathbf{r}, \mathbf{R})
\\* 
\end{eqnarray}
where  $ \lambda_i  $ are dimensionles parameters that we set to the values $0$ or $1$ according to the limits that we are looking for.

In this framework,  
\begin{equation} \label{a3S} 
H_1(t, \mathbf{r}, \mathbf{R})=\frac{1}{2m}\bm{\Pi}(\mathbf{r},t, \mathbf{R})^2 
\end{equation}
is the spinless motion kinetic energy of the electron, 
\begin{equation}
H_2(t, \mathbf{r}, \mathbf{R})=V(\mathbf{r},t, \mathbf{R})
\end{equation}
is the interaction between the electron charge and the scalar potential,
\begin{equation}
H_3(t, \mathbf{r}, \mathbf{R})=
-\frac{e\hbar}{2mc} \bm{\sigma}\!\cdot\!\mathbf{B}(\mathbf{r},t, \mathbf{R})
\end{equation}
is the interaction between the electron's magnetic moment and the magnetic field, while
\begin{equation} \label{a4S} 
H_4(t, \mathbf{r}, \mathbf{R})
=
\, \bm{\mathcal{O}}(\mathbf{r},t, \mathbf{R}, \bm{\sigma}) \!\cdot\! \bm{\Pi}(\mathbf{r},t, \mathbf{R})
\end{equation}
captures either the spin-orbit interaction within the non-relativistic limit by setting  \
${ \displaystyle
\bm{\mathcal{O}} (\mathbf{r},t,  \mathbf{R}, \bm{\sigma}) \equiv  
\frac{\hbar}{4m^2c^2} 
\bm{\sigma} \times \nabla V(\mathbf{r},t  \mathbf{R})
}$,
\
or, in the fully relativistic motion,
 by setting  \
${
\bm{\mathcal{O}} (\mathbf{r},t,  \mathbf{R}, \bm{\sigma}) \equiv 
c \, \bm{\alpha}	 
}$,
with $\bm{\alpha} $ being the Dirac matrices,
corresponds to the kinetic energy of the electron.
Finally, 
\begin{equation}
H_5(t, \mathbf{r}, \mathbf{R})
=
\beta m c^2
\end{equation}
gives the rest energy of the electron.
 
In this framework, the non-Hermitian boundary contribution 
\thinspace 
${  \displaystyle
\bm{\mathcal{S}}(t, \mathbf{R}) 
= 
\left\langle \Psi(t, \mathbf{R}) \, \vert \,  
\left( 
H(t, \mathbf{R})^+ 
- 
H(t, \mathbf{R}) 
\right)  
\bm{\nabla_{\mathbf{R}}} 
\Psi(t, \mathbf{R}) \right\rangle 
}$,
\thinspace 
 is given by
\begin{eqnarray} \label{a5S}   \nonumber
\bm{\mathcal{S}}(t, \mathbf{R})   
 &=&
 \lambda_1 \,
\bm{\mathcal{S}}_1(t, \mathbf{R}) 
 \ 
 +
 \
 \lambda_2  \,
 \bm{\mathcal{S}}_2(t, \mathbf{R}) 
\\*  \nonumber 
&&
+
 \
 \lambda_3  \,
 \bm{\mathcal{S}}_3(t, \mathbf{R}) 
 \ 
 +
 \
 \lambda_4  \,
 \bm{\mathcal{S}}_4(t, \mathbf{R}) 
 \ 
 +
 \
 \lambda_5  \,
 \bm{\mathcal{S}}_5(t, \mathbf{R}) 
 \\* 
\end{eqnarray}  
where, in position representation and for real potentials, each term is given by
\begin{widetext}
\begin{equation} \label{a6S}
\bm{\mathcal{S}}_1(t, \mathbf{R}) 
=
\left\langle \, H_{1} \Psi \vert 
\bm{\nabla_{\mathbf{R}}}
\Psi \right\rangle
-
\left\langle \Psi  \vert H_{1} 
\bm{\nabla_{\mathbf{R}}}
\Psi \right\rangle
=
\frac{1}{2m} 
\iiint_V 
\left( \frac{}{}  
(\, \bm{\Pi}^2 \Psi)^{\dagger}  \, \bm{\nabla_{\mathbf{R}}} \Psi
-
\Psi^{\dagger} \, \bm{\Pi}^2 \, \bm{\nabla_{\mathbf{R}}} \Psi \,
\right) 
dV,
\end{equation}

\begin{equation} \label{a6S1}
\bm{\mathcal{S}}_2(t, \mathbf{R}) 
=
\left\langle \, H_{2} \Psi \vert 
\bm{\nabla_{\mathbf{R}}}
\Psi \right\rangle
-
\left\langle \Psi  \vert H_{2} 
\bm{\nabla_{\mathbf{R}}}
\Psi \right\rangle
=
\iiint_V 
\left( \frac{}{}  
(\, V \, \Psi)^{\dagger}  \, \bm{\nabla_{\mathbf{R}}} \Psi
-
\Psi^{\dagger} \, V  \, \bm{\nabla_{\mathbf{R}}} \Psi \,
\right) 
dV
=0,
\end{equation}

\begin{equation} \label{a6S2}
\bm{\mathcal{S}}_3(t, \mathbf{R}) 
=
\left\langle \, H_{3} \Psi \vert 
\bm{\nabla_{\mathbf{R}}}
\Psi \right\rangle
-
\left\langle \Psi  \vert H_{3} 
\bm{\nabla_{\mathbf{R}}}
\Psi \right\rangle
=
-\frac{e\hbar}{2mc} 
\iiint_V 
\left( \frac{}{}  
(\,  \bm{\sigma}\!\cdot\!\mathbf{B}  \, \Psi)^{\dagger}  \, \bm{\nabla_{\mathbf{R}}} \Psi
-
\Psi^{\dagger} \,  \bm{\sigma}\!\cdot\!\mathbf{B}  \, \bm{\nabla_{\mathbf{R}}} \Psi \,
\right) 
dV
=0,
\end{equation}

\begin{equation} \label{a6S3}
\bm{\mathcal{S}}_4(t, \mathbf{R}) 
=
\left\langle \, H_{4} \Psi \vert 
\bm{\nabla_{\mathbf{R}}}
\Psi \right\rangle
-
\left\langle \Psi  \vert H_{4} 
\bm{\nabla_{\mathbf{R}}}
\Psi \right\rangle
=
\iiint_V 
\left( \frac{}{}  
(\,  \bm{\mathcal{O}} \!\cdot\! \bm{\Pi}  \, \Psi)^{\dagger}  \, \bm{\nabla_{\mathbf{R}}} \Psi
-
\Psi^{\dagger} \,   \bm{\mathcal{O}} \!\cdot\! \bm{\Pi}  \, \bm{\nabla_{\mathbf{R}}} \Psi \,
\right) 
dV,
\end{equation}

\begin{equation} \label{a6S4}
\bm{\mathcal{S}}_5(t, \mathbf{R}) 
=
\left\langle \, H_{5} \Psi \vert 
\bm{\nabla_{\mathbf{R}}}
\Psi \right\rangle
-
\left\langle \Psi  \vert H_{5} 
\bm{\nabla_{\mathbf{R}}}
\Psi \right\rangle
=
m c^2
\iiint_V 
\left( \frac{}{}  
(\,  \beta  \, \Psi)^{\dagger}  \, \bm{\nabla_{\mathbf{R}}} \Psi
-
\Psi^{\dagger} \,  \beta  \, \bm{\nabla_{\mathbf{R}}} \Psi \,
\right) 
dV
=0
\end{equation}
where we have used that, ${ (V \, \Psi)^{\dagger}= \Psi^{\dagger} \, V }$, and  \ 
${( \bm{\sigma}\!\cdot\!\mathbf{B}  \, \Psi)^{\dagger} =
\Psi^{\dagger} \, \bm{\sigma}\!\cdot\!\mathbf{B}  }$  \
as well as  \
${( \beta   \, \Psi)^{\dagger} =
\Psi^{\dagger} \,  \beta  }$.
We next move to find further simplified analytic forms of  \  ${ \bm{\mathcal{S}}_1(t, \mathbf{R})  }$  \ and  \  ${ \bm{\mathcal{S}}_4(t, \mathbf{R})  }$  respectively.

\subsection{Analytic form of  $ \mathbf{S}_1(t, \mathbf{R}) $ }

By using the Cartesian coordinates, we evaluate each component separately, thus,  the $x$ component of Eq.~(\ref{a6S}) is given from
\begin{equation} \label{a9S}
\bm{\mathcal{S}}_{1, x}(t, \mathbf{R}) =
\frac{1}{2m} 
\mathbf{e}_x  \iiint_V
\left( \frac{}{}   
(\, \bm{\Pi}^2 \Psi )^{\dagger}  \,  \frac{\partial \Psi}{\partial R_x} 
-
\Psi^{\dagger} \, \bm{\Pi}^2 \, \frac{\partial \Psi}{\partial R_x} 
\, \right) 
dV,
\end{equation} 
where we have made use of the constant direction of the Cartesian unit vector \thinspace $\mathbf{e}_x $. Using the explicit form of the kinematic momentum operator 
\thinspace
${ \displaystyle \bm{\Pi} = -i \hbar \nabla - \frac{e}{c} \mathbf{A}(\mathbf{r},t) }$, 
\  Eq.~(\ref{a9S}) \thinspace takes the form
\begin{equation} \label{a10S}
\bm{\mathcal{S}}_{1, x}(t, \mathbf{R})
=
\frac{1}{2m} 
\mathbf{e}_x  \iiint_V 
\left( \frac{}{} 
(-i \hbar \nabla - \frac{e}{c} \mathbf{A})  \cdot  \bm{\Pi} \Psi \,)^{\dagger}  \,  \frac{\partial \Psi}{\partial R_x} 
\, - \, 
\Psi^{\dagger} \, (-i \hbar \nabla - \frac{e}{c} \mathbf{A}) \cdot \bm{\Pi} \,  \frac{\partial \Psi}{\partial R_x}  \,
\right) 
dV 
\end{equation}
that leads to
\begin{eqnarray} \label{a11S} \nonumber
\bm{\mathcal{S}}_{1, x}(t, \mathbf{R})
&=&
\frac{1}{2m} 
\mathbf{e}_x (i \hbar) \iiint_V  
\nabla \cdot
\left(\frac{}{}  
(\,\bm{\Pi} \Psi \,)^{\dagger}  \, \frac{\partial \Psi}{\partial R_x}
\, +  \,
\Psi^{\dagger}  \, \bm{\Pi} \, \frac{\partial \Psi}{\partial R_x} \,
\right) 
dV 
\nonumber \\*[6pt]
&& 
-
\frac{1}{2m} 
\mathbf{e}_x (i \hbar) \iiint_V  
\left( \frac{}{} 
(\,\bm{\Pi} \Psi \,)^{\dagger}  \,\cdot\nabla 
( \frac{\partial \Psi}{\partial R_x} )
\, + \,
\nabla \Psi^{\dagger} \cdot \bm{\Pi} \, \frac{\partial \Psi}{\partial R_x}  \,
\, \right) 
dV 
\nonumber \\*[6pt] 
&&
- \frac{1}{2m} 
\mathbf{e}_x  \left( \frac{e}{c}\right)  \iiint_V 
\left( \frac{}{} 
(\,\mathbf{A}  \cdot  \bm{\Pi} \Psi \,)^{\dagger}  \, 
\frac{\partial \Psi}{\partial R_x}
\, - \, 
\Psi^{\dagger} \mathbf{A}  \cdot  \bm{\Pi} \, 
\frac{\partial \Psi}{\partial R_x} 
\,
\, \right) 
dV 
\end{eqnarray} 
which gives
\begin{eqnarray} \label{a12S} \nonumber
\bm{\mathcal{S}}_{1, x}(t, \mathbf{R})
&=&
\frac{i \hbar}{2m} 
\mathbf{e}_x  \oiint_S  
\left( \frac{}{} 
(\,\bm{\Pi} \Psi \,)^{\dagger}  \, \frac{\partial \Psi}{\partial R_x}
+
\Psi^{\dagger}  \bm{\Pi} \,  \frac{\partial \Psi}{\partial R_x}  \,
\right) 
\! \cdot \! d\mathbf{s}
\nonumber \\*[6pt]
&& 
+
\frac{1}{2m} 
\mathbf{e}_x \iiint_V  
\left( \frac{}{} 
(\,\bm{\Pi} \Psi \,)^{\dagger} \cdot 
(-i \hbar \, \nabla \frac{\partial \Psi}{\partial R_x})
\, - \, 
(-i \hbar \, \nabla \Psi)^{\dagger} \cdot \bm{\Pi} \, \frac{\partial \Psi}{\partial R_x}  \,
\right) 
dV 
\nonumber \\*[6pt] 
&&
- 
\frac{1}{2m} 
\mathbf{e}_x 
\left( \frac{e}{c}\right)  
\iiint_V 
\mathbf{A} \! \cdot \!
\left( \frac{}{} 
(\,\bm{\Pi} \Psi \,)^{\dagger}  \, \frac{\partial \Psi}{\partial R_x}
\, - \,
\Psi^{\dagger} \bm{\Pi} \, \frac{\partial \Psi}{\partial R_x} \,
\right) 
dV. 
\end{eqnarray}
Replacing now 
\ ${ \displaystyle -i \hbar \nabla =\bm{\Pi} + \frac{e}{c} \mathbf{A}  }$ \ in the second term of the right hand side of \thinspace Eq.~(\ref{a12S}) \thinspace we find
\begin{eqnarray} \label{a13S} \nonumber
\bm{\mathcal{S}}_{1, x}(t, \mathbf{R})
&=&
\frac{i \hbar}{2m} 
\mathbf{e}_x  \oiint_S  
\left( \frac{}{} 
(\,\bm{\Pi} \Psi \,)^{\dagger}  \, \frac{\partial \Psi}{\partial R_x}
\, + \, 
\Psi^{\dagger}  \bm{\Pi} \, \frac{\partial \Psi}{\partial R_x} \,
\right) 
\! \cdot \! d\mathbf{s}
\nonumber \\* [6pt]
&& 
+
\frac{1}{2m} 
\mathbf{e}_x \iiint_V  
\left(  
(\,\bm{\Pi} \Psi \,)^{\dagger} \!\cdot\! 
( \bm{\Pi} + \frac{e}{c} \mathbf{A} ) 
(\frac{\partial \Psi}{\partial R_x})
\, - \,
( 
(\bm{\Pi} + \frac{e}{c} \mathbf{A} )
\Psi )^{\dagger} \cdot \bm{\Pi} \, \frac{\partial \Psi}{\partial R_x}  \,
\right) 
dV 
\nonumber \\*[6pt] 
&&
- 
\frac{1}{2m} 
\mathbf{e}_x 
\left(  \frac{e}{c} \right) 
\iiint_V 
\mathbf{A}  \cdot 
\left(\frac{}{}  
(\,\bm{\Pi} \Psi \,)^{\dagger}  \, \frac{\partial \Psi}{\partial R_x}
\, - \, 
\Psi^{\dagger} \bm{\Pi} \, \frac{\partial \Psi}{\partial R_x}  \,
\right) 
dV 
\end{eqnarray}
which finally gives 
\begin{equation} \label{a14S}
\bm{\mathcal{S}}_{1, x}(t, \mathbf{R})
=
\frac{i \hbar}{2m} 
\mathbf{e}_x  \oiint_S  
\left(\frac{}{}  
(\,\bm{\Pi} \Psi \,)^{\dagger}  \, \frac{\partial \Psi}{\partial R_x}
\, + \,
\Psi^{\dagger} \, \bm{\Pi} \, \frac{\partial \Psi}{\partial R_x} \,
\, \right) 
\!\cdot\! d\mathbf{s}.
\end{equation}
Taking now into account that \ ${ d\mathbf{s}=\mathbf{n}\,ds }$, \ where $\mathbf{n}$ is the unit vector that is locally normal to the surface $S$, \thinspace Eq.~(\ref{a14S}) \thinspace can be recast in the form
\begin{equation} \label{a15S}
\bm{\mathcal{S}}_{1, x}(t, \mathbf{R})
=
\frac{i \hbar}{2} 
\mathbf{e}_x  \oiint_S  
\mathbf{n} \!\cdot\!
\left(  
(\, \frac{\bm{\Pi}}{m} \Psi \,)^{\dagger} 
\, + \,
\Psi^{\dagger}  \frac{\bm{\Pi}}{m} \,
\right)
 \frac{\partial \Psi}{\partial R_x}
\, ds
\end{equation}
where we have used \ 
${ \displaystyle \mathbf{n} \!\cdot\!\nabla= n_x \frac{\partial}{\partial_x}+ n_{\psi} \frac{\partial}{\partial_{\psi}}  + n_z \frac{\partial}{\partial_z} }$, in order to restructure the term \linebreak
${ \displaystyle
\Psi^{\dagger} 
	\left(  
	\bm{\Pi} \, 
	 \frac{\partial \Psi}{\partial R_x}
	\right) 
	\!\cdot\! d\mathbf{s} }$
\ in the form \ 
${  \displaystyle
\mathbf{n} \!\cdot\!
	\left( 
	\Psi^{\dagger}  \bm{\Pi} \,
	\right)
	 \frac{\partial \Psi}{\partial R_x}
	\, ds 
}$. 
By adding all of the Cartesian components \
${ 
\bm{\mathcal{S}}_{1}(t, \mathbf{R})	
=
\bm{\mathcal{S}}_{1, x}(t, \mathbf{R})
+
\bm{\mathcal{S}}_{1, y}(t, \mathbf{R})
+
\bm{\mathcal{S}}_{1, z}(t, \mathbf{R})	
}$ \
we find the form of the 
${ \bm{\mathcal{S}}_{1}(t, \mathbf{R})	 }$
that is given by
\begin{equation} \label{a16S}
\bm{\mathcal{S}}_{1}(t, \mathbf{R})	
=
\frac{i \hbar}{2} 
\oiint_S  
\mathbf{n} \!\cdot\!
\left(  
(\, \frac{\bm{\Pi}}{m} \Psi \,)^{\dagger} 
\, + \, 
\Psi^{\dagger}  \frac{\bm{\Pi}}{m} \,
\right)
\bm{\nabla_{\mathbf{R}}}
 \Psi 
\, ds.
\end{equation}

\subsection{Analytic form of  $ \mathbf{S}_4(t, \mathbf{R}) $ }

By using the Cartesian coordinates, the $x$ component of Eq.~(\ref{a6S3}) is given from
\begin{equation}
\bm{\mathcal{S}}_{4, x}(t, \mathbf{R}) 
=
\mathbf{e}_x 
\iiint_V 
\left( \frac{}{}  
(\,  \bm{\mathcal{O}} \!\cdot\! \bm{\Pi}  \, \Psi)^{\dagger}  \, 
 \frac{\partial \Psi}{\partial R_x}
-
\Psi^{\dagger} \,   \bm{\mathcal{O}} \!\cdot\! \bm{\Pi}  \,  
 \frac{\partial \Psi}{\partial R_x} \,
\right) 
dV,
\end{equation}
which,  by using 
 \thinspace 
${ {\bm{\mathcal{O}}^{ \dagger}}=
\bm{\mathcal{O}}	
 }$
\thinspace
and 
\ ${(
\bm{\mathcal{O}}		
\,\Psi)^{\dagger}=\Psi^{\dagger}\, \bm{\mathcal{O}}}$, \thinspace as well as
\linebreak
${  \bm{\mathcal{O}}
\!\cdot\!\bm{\Pi}
=
\bm{\Pi}\!\cdot\! \bm{\mathcal{O}}
+
 i \hbar (\nabla\!\cdot\!  \bm{\mathcal{O}} ) }$,
and the following properties:
\begin{eqnarray} \label{a19S} \nonumber
(\bm{\mathcal{O}} \!\cdot\!\bm{\Pi} \,\Psi)^{\dagger}
&=&
(\bm{\Pi}\!\cdot\!  \bm{\mathcal{O}} \,\Psi)^{\dagger} \, + \,
(\,i \hbar (\nabla\!\cdot\! \bm{\mathcal{O}} ) \,\Psi)^{\dagger}
\nonumber \\*[4pt]
&=&
+
i\hbar 
\left( 
\nabla\!\cdot\!(\bm{\mathcal{O}} \,\Psi)^{\dagger}
\right) 
\, - \,
\frac{e}{c} \mathbf{A}\!\cdot\!(\bm{\mathcal{O}} \,\Psi)^{\dagger}
\, - \,
i \hbar \,\Psi^{\dagger}\,(\nabla\!\cdot\! \bm{\mathcal{O}})
\nonumber \\*[4pt]
&=& 
+
i\hbar
\left( 
\nabla\!\cdot\!(\Psi^{\dagger} \bm{\mathcal{O}})
\right) 
\, - \,
\frac{e}{c} \mathbf{A}\!\cdot\!(\Psi^{\dagger} \bm{\mathcal{O}} )
\, - \,
i \hbar \,\Psi^{\dagger}\,(\nabla\!\cdot\! \bm{\mathcal{O}})
\nonumber \\*[4pt]
&=& +i\hbar\nabla\Psi^{\dagger}\!\cdot\! \bm{\mathcal{O}}
\, + \,
i \hbar \,\Psi^{\dagger}\,(\nabla\!\cdot\! \bm{\mathcal{O}})
\, - \,
\frac{e}{c} \mathbf{A}\!\cdot\!(\Psi^{\dagger} \bm{\mathcal{O}})
\, - \,
i \hbar \,\Psi^{\dagger}\,(\nabla\!\cdot\! \bm{\mathcal{O}})
\nonumber \\*[4pt]
&=&
+i\hbar\nabla\Psi^{\dagger}\!\cdot\!  \bm{\mathcal{O}}
\, - \,
\frac{e}{c} \Psi^{\dagger} \, \mathbf{A}\!\cdot\! \bm{\mathcal{O}}
\end{eqnarray}
together with
\begin{eqnarray} \label{a20S} \nonumber
\bm{\mathcal{O}}
\!\cdot\!\bm{\Pi}\,
\frac{\partial \Psi}{\partial R_x}
&=& 
\bm{\Pi} \!\cdot\!
\bm{\mathcal{O}}
\,
\frac{\partial \Psi}{\partial R_x}
\, + \,
i \hbar (\nabla\!\cdot\! \bm{\mathcal{O}})\,
\frac{\partial \Psi}{\partial R_x}
\nonumber \\*[4pt]  \nonumber 
&=& 
-i \hbar\nabla\!\cdot\!(
\bm{\mathcal{O}}
\, \frac{\partial \Psi}{\partial R_x})
\, - \,
\frac{e}{c} \mathbf{A}\!\cdot\!(\bm{\mathcal{O}} \, \frac{\partial \Psi}{\partial R_x})
\, + \, 
i \hbar (\nabla\!\cdot\! \bm{\mathcal{O}} ) \, \frac{\partial \Psi}{\partial R_x} ,
\\*
\end{eqnarray}
we find
\begin{eqnarray} \label{a21S} \nonumber
\bm{\mathcal{S}}_{4, x}(t, \mathbf{R}) 
&=&
\mathbf{e}_x  \iiint_V 
\left( \frac{}{}
i\hbar\nabla\Psi^{\dagger} \cdot  \bm{\mathcal{O}} \, \frac{\partial \Psi}{\partial R_x}
\, - \,
\frac{e}{c} \Psi^{\dagger} \, \mathbf{A} \cdot \bm{\mathcal{O}} \,
\frac{\partial \Psi}{\partial R_x}
\right) 
dV
\nonumber \\*[6pt]
&-&
\mathbf{e}_x  \iiint_V 
\left( 
-i \hbar\Psi^{\dagger}\,\nabla\!\cdot\!(\bm{\mathcal{O}} \, \frac{\partial \Psi}{\partial R_x})
\, - \, 
\frac{e}{c} \Psi^{\dagger}\,\mathbf{A}\!\cdot\!(\bm{\mathcal{O}} \, \frac{\partial \Psi}{\partial R_x})
\, + \, 
i \hbar \Psi^{\dagger}\,(\nabla\!\cdot\! \bm{\mathcal{O}}) \, \frac{\partial \Psi}{\partial R_x}
\right) dV
\nonumber \\*[6pt]
&=&  
\mathbf{e}_x (i\hbar) \iiint_V 
\nabla \!\cdot\! 
\left( 
\Psi^{\dagger}\, \bm{\mathcal{O}} \, \frac{\partial \Psi}{\partial R_x}
\right) 
dV
\, - \, 
\mathbf{e}_x (i\hbar) \iiint_V 
\Psi^{\dagger}\,(\nabla\!\cdot\! \bm{\mathcal{O}}) \, \frac{\partial \Psi}{\partial R_x}
dV
\nonumber \\*[6pt]
&=&  
\mathbf{e}_x 
(i\hbar) 
\oiint_S  
\left( 
\Psi^{\dagger}\, \bm{\mathcal{O}} \, \frac{\partial \Psi}{\partial R_x}
\right)\!\cdot\!d\mathbf{s} 
\end{eqnarray} 
where we have used \ ${ \nabla \! \cdot \! \bm{\mathcal{O}}=0 }$,  \
which is true either when taking into account the spin-orbit interaction
\[
\nabla \cdot \bm{\mathcal{O}}=\frac{\hbar}{4m^2c^2} 
\nabla\cdot
\left( 
\bm{\sigma} \times \nabla V(\mathbf{r},t)
\right)
=
\frac{\hbar}{4m^2c^2}
\left( \,
\nabla V(\mathbf{r},t)\!\cdot\!
\left( \nabla \times \bm{\sigma} \right)
-
\bm{\sigma}\!\cdot\!
\left( \nabla \times \nabla V(\mathbf{r},t)  \right)
\,\right) = 0,
\] 
or when taking into account the relativistic kinetic energy
\[
\nabla \cdot \bm{\mathcal{O}}=
c \, \nabla \! \cdot \! \bm{\alpha}=0.	 
\]

By adding all of the Cartesian components \
${ 
	\bm{\mathcal{S}}_{4}(t, \mathbf{R})	
	=
	\bm{\mathcal{S}}_{4, x}(t, \mathbf{R})
	+
	\bm{\mathcal{S}}_{4, y}(t, \mathbf{R})
	+
	\bm{\mathcal{S}}_{4, z}(t, \mathbf{R})	
}$, \
we find the form of the 
${ \bm{\mathcal{S}}_{4}(t, \mathbf{R})	 }$
that is given from
\begin{equation} \label{a22S}
\bm{\mathcal{S}}_{4}(t, \mathbf{R}) 
= 
i\hbar
\oiint_S  
\mathbf{n}\!\cdot\! ( \Psi^{\dagger}\,\bm{\mathcal{O}} ) 
\,  \bm{\nabla_{\mathbf{R}}} \Psi
ds. 
\end{equation}
By then using \ ${(\bm{\mathcal{O}}\,\Psi)^{\dagger}=\Psi^{\dagger}\,\bm{\mathcal{O}}}$ \ in \ Eq.~(\ref{a22S}) \  we restructure it in the symmetrical form 
\begin{equation} \label{a23S}
\bm{\mathcal{S}}_{4}(t, \mathbf{R}) 
 = 
\frac{i \hbar}{2} \oiint_S  
\mathbf{n}\!\cdot\! 
\left(  (\bm{\mathcal{O}}\Psi)^{\dagger} 
\, +  \,
\Psi^{\dagger}\bm{\mathcal{O}}
\right) 
\, \bm{\nabla_{\mathbf{R}}} \Psi
ds. 
\end{equation}

\subsection{Analytic form of  $ \mathbf{S}(t, \mathbf{R}) $ }

In this framework, the non-Hermitian boundary term contribution 
${  \bm{\mathcal{S}}(t, \mathbf{R})  }$
of the DHFT  is evaluated by 
\begin{equation}
 \bm{\mathcal{S}}(t, \mathbf{R})
 =
 \lambda_1 \,
 \left(
 \frac{i \hbar}{2} 
 \oiint_S  
 \mathbf{n} \!\cdot\!
 \left(  
 (\, \frac{\bm{\Pi}}{m} \Psi \,)^{\dagger} 
 \, + \, 
 \Psi^{\dagger}  \frac{\bm{\Pi}}{m} \,
 \right)
 \bm{\nabla_{\mathbf{R}}}
 \Psi 
 \, ds
 \right) 
 \ 
 +
 \
 \lambda_4 \,
 \left(
 \frac{i \hbar}{2} \oiint_S  
 \mathbf{n}\!\cdot\! 
 \left(  (\bm{\mathcal{O}}\Psi)^{\dagger} 
 \, +  \,
 \Psi^{\dagger}\bm{\mathcal{O}}
 \right) 
 \, \bm{\nabla_{\mathbf{R}}} \Psi
 ds
 \right)
\end{equation}
where the $\lambda_i $ accounts for the different energy scales of the electron, thus for different types of Hamiltonians.
By noting that, \
${ \displaystyle
\frac{i}{\hbar}[\, H_1(t, \mathbf{r}, \mathbf{R})  \,,\mathbf{r}\,] 
=
\frac{\bm{\Pi}}{m}
}$ \
as well as \
${ \displaystyle
\frac{i}{\hbar}[\, H_4(t, \mathbf{r}, \mathbf{R})  \,,\mathbf{r}\,] 
=
\bm{\mathcal{O}}
}$,
\  and the fact that 
\ ${  
[\, H_2(t, \mathbf{r}, \mathbf{R})  \,,\mathbf{r}\,]
=
[\, H_3(t, \mathbf{r}, \mathbf{R})  \,,\mathbf{r}\,] 
=
[\, H_5(t, \mathbf{r}, \mathbf{R})  \,,\mathbf{r}\,] 
=
0
}$,
and by also using the definition of the standard velocity operator
 \begin{equation}
\frac{i}{\hbar}\left[ H(t, \mathbf{r}, \mathbf{R}),\mathbf{r} \right]
=
\mathbf{v},
 \end{equation}
we find that  the non-Hermitian boundary contribution 
${  \bm{\mathcal{S}}(t, \mathbf{R}) }$ \
of the DHFT is in general evaluated  by 
\begin{equation} \label{a24S} 
 \bm{\mathcal{S}}(t, \mathbf{R})
=
\frac{i \hbar}{2} \oiint_S 
\mathbf{n}\!\cdot\! 
\left( \,
( \mathbf{v} \, \Psi^{+} 
+ \Psi^{+} \, \mathbf{v}  
\, \right) \! 
\, \bm{\nabla_{\mathbf{R}}} \Psi
\, dS,
\end{equation}
 either for (i) a spinless non-relativistic motion,  (ii) a fully relativistic (Dirac) spinful motion, or a (iii) a spin-full motion in the non-relativistic limit; with the only requirement that one needs to use the appropriate Hamiltonian \ ${ H(t, \mathbf{r}, \mathbf{R}) }$ \ generating  the time-evolution.	 The above final form is indeed the flux (across the 2D boundary) of a generalized current (see main text).

\end{widetext}

\section{DHFT applied to a state with ansatz form 
$\displaystyle \left| \Psi(t, \mathbf{R}) \right\rangle 
=
e^{ 
\displaystyle i \Theta(t, \mathbf{R})}
\left| 
\Phi(t, \mathbf{R})
\right\rangle $.} 
\label{DHFT}

We consider as solution of \thinspace Eq.(\ref{e1h})  \thinspace
a quantum state that has the ansatz form
\begin{equation}  \label{a1} 
 \left| \Psi(t, \mathbf{R}) \right\rangle 
=
e^{ \displaystyle i \Theta(t, \mathbf{R})}
\left| 
\Phi(t, \mathbf{R})
\right\rangle,
\end{equation}
which satisfies the identities 
\begin{eqnarray}   \label{a20}   \nonumber
\displaystyle \left| \bm{\nabla_{\mathbf{R}}}\Psi(t, \mathbf{R})
\right\rangle  
& =  & 
e^{ \displaystyle i \Theta(t, \mathbf{R})}
\, i \, \bm{\nabla_{\mathbf{R}}}
\Theta(t, \mathbf{R})
\left| 
\Phi(t, \mathbf{R})
\right\rangle 
\\*  
&&
+ \,
e^{ \displaystyle  i \Theta(t, \mathbf{R})}
\left| 
\bm{\nabla_{\mathbf{R}}}
\Phi(t, \mathbf{R})
\right\rangle
\end{eqnarray}
and
\begin{eqnarray}   \label{a21} \nonumber 
\displaystyle
\left| 
\frac{\partial\Psi(t, \mathbf{R}) }
{\partial t} 
\right\rangle  
& = & 
e^{ \displaystyle i \Theta(t, \mathbf{R})}
\, i \, 
\frac{\partial \Theta(t, \mathbf{R})}{\partial t}
\left| 
\Phi(t, \mathbf{R})
\right\rangle 
\\*  
&&
+ \,
e^{ \displaystyle  i \Theta(t, \mathbf{R})}
\left| 
\frac{\partial \Phi(t, \mathbf{R}) }{\partial t}
\right\rangle.
\end{eqnarray}
By substituting  Eqs.~ (\ref{a20}) and (\ref{a21})  in all members of \thinspace Eq.~(\ref{e8h}) \thinspace of the main text,  
we find that the DHFT retains its structure form but the involved quantities are now evaluated with respect to the  state $\left| 
\Phi(t, \mathbf{R})
\right\rangle $, namely
\begin{eqnarray}  \label{e14ha}  \nonumber  
\left\langle \bm{O}(t,\mathbf{R}) \right\rangle
&=&
\bm{\nabla_{\mathbf{R}}} E_{\Phi}(t, \mathbf{R})
\, + \,
\bm{\mathcal{S}}_{\Phi}(t, \mathbf{R}) 
\\* [4pt]
&&
- \hbar \,
\frac{\partial \mathbf{R}}{\partial t}
\! \times 
\bm{\mathcal{B}}_{\Phi}(t, \mathbf{R})
\, - \,  \hbar \,
\bm{\mathcal{E}}_{\Phi}(t, \mathbf{R}),
\end{eqnarray}
where
\begin{equation} \label{e14bha}
\left\langle \bm{O}(t,\mathbf{R}) \right\rangle
=
\left\langle{\Phi}(t, \mathbf{R}) \, 
\right|  
\bm{\nabla_{\mathbf{R}}}
H(t, \mathbf{R}) 
\left| \, 
{\Phi}(t, \mathbf{R}) \right\rangle 
\end{equation}
is the observable in quest,
\begin{equation} \label{a9}
E_{\Phi}(t, \mathbf{R})
=
\left\langle \Phi (t,\mathbf{R}) 
\right|  H(t, \mathbf{R}) \left| 
\Phi (t, \mathbf{R})\right\rangle 
\end{equation}
is the \textquotedblleft energy\textquotedblright, whereas
\begin{equation} \label{e15ha}
\bm{\mathcal{B}}_{\Phi} (t, \mathbf{R})
= i
\left\langle \bm{\nabla_{\mathbf{R}}}
{\Phi}(t, \mathbf{R}) \right| 
\times 
\left|  \bm{\nabla_{\mathbf{R}}}
{\Phi}(t, \mathbf{R}) \right\rangle
\end{equation} 
and
\begin{eqnarray}  \label{e16ha}  \nonumber
\bm{\mathcal{E}}_{\Phi} (t, \mathbf{R})
&=&
i
\left\langle \bm{\nabla_{\mathbf{R}}} 
{\Phi}(t, \mathbf{R}) \,  \vert \, 
\frac{\partial {\Phi}(t, \mathbf{R}) }
{ \partial t} \right\rangle  
\\* [2pt]
&&
- i
\left\langle  
\frac{ \partial {\Phi}(t, \mathbf{R}) }
{\partial t} 
\vert \, 
\bm{\nabla_{\mathbf{R}}} {\Phi}(t, \mathbf{R})  \right\rangle
\end{eqnarray}
are the generalized Berry curvatures,
and finally,  the non-Hermitian boundary term is given by
\begin{eqnarray}  \label{e17ha}  \nonumber
\bm{\mathcal{S}}_{\Phi}  (t, \mathbf{R}) 
&=& 
\left\langle \Phi (t, \mathbf{R}) \, \vert \,  
\left( 
H(t, \mathbf{R})^+ 
- 
H(t, \mathbf{R}) 
\right)  
\bm{\nabla_{\mathbf{R}}} 
\Phi (t, \mathbf{R}) \right\rangle 
\\* [6pt] 
&=&
\frac{i \hbar}{2} 
\oiint_S 
\mathbf{n}\!\cdot\! 
\left( \,
( \mathbf{v} \, \Phi)^{+} 
+ \Phi^{+} \, \mathbf{v}  
\, \right) \! 
\, \bm{\nabla_{\mathbf{R}}} \Phi
\, dS,
\end{eqnarray}
where \thinspace ${ \Phi \equiv \Phi(\mathbf{r}, t, \mathbf{R})=
\left\langle \,  \mathbf{r} \, \vert
\, \Phi (t, \mathbf{R}) \right\rangle  }$.

In deriving \thinspace Eqs.~(\ref{e15ha}) and (\ref{e16ha}) we have used the normalization condition \thinspace 
${  \left\langle {\Psi}(t, \mathbf{R}) \vert      
{\Psi}(t, \mathbf{R})
\right\rangle
=
\left\langle {\Phi}(t, \mathbf{R}) \vert      
{\Phi}(t, \mathbf{R})
\right\rangle=1 }$, 
\thinspace 
which upon differentiation gives the relations 
\[
\left\langle \bm{\nabla_{\mathbf{R}}} {\Phi}(t, \mathbf{R}) \vert      
{\Phi}(t, \mathbf{R})
\right\rangle
+
\left\langle {\Phi}(t, \mathbf{R}) \vert      
\bm{\nabla_{\mathbf{R}}} {\Phi}(t, \mathbf{R})
\right\rangle
=0
\]  
and 
\[
\left\langle  \frac{\partial {\Phi}(t, \mathbf{R})}{\partial t} \vert      
{\Phi}(t, \mathbf{R})
\right\rangle
+
\left\langle {\Phi}(t, \mathbf{R}) \vert      
\frac{\partial {\Phi}(t, \mathbf{R})}{\partial t}
\right\rangle
=0
\]  
that have been used.
Similarly, in deriving \thinspace Eq.~(\ref{e17ha}) 
we have used the fact that the states 
\thinspace ${ \left| \Phi(t, \mathbf{R}) \right\rangle  }$ \thinspace
belong within the domain of definition of the Hamiltonian \thinspace 
${ H(t, \mathbf{R}) }$, 
\thinspace 
that is, the assumed states satisfy
${
\displaystyle 
\frac{d}
{dt}
\left\langle \Psi(t, \mathbf{R}) 
\vert \Psi(t, \mathbf{R})
\right\rangle
=0
}$, 
which gives
\thinspace 
${
\left\langle \Phi (t, \mathbf{R}) \, \vert \,  
\left( 
H(t, \mathbf{R})^+ 
- 
H(t, \mathbf{R}) 
\right)  
\Phi (t, \mathbf{R}) \right\rangle 
=0
}$ 
that has been used.

\subsection{Non adiabatic time-evolution: non-equilibrium Floquet state}

We consider a parameter with arbitrary high magnitude of velocity \thinspace  ${ \displaystyle  
\frac{\partial \mathbf{R}(t)}{ \partial t}
}$ \thinspace and a periodic in time nonadiabatic Hamiltonian 
\thinspace
${ H(t+T, \mathbf{R}(t+T))=H(t, \mathbf{R}(t))
}$, 
\thinspace 
where the time of ${ T }$ is the period of driving and the parameter has the same time periodicity \thinspace 
${ \mathbf{R}(t+T)=\mathbf{R}(t) }$.
We assume as a solution of  \thinspace Eq.(\ref{e1h})  \thinspace  a non-equilibrium Floquet state 
\begin{equation}  \label{a2} 
\left| \Psi_a(t, \mathbf{R}) \right\rangle 
=
e^{-  \displaystyle \frac{i}{\hbar}  \varepsilon_{a} \, t}
\left| 
\Phi_a(t, \mathbf{R})
\right\rangle,
\end{equation}
where \thinspace 
${ \left| 
\Phi_a(t, \mathbf{R})
\right\rangle   }$ \thinspace
is the Floquet mode which is periodic in time \thinspace 
${ \left| 
\Phi_a(t, \mathbf{R})
\right\rangle
=
\left| 
\Phi_a(t+T, \mathbf{R})
\right\rangle
}$
and satisfies the eigenvalue equation
\begin{equation}     \label{a4}
\displaystyle  
H_F(t,  \mathbf{R}) \,
\Phi_a(\mathbf{r}, t, \mathbf{k})
=
\varepsilon_a 
\,
\Phi_a(\mathbf{r}, t, \mathbf{k}),
\end{equation}
${ \varepsilon_a }$ \thinspace is the static quasienergy,
and $ H_F(t,  \mathbf{R}) $  is the  Floquet Hamiltonian  given by
\begin{equation}     \label{a5}
H_F(t,  \mathbf{R})
=
H(t,  \mathbf{R})
-i \hbar \frac{d}{dt}.
\end{equation}

Therefore, by comparing
Eq.~(\ref{a2})  with  Eq.~(\ref{a1}), we replace
\thinspace
${
\displaystyle
\Theta(t, \mathbf{R}) 
\mapsto
- \frac{\varepsilon_{a} \, t}{\hbar} 
}$
\thinspace
as well as
\thinspace  
\[
\left| 
\Phi(t, \mathbf{R})
\right\rangle
\mapsto
\left| 
\Phi_a(t, \mathbf{R})
\right\rangle
\]
\thinspace
in all Eqs.(\ref{a20}) \textendash \thinspace (\ref{e17ha}),
\thinspace resulting to the application of the DHFT into a Floquet state that is given by
\begin{eqnarray}  \label{a11}  \nonumber  
&&
\left\langle \Phi_a(t, \mathbf{R}) \, 
\right|  
\bm{\nabla_{\mathbf{R}}}
H(t, \mathbf{R}) 
\left| \, 
\Phi_a(t, \mathbf{R}) \right\rangle 
=
\bm{\nabla_{\mathbf{R}}} E_{a}(t, \mathbf{R})
\\* [4pt]  \nonumber
&&
\ \ \ \ \ \ \ \ \ \ \ \ \ \ \ \
+ \,
\bm{\mathcal{S}}_{a}(t, \mathbf{R}) 
- \hbar \, 
\frac{\partial \mathbf{R}}{\partial t}
\! \times 
\bm{\mathcal{B}}_{a}(t, \mathbf{R})
\, -  \,
\,  \hbar  \, 
\bm{\mathcal{E}}_{a}(t, \mathbf{R}),
\\* [-2pt]
\end{eqnarray}
where all involved quantities are assumed to be evaluated with respect to 
${  \left| \Phi_a(t, \mathbf{R}) \right\rangle }$.

\subsection{Adiabatic time-evolution: equilibrium eigenstate of the Hamiltonian}    \label{ADHFT}

We consider a parameter with velocity that tends to zero 
\thinspace  ${ \displaystyle  
\frac{\partial \mathbf{R}(t)}{ \partial t} \rightarrow 0
}$, \thinspace and a time-dependent Hamiltonian 
\thinspace
${ H(t, \mathbf{R}(t))}$
\thinspace 
that satisfies the criteria for adiabaticity.
We assume as a solution of  \thinspace Eq.(\ref{e1h})  \thinspace  an adiabatically evolved quantum state with initially well defined energy
\begin{equation}  \label{a3} 
	\left| \Psi_n(t, \mathbf{R}) \right\rangle 
=
e^{ \displaystyle i \Theta_n(t, \mathbf{R})}
\left| 
n(t, \mathbf{R})
\right\rangle, 
\end{equation}
where \thinspace 
${ \left| 
n(t, \mathbf{R})
\right\rangle  }$
\thinspace is the instantaneous eigenstate of the Hamiltonian satisfying the eigenvalue equation 
\begin{equation}  \label{a6}
H(t, \mathbf{R}) \left| 
n(t, \mathbf{R})
\right\rangle 
=
E_n(t, \mathbf{R}) \left| 
n(t, \mathbf{R})
\right\rangle, 
\end{equation}
and
\thinspace ${\Theta_n(t, \mathbf{R}) }$ \thinspace is the total (sum of dynamic and geometric) phase of the wavefunction satisfying the equation 
\[ 
\displaystyle -\hbar  \frac{d\Theta_n(t, \mathbf{R})}{dt}
=
E_n(t, \mathbf{R})
- i \hbar 
\left\langle n(t, \mathbf{R}) \vert \,
\frac{dn(t, \mathbf{R})}{dt}
\right\rangle. 
\]

Therefore, by comparing
Eq.~(\ref{a3})  with  Eq.~(\ref{a1}), we replace
\thinspace  
${ \displaystyle
\Theta(t, \mathbf{R}) 
\mapsto
\Theta_n(t, \mathbf{R}) 
}$
\thinspace
as well as
\[
\left| 
\Phi(t, \mathbf{R})
\right\rangle
\mapsto
\left| 
n(t, \mathbf{R})
\right\rangle
\]
in all Eqs.(\ref{a20}) \textendash \thinspace (\ref{e17ha}),
\thinspace 
resulting to the application of the DHFT  into
an adiabatically evolved eigenstate of the Hamiltonian that is given by
 \begin{eqnarray}  \label{a12}  \nonumber  
 &&
 \left\langle n(t, \mathbf{R}) \, 
 \right|  
 \bm{\nabla_{\mathbf{R}}}
 H(\mathbf{R}) 
 \left| \, 
 n(t, \mathbf{R}) \right\rangle 
 =
 \bm{\nabla_{\mathbf{R}}} E_n(t, \mathbf{R})
 \, + \,
 \bm{\mathcal{S}}_n(t, \mathbf{R}) 
 \\* [4pt]  \nonumber
 &&
 \ \ \ \ \ \ \ \ \ \ \ \ \ \ \ \ \ \ \ \ \ \ \ \ \ \ \ \ \ \ \ \ 
 - \hbar \, 
 \frac{\partial \mathbf{R}}{\partial t}
 \! \times 
 \bm{\mathcal{B}}_n(t, \mathbf{R})
 \, - 
 \,  \hbar  \, 
 \bm{\mathcal{E}}_n(t, \mathbf{R}),
 \\* [-2pt]
 \end{eqnarray}
 where all involved quantities are assumed to be evaluated with respect to 
 ${  \left| n(t, \mathbf{R}) \right\rangle }$.
We point out that, when the Hamiltonian does not have explicit time-dependence  ${ H=H(\mathbf{R})}$,  the instantaneous eigenstates  will also not  have explicit time dependence
\thinspace 
${ 
\left| 
n(\mathbf{R})
\right\rangle  }$
\thinspace
according to the eigenvalue equation Eq.~(\ref{a6}), thus,  the Berry curvature 
${  \bm{\mathcal{E}}_n (\mathbf{R})    }$
will have zero value by definition according to  
Eq.~(\ref{e16ha}),  where 
${  \left| 
\Phi(t, \mathbf{R})
\right\rangle
\mapsto
\left| 
n(\mathbf{R})
\right\rangle }$
is assumed.

\subsection{Static Hamiltonian: stationary  eigenstate}  \label{a22}

We consider a static parameter 
\thinspace  ${ \displaystyle  
\frac{\partial \mathbf{R}}{ \partial t}= 0
}$ \thinspace and a static Hamiltonian 
\thinspace
${ H( \mathbf{R})}$.
We assume as a solution of  \thinspace Eq.(\ref{e1h})  \thinspace  
a stationary eigenstate of the Hamiltonian with well defined energy 
\begin{equation}   \label{a7}
\left| \Psi(t, \mathbf{R}) \right\rangle 
=
e^{ \displaystyle i \Theta_n(t, \mathbf{R})}
\left| 
n(\mathbf{R})
\right\rangle, 
\end{equation}
where \thinspace 
${ \left| 
n(\mathbf{R})
\right\rangle  }$
\thinspace is the static eigenstate of the static Hamiltonian satisfying the eigenvalue equation 
\begin{equation}
H(\mathbf{R}) \left| 
n(\mathbf{R})
\right\rangle 
=
E_n(\mathbf{R}) \left| 
n(\mathbf{R})
\right\rangle 
\end{equation}
and
\thinspace ${\Theta_n(t, \mathbf{R}) }$ \thinspace is the  dynamic phase of the wavefunction satisfying the equation 
\[ 
\displaystyle -\hbar  \frac{d\Theta_n(t, \mathbf{R})}{dt}
=
E_n(\mathbf{R}). 
\]

Therefore, by comparing
Eq.~(\ref{a7})  with  Eq.~(\ref{a1}), we replace
\thinspace  
${ \displaystyle
	\Theta(t, \mathbf{R}) 
	\mapsto
	\Theta_n(t, \mathbf{R}) 
}$
\thinspace
as well as
\[
\left| 
\Phi(t, \mathbf{R})
\right\rangle
\mapsto
\left| 
n(\mathbf{R})
\right\rangle
\]
in all Eqs.(\ref{a20}) \textendash \thinspace (\ref{e17ha}),
\thinspace 
resulting to the application of the DHFT  into
a  a stationary eigenstate of the Hamiltonian that is given by
\begin{equation} \label{a8}
\left\langle n(\mathbf{R}) \, 
\right|  
\bm{\nabla_{\mathbf{R}}}
H(\mathbf{R}) 
\left| \, 
n(\mathbf{R}) \right\rangle 
= 
\bm{\nabla_{\mathbf{R}}} E_n(\mathbf{R})
\, + \,
\bm{\mathcal{S}}_n(\mathbf{R}),
\end{equation}
where all involved quantities are assumed to be evaluated with respect to 
${  \left| n( \mathbf{R}) \right\rangle }$.
We note that the Berry curvature 
${  \bm{\mathcal{E}}_n (\mathbf{R})    }$
is zero by definition according to  
Eq.~(\ref{e16ha})  owing to
${  
	\left| 
	\Phi(t, \mathbf{R})
	\right\rangle
	\mapsto
	\left| 
	n(\mathbf{R})
	\right\rangle 
}$
that does not have explicit time dependence,
whereas the Berry curvature 
${  \bm{\mathcal{B}}_n (\mathbf{R})    }$
is not zero but  it does not contribute into the DHFT   \thinspace Eq.~(\ref{e14ha}) due to zero velocity of the parameter 
\thinspace  ${ \displaystyle  
	\frac{\partial \mathbf{R}}{ \partial t}
	= 0  }$.
We point out that Eq.~(\ref{a8}) is the form of the generalized static HF theorem that was derived in Ref. \onlinecite{esteve2010generalization}.

\section{Discrete symmetry considerations}  \label{ab}

It is useful to investigate the discrete symmetries that are satisfied by the curvatures and the non-Hermitian contribution, Eqs.(\ref{e15ha}) \textendash \thinspace (\ref{e17ha}),  that enters into the DHFT \thinspace Eq.(\ref{e14ha}). 
These symmetries gives valuable information for the collective transport quantities when they are calculated by means of the DHFT 
\thinspace Eq.(\ref{e14ha}).
Specifically, we are assuming solutions of 
\thinspace Eq.(\ref{e1h}) \thinspace
that have the ansatz form \thinspace Eq.(\ref{a1})  \thinspace
and we are  deriving discrete symmetries that come out from the eigenvalue equation  \thinspace (\ref{a4})  \thinspace  
for a nonequilibrium Floquet state, and from
\thinspace (\ref{a6})  \thinspace for an adiabatically evolved eigenstate of the Hamiltonian.

\subsection{Instantaneous eigenstates of the Hamiltonian}
\label{ins}

From the symmetries of the instantaneous eigenvalue equation 
\begin{equation}  \label{e17ah}
H(\mathbf{r}, t, \mathbf{R}) \,
\psi_n(\mathbf{r}, t, \mathbf{R})
=
E_n(t, \mathbf{R})
\, 
\psi_n(\mathbf{r}, t, \mathbf{R}),
\end{equation}
\thinspace 
the symmetries of the curvatures themselves can be deduced, which gives valuable information about the topology of the projective Hilbert space of the energy band \thinspace ${ E_n(t, \mathbf{R}) }$.
In the latter eigenvalue equation, the time $t$ and the parameter momentum ${\mathbf{R}}$ are assumed to be parameters. 
We identify the parameter entering into the DHFT as the
wave vector 
\thinspace ${ \mathbf{R} \equiv \mathbf{k} }$, \thinspace  
and we assume that the instantaneous  wavefunction has the ansatz form 
\linebreak
${  \psi_n(\mathbf{r}, t, \mathbf{k})=
e^{\displaystyle i \mathbf{k} \! \cdot \! \mathbf{r} }
u_n(\mathbf{r}, t, \mathbf{k})
}$. \thinspace  For simplicity reasons we assume spinless motion,
that is, the Hamiltonian is a scalar operator and the
wavefunctions \thinspace ${ 
u_n(\mathbf{r}, t, \mathbf{k})
}$ \thinspace  
are scalar quantities. We further assume that the initial Hamiltonian does  not have explicit wave vector (parameter) dependence, rather than the parameter is involved only due to the assumed form of the considered state
${  \psi_n(\mathbf{r}, t, \mathbf{k})=
e^{\displaystyle i \mathbf{k} \! \cdot \! \mathbf{r} }
u_n(\mathbf{r}, t, \mathbf{k}).
}$ 
Therefore, the wavefunction  \thinspace ${ 
\psi_n(\mathbf{r}, t, \mathbf{k})
}$ \thinspace satisfy the eigenvalue equation 
\begin{equation}  \label{e17bh}
H(\mathbf{r}, t) \,
\psi_n(\mathbf{r}, t, \mathbf{k})
=
E_n(t, \mathbf{k})
\, 
\psi_n(\mathbf{r}, t, \mathbf{k})
\end{equation}
and the wavefunction 
\thinspace ${ 
u_n(\mathbf{r}, t, \mathbf{k})
}$ \thinspace  the corresponding
\begin{equation}  \label{e17ch}
e^{\displaystyle - i \mathbf{k} \! \cdot \! \mathbf{r} }
H(\mathbf{r}, t) \,
e^{\displaystyle i \mathbf{k} \! \cdot \! \mathbf{r} }
\,
u_n(\mathbf{r}, t, \mathbf{k})
=
E_n(t, \mathbf{k})
\, 
u_n(\mathbf{r}, t, \mathbf{k}),
\end{equation}
where \thinspace 
 ${ 
 e^{\displaystyle - i \mathbf{k} \! \cdot \! \mathbf{r} }
 H(\mathbf{r}, t) \,
 e^{\displaystyle i \mathbf{k} \! \cdot \! \mathbf{r} }
 =
H_k(\mathbf{r}, t, \mathbf{k}) 
}$.

We further assume adiabatic time-evolution, that is, 
${ 	\left| \Psi_n(t, \mathbf{R}) \right\rangle 
=
e^{ \displaystyle i \Theta_n(t, \mathbf{R})}
\left| 
\psi_n(t, \mathbf{R})
\right\rangle    }$
satisfy the time-dependent Schrödinger  equation 
${ \displaystyle  
i \hbar \frac{d}{dt}\left| \Psi_n(t, \mathbf{R}) \right\rangle
=
H( \mathbf{r}, t) 
\left| \Psi_n(t, \mathbf{R}) \right\rangle,   }$

${  \displaystyle -\hbar  \frac{d\Theta_n(t, \mathbf{k})}{dt}
=
E_n(t, \mathbf{k})
- i \hbar 
\left\langle \psi_n(t, \mathbf{k}) \vert \,
\frac{d \psi_n(t, \mathbf{k})}{dt}
\right\rangle    }$

\subsubsection{Time-reversal symmetry}

A  time-reversal symmetric (spinless) Hamiltonian 
${ H_k(\mathbf{r}, t, \mathbf{k}) }$, 
under the action of the anti-unitary operator \thinspace ${ \mathcal{O}_{\text{T.R.}} }$ \thinspace 
that performs the inversions
\[ 
(i, \mathbf{r}, t, \mathbf{k}) \longmapsto
(-i, \mathbf{r}, -t, -\mathbf{k})
\]
remains invariant, 
\[
\mathcal{O}_{\text{T.R.}}  
\left( 
e^{\displaystyle - i \mathbf{k} \! \cdot \! \mathbf{r} }
\,
H(\mathbf{r}, t) \,
e^{\displaystyle i \mathbf{k} \! \cdot \! \mathbf{r} }
\,
\right) 
\mathcal{O}_{\text{T.R.}} ^{-1}
=
e^{\displaystyle - i \mathbf{k} \! \cdot \! \mathbf{r} }
\,
H(\mathbf{r}, t) \,
e^{\displaystyle i \mathbf{k} \! \cdot \! \mathbf{r} },
\]
that is, ${ H_k(\mathbf{r}, t, \mathbf{k})=H(\mathbf{r}, -t, -\mathbf{k})^{\displaystyle *} }$,
or simply ${ H(\mathbf{r}, t)=H(\mathbf{r}, -t)^{\displaystyle *} }$.
As a result, the spinless eigenfunctions of a time-reversal invariant Hamiltonian satisfy the symmetries
${
\Psi_n(\mathbf{r}, t, \mathbf{k})
=
\Psi_n(\mathbf{r}, -t, -\mathbf{k})^{\displaystyle *}
}$
and 
${
u_n(\mathbf{r}, t, \mathbf{k})
=
u_n(\mathbf{r}, -t, -\mathbf{k})^{\displaystyle *},
}$
up to a 
\thinspace
${ U(1)}$ phase  \thinspace 
${ e^{ \displaystyle i \lambda}
}$ \thinspace which is neglected for simplicity (the curvatures and the non-Hermitian boundary term are invariant with respect to such phases).  
In this respect, the curvatures 
evaluated with respect to 
${ u_n(\mathbf{r}, t, \mathbf{k}) = \left\langle \mathbf{r} | n(t, \mathbf{k}) \right\rangle  }$
transform like 
\begin{eqnarray} \label{e15ah} \nonumber 
\bm{\mathcal{B}}_n (t, \mathbf{k})
&=& 
i
\left\langle \bm{\nabla_{\mathbf{k}}}
u_n(t, \mathbf{k}) \right| 
\times 
\left|  \bm{\nabla_{\mathbf{k}}}
u_n(t, \mathbf{k}) \right\rangle
\\* [6pt] \nonumber
&=&
i
\left\langle \bm{\nabla_{\mathbf{k}}}
u_n(-t, -\mathbf{k})^{\displaystyle *}
\right| 
\times 
\left|  \bm{\nabla_{\mathbf{k}}}
u_n(-t, -\mathbf{k})^{\displaystyle *}
\right\rangle
\\*[6pt]  \nonumber
&=&
-
i
\left\langle \bm{\nabla_{\mathbf{k}}}
u_n(-t, -\mathbf{k})
\right| 
\times 
\left|  \bm{\nabla_{\mathbf{k}}}
u_n(-t, -\mathbf{k})
\right\rangle  
\\* [6pt]   \nonumber  
&=&
-
i
\left\langle \bm{\nabla_{(-\mathbf{k})}}
u_n(-t, -\mathbf{k})
\right| 
\times 
\left|  \bm{\nabla_{(-\mathbf{k})}}
u_n(-t, -\mathbf{k})
\right\rangle
\\* [6pt]
&=&
-
\bm{\mathcal{B}}_n (-t, -\mathbf{k}) 
\end{eqnarray}
and
\begin{eqnarray} \label{e16ah} 
\nonumber 
\bm{\mathcal{E}}_n (t, \mathbf{k})
&=&
i
\left\langle \bm{\nabla_{\mathbf{k}}} 
u_n(t, \mathbf{k}) \,  \vert \, 
\frac{\partial u_n(t, \mathbf{k}) }
{ \partial t} \right\rangle  
\\*[2pt]  \nonumber 
&&
-
i
\left\langle  
\frac{ \partial u_n(t, \mathbf{k}) }
{\partial t} 
\vert \, 
\bm{\nabla_{\mathbf{k}}} u_n (t, \mathbf{k})  \right\rangle
\\*[8pt]  \nonumber 
&=&
i
\left\langle \bm{\nabla_{\mathbf{k}}} 
u_n(-t, -\mathbf{k})^{\displaystyle *} 
\,  \vert \, 
\frac{\partial 
	u_n(-t, -\mathbf{k})^{\displaystyle *} }
{ \partial t} \right\rangle  
\\*[2pt]  \nonumber 
&&
-
i
\left\langle  
\frac{ \partial 
	u_n(-t, -\mathbf{k})^{\displaystyle *} }
{\partial t} 
\vert \, 
\bm{\nabla_{\mathbf{k}}} 
u_n(-t, -\mathbf{k})^{\displaystyle *}  \right\rangle
\\*[8pt]  \nonumber 
&=& \!
- 
i
\left\langle \bm{\nabla_{\mathbf{k}}} 
u_n(-t, -\mathbf{k})
\,  \vert \, 
\frac{\partial 
	u_n(-t, -\mathbf{k}) }
{ \partial t} \right\rangle
\\*[2pt]  \nonumber 
&&
+
i
\left\langle  
\frac{ \partial 
	u_n(-t, -\mathbf{k})}
{\partial t} 
\vert \, 
\bm{\nabla_{\mathbf{k}}} 
u_n(-t, -\mathbf{k}) \right\rangle
\\*[8pt]   \nonumber
&=& \!
-
i
\left\langle \bm{\nabla_{(-\mathbf{k})}} 
u_n(-t, -\mathbf{k})
\,  \vert \, 
\frac{\partial 
	u_n(-t, -\mathbf{k}) }
{ \partial (-t)} \right\rangle
\\*[2pt]  \nonumber 
&&
+
i
\left\langle  
\frac{ \partial 
	u_n(-t, -\mathbf{k})}
{\partial (-t)} 
\vert \, 
\bm{\nabla_{(-\mathbf{k})}} 
u_n(-t, -\mathbf{k}) \right\rangle
\\*[8pt]   
&=&
-
\bm{\mathcal{E}}_n (-t, -\mathbf{k}).
\end{eqnarray}

Similarly, the non-Hermitian term transform like
\begin{eqnarray}   \nonumber   \label{a23qa}
\bm{\mathcal{S}}_n  (t, \mathbf{k}) 
&=&
\left\langle \Psi_n (t, \mathbf{k}) \, \vert \,  
\left( 
H(\mathbf{r}, t)^+ 
- 
H(\mathbf{r}, t) 
\right)  
\bm{\nabla_{\mathbf{k}}} 
\Psi_n (t, \mathbf{k}) \right\rangle 
\\*[8pt]  \nonumber 
&&
\! \!  \!  \!  \!  \!  \!  \! \!  \!  \!  \!  \!  \!  \!  \!  \! \!  \!  \!  \!   \!  \! \! \! \! \! \!
=
\left\langle \Psi_n (-t, -\mathbf{k})^{\displaystyle *}  \vert \
\left(  \!
H(\mathbf{r}, t)^+ 
- 
H(\mathbf{r}, t) 
\right)  \!
\bm{\nabla_{\mathbf{k}}} 
\Psi_n (-t, -\mathbf{k})^{\displaystyle *} \right\rangle 
\\*[8pt]  \nonumber 
&&
\! \!  \!  \!  \!  \!  \!  \! \!  \!  \!  \!  \!  \!  \! \!  \!  \!  \!  \!  \!  \! \!  \!  \! \!  \! \!
=
\left\langle \Psi_n (-t, -\mathbf{k})^{\displaystyle *} \vert 
\left( \!
{H(\mathbf{r}, t)^{\displaystyle *}}^+
\! \! \!  -  \! 
H(\mathbf{r}, t)^{\displaystyle *}  
\right)  \!
\bm{\nabla_{\mathbf{k}}} 
\Psi_n (-t, -\mathbf{k})^{\displaystyle *} \right\rangle 
\\*[8pt]  \nonumber 
&&
\! \!  \!  \!  \!  \!  \!  \! \!  \!  \!  \!  \!  \!  \!  \!  \! \!  \!  \!  \!   \!  \! \! \! \! \! \!
=
\left\langle    
\left( 
H(\mathbf{r}, t)^+
\! \! - \! 
H(\mathbf{r}, t) 
\right)  
\bm{\nabla_{\mathbf{k}}} 
\Psi_n (-t, -\mathbf{k}) 
\, \vert \,
\Psi_n (-t, -\mathbf{k})
\right\rangle 
\\* [8pt]   \nonumber
&&
\! \!  \!  \!  \!  \!  \!  \! \!  \!  \!  \!  \!  \!  \!  \!  \! \!  \!  \!  \!   \!  \! \! \! \! \! \!
=
-
\left\langle    
\left( \!
H(\mathbf{r}, t)^+
\! \! - \! 
H(\mathbf{r}, t) 
\right)  \!
\bm{\nabla_{(-\mathbf{k})}} 
\Psi_n (-t, -\mathbf{k}) 
\, \vert \,
\Psi_n (-t, -\mathbf{k})
\right\rangle 
\\* [8pt]   
&&
\! \!  \!  \!  \!  \!  \!  \! \!  \!  \!  \!  \!  \!  \!  \!  \! \!  \!  \!  \!   \!  \! \! \! \! \! \!
=
- \bm{\mathcal{S}}_n  (-t, -\mathbf{k})^{\displaystyle *}
=
- \bm{\mathcal{S}}_n  (-t, -\mathbf{k}).
\end{eqnarray} 
In the last step of the above equation we have used that 
the non-Hermitian contribution satisfies the relation 
\[
\bm{\mathcal{S}}_n(t, \mathbf{k})
=
\bm{\mathcal{S}}_n(t, \mathbf{k})^{\displaystyle *}
+
\left( 
\left\langle \psi_n 
\right|  
\bm{\nabla_{\mathbf{k}}}
H 
\left.  
\psi_n \right\rangle 
-
\left\langle \bm{\nabla_{\mathbf{k}}}H \,  \psi_n 
\right|  
\left.  \!
\psi_n \right\rangle 
\right), 
\]
thus, the non-Hermitian contribution \thinspace ${  \bm{\mathcal{S}}_n(t, \mathbf{k})  }$  \thinspace evaluated with respect to an instantaneous Bloch state 
\thinspace $ \left| \psi_n(t, \mathbf{k}) \right\rangle $ \thinspace 
is always a real quantity owing to  \thinspace
${ \bm{\nabla_{\mathbf{k}}}H(\mathbf{r}, t)=0  }$. 

All above
time-reversal discrete symmetry transformation properties remain the same in spinful motions, provided that the Hamiltonian satisfies 
\thinspace ${  H(\mathbf{r}, t)=T H(\mathbf{r}, -t) T^{-1}  }$, \thinspace where 
${T= - i K \sigma_{y}= K
\left(\!\begin{array}{cc}
0 & -1  \\  
1 & 0
\end{array}\!\right) 
}$ 
is the so-called time-reversal operator, ${  K }$ is the complex conjugated operator and  ${ \sigma_{y} }$ the Pauli matrice.
For spinful motions 
the cell periodic wavefunction becomes a two-component spinor  
\thinspace 
${ u_n(\mathbf{r}, t, \mathbf{k})
	\equiv  
	\left( 
	a_n (\mathbf{r}, t,\mathbf{k}),
	\ 
	b_n (\mathbf{r}, t,\mathbf{k})
	\right)^{\text{T}} 
}$, \thinspace		
where  ${ a_n (\mathbf{r}, t,\mathbf{k})  }$
and  ${ b_n (\mathbf{r}, t,\mathbf{k})  }$ are cell periodic scalar wavefunctions.
For time-reversal Hamiltonians, the spinor wavefunctions 
satisfy 
\thinspace
${
\Psi_n(\mathbf{r}, t, \mathbf{k})
=
T \Psi_n(\mathbf{r}, -t, -\mathbf{k})
}$
\thinspace
and 
\thinspace
${
u_n(\mathbf{r}, -t, \mathbf{k})
=
T u_n(\mathbf{r}, -t, -\mathbf{k})
}$,
\thinspace that is, the components of the spinor wavefunction transform like
\thinspace
${
	a_n(\mathbf{r}, t, \mathbf{k})
	=
	-b_n(\mathbf{r}, -t, -\mathbf{k})^{\displaystyle *}
}$
\thinspace
and
\thinspace
${
	b_n(\mathbf{r}, t, \mathbf{k})
	=
	a_n(\mathbf{r}, -t, -\mathbf{k})^{\displaystyle *}
}$.
By using the transformation of the spinors' components in the definition of the Berry curvatures, it is straightforward  to verify that 
Eqs.~(\ref{e15ah}) - (\ref{e16ah}) also hold  for spinful electrons under time-reversal symmetry.  Similarly, by using 
\thinspace ${  H(\mathbf{r}, t)=T H(\mathbf{r}, -t) T^{-1}  }$ \thinspace
as well as \thinspace
${
	\Psi_n(\mathbf{r}, t, \mathbf{k})
	=
	T \Psi_n(\mathbf{r}, t, -\mathbf{k})
}$
\thinspace
it is easy to show that \thinspace Eq.~(\ref{a23qa}) \thinspace  is also valid for spinful electrons.

\subsubsection{Space-inversion symmetry}  \label{ami}
In a  space-inversion symmetric spinless Hamiltonian ${ H_k(\mathbf{r}, t, \mathbf{k}) }$, under the action of the unitary operator \thinspace ${ \mathcal{O}_{\text{S.I.}} }$ \thinspace 
that performs the inversions
\[ 
(i, \mathbf{r}, t, \mathbf{k}) \longmapsto
(i, -\mathbf{r}, t, -\mathbf{k})
\]
remains invariant,
\[
\mathcal{O}_{\text{S.I.}} 
\left( 
e^{\displaystyle - i \mathbf{k} \! \cdot \! \mathbf{r} }
\,
H(\mathbf{r}, t) \,
e^{\displaystyle i \mathbf{k} \! \cdot \! \mathbf{r} }
\,
\right) 
\mathcal{O}_{\text{S.I.}}^{-1}
=
e^{\displaystyle - i \mathbf{k} \! \cdot \! \mathbf{r} }
\,
H(\mathbf{r}, t) \,
e^{\displaystyle i \mathbf{k} \! \cdot \! \mathbf{r} },
\]
that is, ${ H_k(\mathbf{r}, t, \mathbf{k})=H(-\mathbf{r}, t, -\mathbf{k}) }$,
or  ${ H(\mathbf{r}, t)=H(-\mathbf{r}, t) }$.
As a result, the eigenfunctions of this space-inversion symmetric Hamiltonian satisfy the symmetries
${
\Psi_n(\mathbf{r}, t, \mathbf{k})
=
\Psi_n(-\mathbf{r}, t, -\mathbf{k})
}$
and 
${ u_n(\mathbf{r}, t, \mathbf{k})
=
u_n(-\mathbf{r}, t, -\mathbf{k})
}$, and  the curvatures transform like
\begin{eqnarray} \label{e15bh} \nonumber 
\bm{\mathcal{B}}_n (t, \mathbf{k})
&=& 
i
\left\langle \bm{\nabla_{\mathbf{k}}}
u_n(t, \mathbf{k}) \right| 
\times 
\left|  \bm{\nabla_{\mathbf{k}}}
u_n(t, \mathbf{k}) \right\rangle
\\*  [6pt]  \nonumber
&=&
i
\left\langle \bm{\nabla_{\mathbf{k}}}
u_n(t, -\mathbf{k})
\right| 
\times 
\left|  \bm{\nabla_{\mathbf{k}}}
u_n(t, -\mathbf{k})
\right\rangle
\\*[6pt]  \nonumber 
&=&
i
\left\langle \bm{\nabla_{(-\mathbf{k})}}
u_n(t, -\mathbf{k})
\right| 
\times 
\left|  \bm{\nabla_{(-\mathbf{k})}}
u_n(t, -\mathbf{k})
\right\rangle
\\*[6pt]  
&=&
\bm{\mathcal{B}}_n (t, -\mathbf{k}) 
\end{eqnarray}
and
\begin{eqnarray} \label{e16bh} 
\nonumber 
\bm{\mathcal{E}}_n (t, \mathbf{k})
&=&
i
\left\langle \bm{\nabla_{\mathbf{k}}} 
u_n(t, \mathbf{k}) \,  \vert \, 
\frac{\partial u_n(t, \mathbf{k}) }
{ \partial t} \right\rangle
\\*[2pt]  \nonumber 
&&
-
i
\left\langle  
\frac{ \partial u_n(t, \mathbf{k}) }
{\partial t} 
\vert \, 
\bm{\nabla_{\mathbf{k}}} u_n (t, \mathbf{k})  \right\rangle
\\*[8pt]  \nonumber 
&=&
i
\left\langle \bm{\nabla_{\mathbf{k}}} 
u_n(t, -\mathbf{k}))
\,  \vert \, 
\frac{\partial 
	u_n(t, -\mathbf{k})) }
{ \partial t} \right\rangle
\\*[2pt]  \nonumber 
&&  
-
i
\left\langle  
\frac{ \partial 
	u_n(t, -\mathbf{k}))}
{\partial t} 
\vert \, 
\bm{\nabla_{\mathbf{k}}} 
u_n(t, -\mathbf{k})) \right\rangle
\\*[8pt]  \nonumber 
&=&
- 
i
\left\langle \bm{\nabla_{(-\mathbf{k})}} 
u_n(t, -\mathbf{k}))
\,  \vert \, 
\frac{\partial 
	u_n(t, -\mathbf{k})) }
{ \partial t} \right\rangle
\\*[2pt]  \nonumber 
&&  
+
i
\left\langle  
\frac{ \partial 
	u_n(t, -\mathbf{k}))}
{\partial t} 
\vert \, 
\bm{\nabla}_{(-\mathbf{k})} 
u_n(t, -\mathbf{k})) \right\rangle
\\*[8pt]   
&=&
-
\bm{\mathcal{E}}_n (t, -\mathbf{k}).
\end{eqnarray}
Analogously, the non-Hermitian term transform like
\begin{eqnarray}   \nonumber   \label{a23}
\bm{\mathcal{S}}_n  (t, \mathbf{k}) 
&=&
\left\langle \Psi_n (t, \mathbf{k}) \, \vert \,  
\left( 
H(\mathbf{r}, t)^+ 
- 
H(\mathbf{r}, t) 
\right)  
\bm{\nabla_{\mathbf{k}}} 
\Psi_n (t, \mathbf{k}) \right\rangle 
\\*[8pt]  \nonumber 
&&
\! \!  \!  \!  \!  \!  \!  \! \!  \!  \!  \!  \!  \!
=
\left\langle \Psi_n (t, -\mathbf{k}) \, \vert \,  
\left( 
H(\mathbf{r}, t)^+ 
- 
H(\mathbf{r}, t) 
\right)  
\bm{\nabla_{\mathbf{k}}} 
\Psi_n (t, -\mathbf{k}) \right\rangle 
\\*[8pt]  \nonumber 
&&
\!  \!  \!  \!  \!  \!  \!  \!  \!  \!  \!  \! \!  \!  \!  \!  \!  \!  \! \!  \!  \!  \!  \!  \!
= 
-
\left\langle \Psi_n (t, -\mathbf{k}) \, \vert \,  
\left( 
H(\mathbf{r}, t)^+ 
- 
H(\mathbf{k}, t) 
\right)  
\bm{\nabla_{(-\mathbf{k})}} 
\Psi_n (t, -\mathbf{k}) \right\rangle
\\* [8pt]  
&&
\! \!  \!  \!  \!  \!  \!  \! \!  \!  \!  \!  \!  \!
=
- \bm{\mathcal{S}}_n  (t, -\mathbf{k}).  
\end{eqnarray}


\subsubsection{Mirror inversion symmetry}   \label{ms}

We assume that the Hamiltonian is invariant under the action of the unitary operator \thinspace ${ \mathcal{O}_{\text{M.I.}} }$ \thinspace
\[
\mathcal{O}_{\text{M.I.}} 
\left( 
e^{\displaystyle - i \mathbf{k} \! \cdot \! \mathbf{r} }
\,
H(\mathbf{r}, t) \,
e^{\displaystyle i \mathbf{k} \! \cdot \! \mathbf{r} }
\,
\right) 
\mathcal{O}_{\text{M.I.}}^{-1}
=
e^{\displaystyle - i \mathbf{k} \! \cdot \! \mathbf{r} }
\,
H(\mathbf{r}, t) \,
e^{\displaystyle i \mathbf{k} \! \cdot \! \mathbf{r} }
\]
that performs the mirror inversions
over the plane \thinspace $ x=0 $ 
\thinspace
\[ 
(i, x, y, z, t, k_x, k_y, k_z) \longmapsto
(i, -x, y, z, t, -k_x, k_y, k_z).
\]
The eigenfunctions satisfy the symmetries
\[
\psi_n(x, y, z, t, k_x, k_y, k_z)
=
\psi_n(-x, y, z, t, -k_x, k_y, k_z)
\]
and 
\[
u_n(x, y, z, t, k_x, k_y, k_z)
=
u_n(-x, y, z, t, -k_x, k_y, k_z)
\]
respectively.
Therefore, the curvatures 
transform like
\begin{eqnarray} \label{e16ch2} \nonumber 
\bm{\mathcal{B}}_n (t, k_x, k_y, k_z)
\cdot 
\bm{e}_x
&=&
\mathcal{B}_{n, \, x} (t, k_x, k_y, k_z)
\\*[8pt]  \nonumber 
&=&
i
\left\langle 
\frac{ \partial
	u_n(t, \mathbf{k})}
{\partial k_y}
\,  \vert \, 
\frac{\partial u_n(t, \mathbf{k}) }
{ \partial k_z} \right\rangle
\\*[2pt]  \nonumber 
&&  
-
i
\left\langle  
\frac{ \partial u_n(t, \mathbf{k}) }
{\partial k_z} 
\vert \, 
\frac{ \partial
u_n(t, \mathbf{k})}
{\partial k_y}
\right\rangle
\\*[8pt]   
&=&
\mathcal{B}_{n, \, x} (t, -k_x, k_y, k_z), 
\end{eqnarray}
and
\begin{eqnarray} \label{e16ch} \nonumber 
\bm{\mathcal{E}}_n (t, k_x, k_y, k_z)
\cdot 
\bm{e}_x
&=&
\mathcal{E}_{n, \, x} (t, k_x, k_y, k_z)
\\*[8pt]  \nonumber 
&=&
i
\left\langle 
\frac{ \partial
	u_n(t, \mathbf{k})}
{\partial k_x}
\,  \vert \, 
\frac{\partial u_n(t, \mathbf{k}) }
{ \partial t} \right\rangle
\\*[2pt]  \nonumber 
&&  
-
i
\left\langle  
\frac{ \partial u_n(t, \mathbf{k}) }
{\partial t} 
\vert \, 
\frac{ \partial
u_n(t, \mathbf{k})}
{\partial k_x}
\right\rangle
\\*[8pt]   
&=&
-
\mathcal{E}_{n, \, x} (t, -k_x, k_y, k_z), 
\end{eqnarray}
and the non-Hermitian term \thinspace ${ \bm{\mathcal{S}}_n (t, \mathbf{k}) }$ \thinspace as
\begin{eqnarray} \label{e16cha} \nonumber 
\bm{\mathcal{S}}_n (t, k_x, k_y, k_z)
\cdot 
\bm{e}_x
&=&
\mathcal{S}_{n, \, x} (t, k_x, k_y, k_z)
\\*[8pt]  \nonumber 
&&
\!  \!  \!  \!  \!  \!  \!  \!  \!  \!  \!  \! \!  \!  \!  \!  \!  \!  \!  \!  \!  \!  \!  \!  \!  
=
\left\langle \Psi_n (t, \mathbf{k}) \, \vert \,  
\left( 
H(\mathbf{r}, t)^+ 
- 
H(\mathbf{r}, t) 
\right)  
\vert \, 
\frac{ \partial
\Psi_n(t, \mathbf{k})}
{\partial k_x}
\right\rangle
\\*[8pt]  \nonumber 
&=&
-
\mathcal{S}_{n, \, x} (t, -k_x, k_y, k_z).
\end{eqnarray}


\subsection{Floquet  eigenstates}
Similar arguments can also be made for  
the Floquet eigenstates.  The Floquet eigenstates
\thinspace  
${ \Phi_{\varepsilon_{a}}(\mathbf{r}, t, \mathbf{k}) 
}$
\thinspace are periodic in time 
\thinspace  
${ \Phi_{\varepsilon_{a}}(\mathbf{r}, t, \mathbf{k}) 
	=
	\Phi_{\varepsilon_{a}}(\mathbf{r}, t+T, \mathbf{k})
}$
\thinspace
and 
satisfy the eigenvalue equation
\begin{eqnarray}  \label{e44abgh}
\left( 
H(\mathbf{r}, t)
-i \hbar \frac{d}{dt}
\right) 
\Phi_{\varepsilon_{a}}(\mathbf{r}, t, \mathbf{k}) 
&=&
H_F(\mathbf{r}, t)
\Phi_{\varepsilon_{a}}(\mathbf{r}, t, \mathbf{k}) 
\nonumber
\\* [2pt]  \nonumber
& =&
\varepsilon_a(\mathbf{k})
\Phi_{\varepsilon_{a}}(\mathbf{r}, t, \mathbf{k}) 
\\*
\end{eqnarray}
where \thinspace ${\varepsilon_a(\mathbf{k}) }$ \thinspace is the 
quasienergy.
We assume that each Floquet state is expressed 
with respect to the Floquet-Bloch state \cite{gomez2013floquet}
\thinspace 
${
\Phi_{\varepsilon_{a}}(\mathbf{r}, t, \mathbf{k})
=
e^{\displaystyle i \mathbf{k} \! \cdot \! \mathbf{r}}
u_a(\mathbf{r}, t, \mathbf{k}),
}$
\thinspace
where \thinspace ${ u_a(\mathbf{r}, t, \mathbf{k}) }$ \thinspace
is periodic both in ${ \mathbf{r} }$ and $t$. 
Accordingly, the wavefunction 
\thinspace ${ 
u_a(\mathbf{r}, t, \mathbf{k})
}$ \thinspace satisfies the instantaneous eigenvalue equation
\begin{equation}  \label{e17cha}
e^{\displaystyle - i \mathbf{k} \! \cdot \! \mathbf{r} }
H_F(\mathbf{r}, t) \,
e^{\displaystyle i \mathbf{k} \! \cdot \! \mathbf{r} }
\,
u_a(\mathbf{r}, t, \mathbf{k})
=
\varepsilon_a(t, \mathbf{k})
\, 
u_a(\mathbf{r}, t, \mathbf{k}),
\end{equation}
which is the counterpart \thinspace Eq.~(\ref{e17ch}).
Therefore, by examining the symmetries that are satisfied by the Hamiltonian \thinspace
${  e^{\displaystyle - i \mathbf{k} \! \cdot \! \mathbf{r} }
H_F(\mathbf{r}, t) \,
e^{\displaystyle i \mathbf{k} \! \cdot \! \mathbf{r} } }$
\thinspace we may infer the symmetries of the 
Floquet-Bloch states, which in turn gives the symmetries of the curvatures themselves that are evaluated with respect to the Floquet-Bloch states.

We can easily find the symmetries by mapping \thinspace Eq.~(\ref{e17ch}) \thinspace to \thinspace Eq.~(\ref{e17cha}) \thinspace with the identification  \thinspace ${  H (\mathbf{r}, t) \mapsto  H_F(\mathbf{r}, t) }$ \thinspace
and  \thinspace ${ u_n(\mathbf{r}, t, \mathbf{k})  \mapsto u_a(\mathbf{r}, t, \mathbf{k}) }$,  \thinspace and this gives for:

\subsubsection{Time-reversal symmetry}
${
\bm{\mathcal{B}}_a (t, \mathbf{k})
=
- \bm{\mathcal{B}}_a (-t, -\mathbf{k})
}$ 
and  \thinspace
${
\bm{\mathcal{E}}_a (t, \mathbf{k})
=
-\bm{\mathcal{E}}_a (-t, -\mathbf{k})
}$,  \thinspace
as well as
\thinspace
${
\bm{\mathcal{S}}_a (t, \mathbf{k})
=
-\bm{\mathcal{S}}_a (-t, -\mathbf{k})
}$.

\subsubsection{Space-inversion symmetry}
${
\bm{\mathcal{B}}_a (t, \mathbf{k})
=
\bm{\mathcal{B}}_a (t, -\mathbf{k})
}$
and \thinspace
${
\bm{\mathcal{E}}_a (t, \mathbf{k})
=
-
\bm{\mathcal{E}}_a (t, -\mathbf{k}),
}$
as well as  \thinspace   \thinspace
${
\bm{\mathcal{S}}_a  (t, \mathbf{k}) 
=
-\bm{\mathcal{S}}_a  (t, -\mathbf{k})
}$.
\\

\subsubsection{Mirror inversion symmetry}
$
\mathcal{B}_{a, \, x} (t, k_x, k_y, k_z)
=
\mathcal{B}_{a, \, x} (t, -k_x, k_y, k_z)
$
and
$
\mathcal{E}_{a, \, x} (t, k_x, k_y, k_z)
=
-
\mathcal{E}_{a, \, x} (t, -k_x, k_y, k_z)
$,
as well as
$
\mathcal{S}_{a, \, x} (t, k_x, k_y, k_z)
=
-
\mathcal{S}_{a, \, x} (t, -k_x, k_y, k_z)
$.

\bibliography{refpola.bib}

\end{document}